\pdfoutput=1
\documentclass[11pt,twoside,letterpaper]{article} 
\usepackage{times,fancyhdr,graphicx}
\usepackage{amsmath}
\usepackage{cite}
\def\snr{SN\,1993J~}
\def\sn1006{SN\,1006~}
\def\sun{\hbox{$\odot$}}
\def\etacar{$\eta$-Carinae~}
\def\aap{A\&A\,  }
\def\aaps{A\&AS  }
\def\aj{AJ  }
\def\apj{ApJ\,  }
\def\apjl{ApJ\,  }
\def\apjs{ApJS  }
\def\iaucirc{IAU circ.  } 
 
\def\mnras{MNRAS\,  }
\def\pasj{PASJ\,  }
\def\pasp{PASP  }

  

\makeatletter
\def\cleardoublepage{\clearpage\if@twoside \ifodd\c@page\else%
    \hbox{}%
    \thispagestyle{empty}%
    \newpage%
    \if@twocolumn\hbox{}\newpage\fi\fi\fi}
\makeatother

\def\figurename{Figure}
\makeatletter
\renewcommand{\fnum@figure}[1]{\figurename~\thefigure.}
\makeatother

\def\tablename{Table}
\makeatletter
\renewcommand{\fnum@table}[1]{\tablename~\thetable.}
\makeatother

\begin{document}
\title{
{\begin{flushleft} \vskip 0.45in
{\normalsize\bfseries\textit{Chapter~1}}
\end{flushleft}
\vskip 0.45in \bfseries\scshape  Interaction of planetary  nebulae
,  Eta-Carinae and supernova remnants  with the Interstellar Medium
}}
\author{\bfseries\itshape  Lorenzo Zaninetti \thanks{Email address: zaninetti@ph.unito.it}\\
Dipartimento di Fisica
           Generale, Via Pietro Giuria 1, 10125 Torino,}
\date{}
\maketitle \thispagestyle{empty} \setcounter{page}{1}

\thispagestyle{fancy} \fancyhead{}
\fancyhead[L]{In: Book Title \\
Editor: Editor Name, pp. {\thepage-\pageref{lastpage-01}}} 
\fancyhead[R]{ISBN 0000000000  \\
\copyright~2007 Nova Science Publishers, Inc.} \fancyfoot{}
\renewcommand{\headrulewidth}{0pt}

\vspace{2in}

\vspace{.08in} \noindent \textbf{Keywords:} Interstellar medium
(ISM) and nebulae in Milky Way , Planetary nebulae,Supernova
remnants

\pagestyle{fancy} \fancyhead{} \fancyhead[EC]{Zaninetti}
\fancyhead[EL,OR]{\thepage} \fancyhead[OC]{Interaction of PN and
SNR with the ISM} \fancyfoot{}
\renewcommand\headrulewidth{0.5pt}

\section*{Abstract}
The image of planetary  nebulae (PN), supernova remnant (SNR) and
Eta-Carinae is made by three different physical processes. The
first process is the expansion of the shell that can be modeled
by the canonical  laws of motion in the spherical case and   by
the momentum conservation when gradients of density are present in
the interstellar medium. The quality of the simulations  is
introduced along one direction as well  along many directions.
 The second process is the
diffusion of particles that radiate from the advancing layer. The
3D diffusion from a sphere , the  1D diffusion with drift and 1D
random walk are analyzed. The third process is the composition of
the image through an integral operation along the line of sight.
The developed framework is applied to three PN which are  A39 ,
the Ring nebula and  the etched hourglass nebula  MyCn 18, the
hybrid object Eta-Carinae , and to two SNR which are SN 1993J and
SN 1006. In all the considered cases a careful  comparison between
the observed and theoretical profiles in intensity is done.

\section{Introduction}

The spherical  explosion in galactic astrophysics 
models two different objects:
\begin{itemize}
\item  
The planetary nebula (PN)  that are characterized by small
terminal  velocities of the order of few $km/s$ 
and  small energy , $E$, released in the expansion ,
$ E \approx 10^{42}$ \mbox{erg}.
\item
The supernova remnant (SNR) that are characterized by high velocities 
$\approx 5000 km/s$ and high energy involved ,
$ E \approx 10^{51}$ \mbox{erg}.
\end{itemize}
These two main classifications  does not cover 
peculiar astrophysical  objects  such as 
the nebula around  \etacar which is characterized  
by high velocity  , $\approx$ 300 $km/s$ and low 
energy involved , 
$ E \approx 10^{39}$ \mbox{erg}.
We now summarize the existing  models  on these
three astrophysical objects.
{\bf PN}
The PN  rarely 
presents   a circular shape generally thought
to be the projection of a sphere on the sky.
In order to explain the properties of  PN,
\cite{Kwok1978} proposed the interacting 
stellar wind (ISW) theory.
Later on \cite{Sabbadin1984}  
proposed the two wind 
model and the two phase model.
More often various types of shapes 
such as elliptical , bipolar or cigar are present,
see~\cite{Balick1987,Schwarz1992,Manchado1996,Guerrero2004,Soker2002a,
Soker2002b}.
The bipolar PNs , for example , are explained by the interaction
of the winds which   originate  from the central star ,
see~\cite{Icke1988, Frank1995, Langer1999, Gonzales2004a}.
Another class of models explains some basic structures 
in PNs through hydrodynamical models, 
see~\cite{Kahn1985,Mellema1991} or 
through  self-organized magnetohydrodynamic (MHD) plasma 
configurations with radial
flow, 
see~\cite{Tsui2008}. 
An attempt 
to make a catalog of line profiles 
 using various shapes  observed in real PNs
was  done by \cite{Morisset2008}.
This ONLINE atlas , available
at 
\newline
 http://132.248.1.102/Atlas$\_$profiles/img/,
is  composed
of  26 photo-ionization models corresponding to 5 geometries,
3 angular density laws and 2 cavity sizes,
 four velocity fields 
for  
a total of  104 PNs,
each of which can be observed from 3 different
directions.
\cite{Matsumoto2006} suggest  that
a planetary nebula is formed and evolves by the interaction
of a fast wind from a central star with 
a slow wind from its progenitor , an  
Asymptotic Giant  Branch (AGB) star.

{\bf \etacar}
The nebula around $\eta$-Carinae  was
discovered by \cite{Thackeray1949}
and  the
name ``the Homunculus'' arises from the fact
that  on the photographic plates it
resembled a small plump man, see \cite{Gaviola1950}.
More details on the various aspects of
 $\eta$-Carinae can be found in \cite{Smith2009}.
The structure  of the Homunculus Nebula
around $\eta$-Carinae has  been  analyzed
with different models,
we cite  some of them:
\begin{itemize}
\item The shape  and kinematics is explained by the interaction
of the winds expelled by the central star at different injection
velocities, see \cite{Icke1988}. 
\item The  possibility that the
nebulae around luminous blue variables (LBVs) are shaped by
interacting winds  has  been analyzed by \cite{Nota1995}. In this
case  a  density contrast profile of the form $\rho$ =
$\rho_0 (1 + 5  \cos^4 \Theta ) $
was used
where $\Theta$ is the angle to
the equatorial plane. \item The origin and evolution of the
bipolar nebula has  been modeled by a  numerical two-dimensional
gasdynamic model
 where  a stellar wind interacts with an
aspherical circumstellar environment, see \cite{Frank1995}. \item
Cooling models form  ballistic
 flows (that is, a
pair of cones each with a spherical base) whose lateral edges
become wrinkled by shear instabilities, see \cite{Dwarkadas1998}.
\item The scaling relations derived from the theory of radiatively
driven winds can model the outflows from luminous blue variable
(LBV) stars, taking account of stellar rotation and the associated
latitudinal variation of the stellar flux due to gravity
darkening. In particular for  a star rotating close to its
critical speed, the decrease in effective gravity near the equator
and the associated decrease in the equatorial wind
 speed results
naturally in a bipolar, prolate interaction front, and therefore
in an asymmetric wind, see  \cite{Dwarkadas2002}.
 \item Two
oppositely ejected jets inflate two lobes (or bubbles)
representing a unified model for the formation of bipolar lobes,
see \cite{Soker2004,Soker2007}. \item A two-dimensional,
time-dependent hydrodynamical simulation of radiative cooling, see
\cite{Gonzales_2004}. \item Launch of  material normal to the
surface of the oblate rotating star with an initial kick velocity
that scales approximately with the local escape speed, see
\cite{Smith_2007}. \item
 A  3D model of wind-wind collision
for  X-ray emission from a supermassive star,
see \cite{Parkin2009}.
\item
Two-dimensional hydrodynamical simulations of
the eruptive events of the 1840s (the great outburst)
and 1890s (the minor outburst), see \cite{Gonzales2010}.
\end{itemize}

{\bf SNR}
The study of
the supernova remnant (SNR)  started  with
\cite{Oort1946}
where an on ongoing  collisional excitation
as a result  of a post-explosion
expansion of the SNR  against the
ambient medium was suggested.
The next six decades
where dedicated
to the deduction of an analytical or numerical
law of expansion.
The target  is  a relationship  for the  instantaneous
radius of expansion, $R$,
of the type
$\propto~ t^m $ where $t$  is time
and $m$ is a parameter that depends on the chosen model.
On adopting  this point of view,  the Sedov expansion
predicts  $R \propto  t^{0.4}$, see \cite{Sedov1959},
and the thin layer approximation  in the
presence of a constant density medium
predicts $R \propto  t^{0.25}$,
see \cite{Dyson1997}.
A  simple approach  to the SNR evolution
in the first $10^4$ yr
assumes an initial  free expansion in  which
$R \propto t$  until the surrounding mass
is of the order of  1 $M_{\sun}$ and a second phase
characterized by the energy conservation in  which according
to the Sedov solution $R \propto t^{2/5}$,
see \cite{Dalgarno1987}.
A third phase characterized by an adiabatic expansion
with $R \propto t^{2/7}$
starts after $10^4$ yr, see \cite{Dalgarno1987}.
A more sophisticated approach given
by \cite{Chevalier1982a,Chevalier1982b}
analyzes  self-similar solutions
with varying inverse power law exponents
for the density profile of the advancing matter,
$R^{-n}$, and ambient  medium,
$R^{-s}$. The  previous assumptions give
a law  of motion
$R \propto  t^{\frac{n-3}{n-s} }$  when
 $n \, > 5$.
Another example  is an analytical solution suggested
by  \cite{Truelove1999} where the radius--time relationship
is regulated by the decrease in density:
as an example, a density  proportional to $R^{-9}$
gives  $R\propto t ^{2/3}$.
With regard to observations,
the radius--time relationship was clarified
when  a decade of
very-long-baseline interferometry (VLBI)
observations of \snr at wavelengths of
3.6, 6, and 18 cm  became available,
see  \cite{Marcaide2009,MartiVidala2011,MartiVidalb2011}.
As a first example, these observations
collected over a 10 year period
can be approximated by  a power law dependence
of the type  $R\,  \propto  t^{0.82}$.
This observational fact  rules  out the
Sedov model and the momentum conservation model.
In this  paper 
we describe  in Section~\ref{sec_objects}
the  observed morphologies of PNs, SNRs and  $\eta-car$.
Section~\ref{sec_motion}  analyzes five  different
laws of motion that model the spherical and aspherical expansion  and 
Section \ref{sec_application_motion} applies the various law  of motion
to well defined astrophysical objects introducing the quality of the
simulation.
Section~\ref{sec_diffusion} reviews old and new formulas
on diffusion ,
Section~\ref{sec_transfer}  
reviews the existing situation
with the radiative transport equation
and Section~\ref{sec_image} contains detailed information
on how to build an image of the astrophysical objects 
here considered.

\section{Astrophysical Objects}
\label{sec_objects}

This section presents the astronomical data of a nearly spherical
PN  known as  A39, a weakly asymmetric PN , the Ring nebula , and
a bipolar PN which is the etched hourglass nebula  MyCn 18 . The
basic data of \etacar which is not classified as a PN due to the
high velocities observed are also reported. The section ends with
two SNR , the symmetric \snr and the weakly asymmetric
\sn1006.

\subsection{A circular  PN , A39}

\begin{figure*}
\begin{center}
\includegraphics[width=10cm]{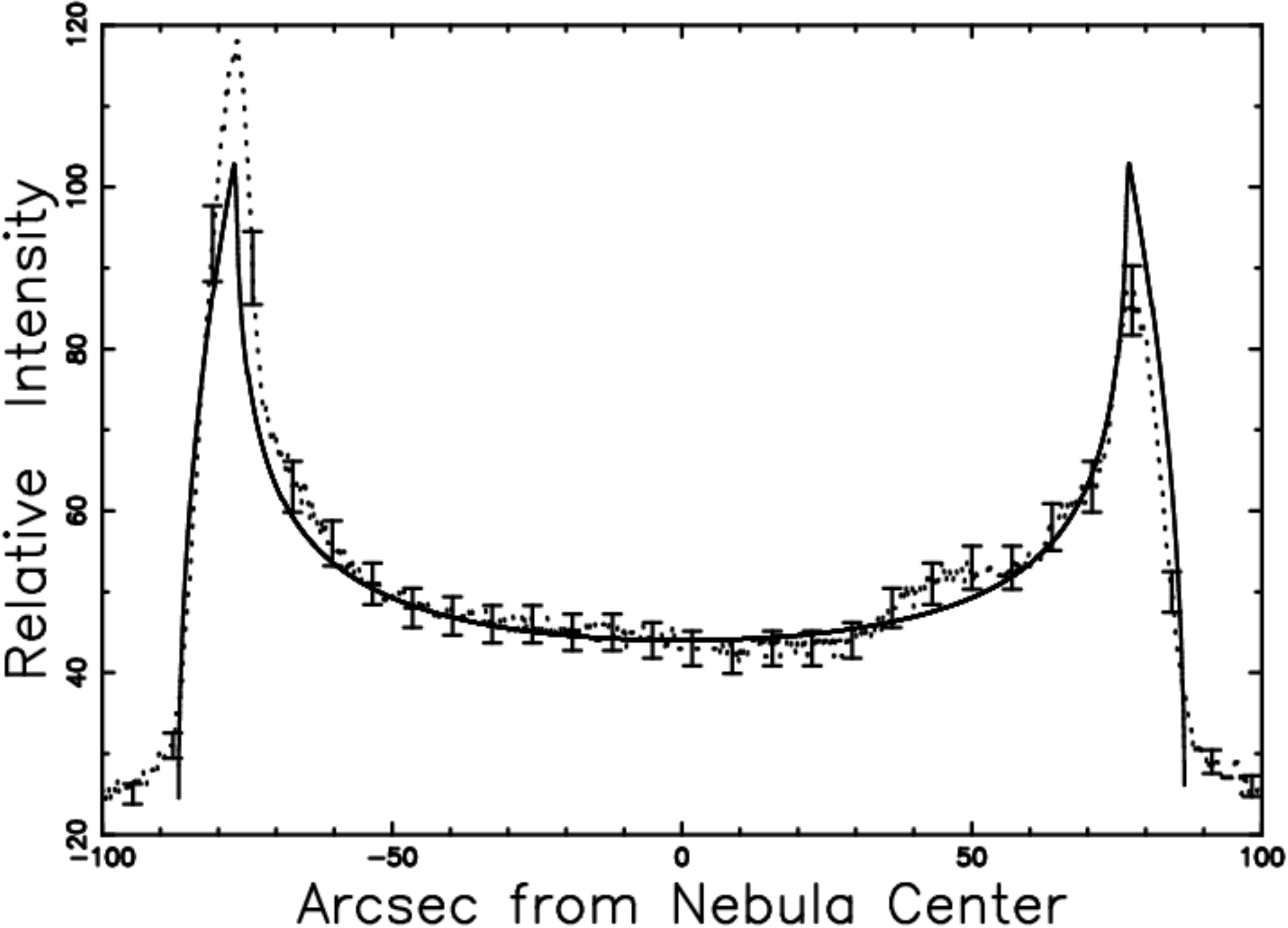}
\end {center}
\caption {
 Cut of the relative intensity of the PN  A39
 crossing the center
 in the east-west direction  (dotted line with some error bar)
 and  the rim model          (full  line)
 fully described in ~Jacoby et al. (2001) .
} \label{cuta39}
    \end{figure*}

The PN A39 is extremely round and therefore can be considered  an
example of spherical symmetry, see for example Figure~1 in
\cite{Jacoby2001} . In A39  the radius  of the shell , $R_{shell}$
is
\begin{equation}
 R_{shell} = 2.42 \times 10^{18} \Theta_{77} D_{21} ~cm
=0.78~pc \quad ,
\end{equation}
where $\Theta_{77}$ is the angular radius in units  of
$77^{\prime\prime}$ and  $D_{21}$ the distance in units of
2.1~kpc ,  see \cite{Jacoby2001}~. The expansion velocity has a
range $[32 \leftrightarrow  37~\frac {km}{s} ]$ according to
\cite{Hippelein1990} and the age  of  the free  expansion is 23000
yr, see \cite{Jacoby2001}. The angular thickness of the shell is
\begin{equation}
\delta\,r_{shell} = 3.17\; 10^{17} \Theta_{10} D_{21} cm =0.103
~pc \quad  ,
\end{equation}
where $\Theta_{10}$ is the thickness  in units  of
$10.1^{\prime\prime}$ and the height above the galactic plane is
1.42 $kpc$ , see \cite{Jacoby2001}. The radial distribution of the
intensity in $[OIII]$ image
 of  A39
after  subtracting  the  contribution of the central star is well
described by a  spherical shell with a $10^{\prime\prime}$  rim
thickness, see Figure \ref{cuta39}  and  \cite{Jacoby2001}.

\subsection{A weakly asymmetric PN , M57}

The Ring nebula , also known as M57 or NGC6720 , presents an
elliptical  shape  characterized by a semi-major axis  of $42
^{\prime\prime}$, a semi-minor  axis  of $29.4 ^{\prime\prime}$
and ellipticity of  0.7, see Table I in \cite {Hiriart2004}. The
distance of the Ring nebula is not very well known ; according to
\cite {Harris1997} the distance is 705 $pc$ . In physical units
the two radii are
\begin{eqnarray}
 R_{shell,minor} = 0.1  \Theta_{29.4} D_{705} ~pc \quad ~~~ semi-minor ~radius
\nonumber \\
 R_{shell,major } = 0.14  \Theta_{42} D_{705} ~pc \quad ~~~ semi-major ~radius
\quad  ,
\end{eqnarray}
where $\Theta_{29.4}$ is the angular minor radius in units  of
$29.4^{\prime\prime}$, $\Theta_{42}$ is the angular major  radius
in units  of $42^{\prime\prime}$  and $D_{705}$ the distance in
units of  705~$pc$. The radial velocity structure in the Ring
Nebula
 was
derived from observations of the $H_2$ (molecular Hydrogen) ~ v =
1- 0 S(1) emission line at 2.122 $\mu m$ obtained by using a
cooled Fabry- Perot etalon and a near-infrared imaging detector ,
see  \cite{Hiriart2004} . The velocity structure of the Ring
Nebula covers  the range $[-30.3 \leftrightarrow 48.8~\frac
{km}{s} ]$~.

\subsection{A strongly asymmetric PN  , MyCn 18  }

MyCn 18 is a PN at a distance of  2.4 $kpc$ and  clearly shows an
hourglass-shaped nebula, see  \cite{Corradi1993,Sahai1999}. On
referring  to Table~1 in~\cite{Dayal2000} we can  fix the
equatorial radius in  $2.80 \times 10^{16}~cm$ , or   $0.09~pc$
, and the radius at $60^{\circ}$ from the equatorial plane $3.16
\times 10^{17}~cm$ or $0.102~pc$~. The determination of the
observed field of velocity of MyCn 18 varies  from an overall
value of 10 $\frac{km}{s}$ as  suggested   by the expansion of
$[OIII]$ , see \cite{Sahai1999} , to a theoretical model by
\cite{Dayal2000} in which the velocity is  9.6 $\frac{km}{s}$ when
the latitude is 0 $^{\circ }  $ (equatorial plane) to 40.9
$\frac{km}{s}$ when the latitude is 60 $^{\circ }  $.

\subsection{Homunculus nebula}

The star \etacar had a great outburst in 1840 and at the moment of
writing presents a bipolar shape called the Homunculus, its
distance is 2250 pc, see \cite{Smith2002}. A more refined
classification distinguishes between the large and little
Homunculus, see \cite{Gonzales_2006}. The Homunculus has been
observed at different
 wavelengths such as the
ultraviolet and infrared  by \cite{Smith_2004,Smith2009}, x-ray by
\cite{Corcoran2004}, $[FeII]\lambda$16435 by \cite{Smith_2005},
ammonia by \cite{Smith_2006}, radio-continuum by
\cite{Gonzales_2006}, near-infrared by \cite{Teodoro_2008} and
scandium and   chromium lines  by \cite{Bautista_2009}.

 Referring to Table 1 in~\cite{Smith2006}, we can
fix the major radius at  22014~AU (0.106~pc) and the equatorial
radius at 2100~AU (0.01~pc). The expansion velocity rises from
$\approx ~ 93~km/s$ at the equator to $\approx ~ 648~km/s$  in the
polar direction, see Table~1 and Figure~4 in~\cite{Smith2006}. The
thickness of the $H_2$ shell is roughly $2-3\%$  of the  polar
radius, see \cite{Smith2006}.

\subsection{A circular SNR, \snr }

\begin{figure*}
\begin{center}
\includegraphics[width=10cm]{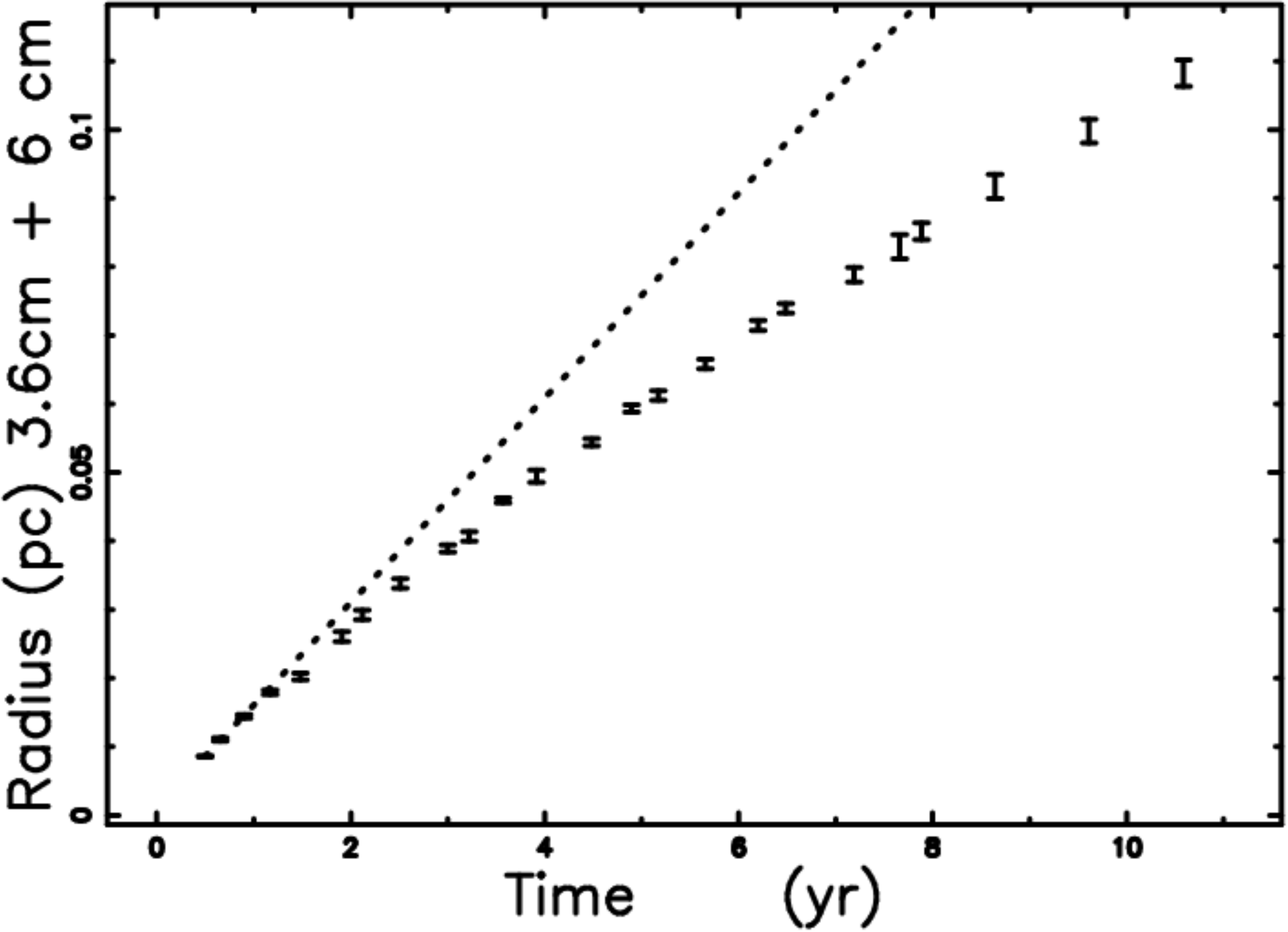}
\end {center}
\caption { Radius in  pc versus year of the SNR \snr with vertical
error bars. The em bands  are  $\lambda=3.6 \, cm $ and $\lambda=6
\, cm $. The data are extracted from Table 1 in Marcaide et al.
2009. The dotted line represents an expansion at a constant
velocity. } \label{1993pc}
    \end{figure*}

The supernova \snr started to be visible in M81 in 1993, see
\cite{Ripero1993}, and presented a circular symmetry for 4000
days, see \cite{Marcaide2009}. Its distance is 3.63~Mpc (the same
as M81), see \cite{Freedman1994}. The expansion of \snr  has been
monitored in various bands over a decade  and Figure  \ref{1993pc}
reports its temporal evolution.

The observed instantaneous velocity decreases from
$v=15437~\frac{km}{s}$ at $t=0.052$\ yr  to $v=8474~ \frac{km}{s}$
at $t=10.53$\ yr. We briefly recall that \cite{Fransson2005} quote
an inner velocity from  the shapes of the lines of  $\approx 7000
\frac{km}{s}$ and an outer  velocity of $\approx 10000
\frac{km}{s}$.

\subsection{A weakly asymmetric  SNR, \sn1006 }

The diameter of the known remnants  spans the range from  3~pc to
60~pc  and    attention is fixed on \sn1006 which started to be
visible in 1006 AD
  and its possible
diameter of 12.7 \mbox{pc}, see \cite{Strom} . More precisely, on
referring to the radio--map of \sn1006  at 1370 \mbox{MHz} by
\cite {Reynolds1986}, it   can be observed that the radius is
greatest in the north--east direction.  From the radio-map
previously mentioned we can  extract the following observed radii,
$R=6.8~pc$ in the polar  direction and $R= 5.89~pc$ in the
equatorial direction.
 Information on the
thickness of emitting layers is contained in  a recent study by
\cite{Bamba2003}
  which  analyzes
Chandra observations (i.e., synchrotron X-rays) from \sn1006. The
observations found that  sources of non-thermal radiation are
likely to be thin sheets with a  thickness of about 0.04 pc
upstream  and 0.2 pc downstream of the surface of maximum
emission, which coincide with the locations of Balmer-line optical
emission , see \cite{Ellison1994} .
The high resolution XMM-Newton Reflection Grating Spectrometer (RGS)
spectrum of SN 1006 gives
two  solutions for the O VII triplet.
One gives  a shell velocity of ~ 6500 $km/s$
  and the second one a shell velocity of 9500 $km/s$, see
  \cite{Vink2005}.
The value here adopted for the
magnetic field can be $H=10\mu~Gauss$   as suggested by
 \cite{Bamba2003}.

\section{Law of motion}
\label{sec_motion} 
This Section presents three  solutions for the law
of motion that describe  a symmetric expansion. The momentum
conservation is then applied in  cases where  the density of the
interstellar medium is not constant but regulated by  exponential
or power law behavior.

\subsection{Spherical Symmetry - Power law solution}
\label{sphericalpowerlaw}

The equation of   the expansion  of the SNR
can  be modeled by a power law  of the type
\begin{equation}
R(t) = R_0 (\frac{t}{t_0})^{\alpha}
\label{rpower}
\quad ,
\end{equation}
where
$R$ is the radius  of the expansion,
$t$ is the time  ,
$R_0$ is the radius  at  $t=t_0$
and  $\alpha$ an exponent which  can be found from the
numerical analysis.
In the case  of  \snr
we have  $\alpha$ =0.828 , $R_0$=0.0087 pc and
$t_0$ = 0.498 yr

\subsection{Spherical Symmetry - Sedov solution}

The momentum conservation is applied  to a conical  section of
radius  $R$ with  a solid   angle   $\Delta\;\Omega$, in polar
coordinates, see   ~\cite{Dalgarno1987}
\begin {equation}
\frac {d}{dt} (\Delta M  R)  = \Delta  F \,,
\end {equation}
where
\begin {equation}
\Delta M  =  \int_0^R \rho   (R,\theta,\phi) dV \,,
\end {equation}
is the mass  of  swept--up  interstellar   medium in the  solid
angle  $\Delta\;\Omega$, $\rho$  the density  of the medium , $P$
the interior pressure
  and the   driving force:
\begin {equation}
\Delta \; F =  PR^2 \Delta \Omega \,.
\end  {equation}
After some algebra the Sedov  solution    is obtained,
see~\cite{Sedov1959,Dalgarno1987}
\begin{equation}
R(t)= \left ({\frac {25}{4}}\,{\frac {{\it E}\,{t}^{2}}{\pi
\,\rho}} \right )^{1/5} \quad , \label{sedov}
\end{equation}
where $E$ is the  energy injected in the process and  $t$ the
time.

Another slightly different solution is formula~(7.56) in
\cite{Dyson1997}
\begin{equation}
R(t)= \left ({\frac {25}{3}}\,{\frac {{\it E}\,{t}^{2}}{\pi
\,\rho}} \right )^{1/5} \quad , \label{sedovdue}
\end{equation}
where the difference is due to the adopted approximations.

Our astrophysical  units are: time ($t_4$), which is expressed  in
$10^4$ \mbox{yr} units; $E_{42}$, the  energy in  $10^{42}$
\mbox{erg}; and    $n_0$  the number density  expressed  in
particles~$\mathrm{cm}^{-3}$~ (density~$\rho=n_0$m, where
m=1.4$m_{\mathrm {H}}$). With these   units equation~(\ref{sedov})
becomes
\begin{equation}
R(t) \approx  0.198 \left ({\frac {{\it E_{42}}\,{t_4}^{2}}{n_0}}
\right)^{1/5}~pc \quad . \label{sedovastro}
\end{equation}
The expansion velocity is
\begin{equation}
V(t) = \frac {2}{5} \frac {R(t)}{t} \label{velocity} \quad ,
\end{equation}
which  expressed in astrophysical units is
\begin{equation}
V (t) \approx 7.746 \,{\frac {\sqrt [5]{{\it E_{42}}}}{\sqrt
[5]{{\it n_0}}{{\it t_4}}^{3/5}}}~  \frac{km}{s} \quad .
\label{velocityastro}
\end {equation}
Equations~(\ref{sedovastro})  and (\ref{velocityastro}) represent
a system of two equations in two unknowns : $t_4$ and  $E_{42}$ .
By inserting for example  $R=0.78~pc$ in
equation~(\ref{sedovastro}) we find
\begin{equation}
t_4= 77.15\,{\frac {1}{\sqrt {{\it E_{42}}}}} \label{t4equation}
\quad  ,
\end{equation}
and    inserting $V=35~km\;s^{-1}$ in
equation~(\ref{velocityastro}) we obtain
\begin{equation}
0.3954\,\sqrt {{\it E_{42}}}=35 \quad .
\end{equation}
The previous equation is solved for $E_{42} = 7833.4$ that
according to equation~(\ref{t4equation}) means $t_4$=.87173. These
two parameters allows a rough evaluation of the mechanical
luminosity $L=\frac{E}{t}$ that turns out to be $L\approx 2.847
\;10^{34} ergs\;s^{-1} $. This value should be  bigger than  the
observed luminosities in the various bands. As an example the
X-ray  luminosity of PNs , $L_X$, in the wavelength band 5-28~\AA~
has a range $[10^{30.9} \leftrightarrow  10^{31.2} ergs\;s^{-1}
]$ , see Table 3 in \cite{Steffen2008}.

Due to the fact  that is difficult to compute the volume  in an
asymmetric expansion the Sedov solution is adopted only in this
paragraph.


\subsection{Spherical Symmetry - Momentum Conservation}

The thin layer approximation assumes that all the swept-up gas
accumulates infinitely in a thin shell just after the shock front.
The conservation of the radial momentum requires that
\begin{equation}
\frac{4}{3} \pi R^3 \rho \dot {R} = M_0 \quad ,
\end{equation}
where $R$ and $\dot{R}$   are  the radius and the velocity of the
advancing shock , $\rho$ the density of the ambient medium , $M_0$
the momentum evaluated at $t=t_0$ , $R_0$ the initial radius and
$\dot {R_0}$  the  initial velocity , see
\cite{Dyson1997,Padmanabhan_II_2001}. The law of motion is
\begin{equation}
R = R_0 \left  ( 1 +4 \frac{\dot {R_0}} {R_0}(t-t_0) \right
)^{\frac{1}{4}} \label{radiusm} \quad .
\end{equation}
and the velocity
\begin{equation}
\dot {R} = \dot {R_0} \left ( 1 +4 \frac{\dot {R_0}}
{R_0}(t-t_0)\right )^{-\frac{3}{4}} \label{velocitym} \quad .
\end{equation}
From equation (\ref{radiusm}) we can extract $\dot {R_0}$ and
insert it in equation (\ref{velocitym})
\begin{equation}
\dot {R} =\frac{1}{4(t-t_0)}  \frac{R^4-R_0^4}{R_0^3} \left  (
1+\frac{R^4-R_0^4}{R_0^4} \right )^{-\frac{3}{4}}
\label{velocitym2} \quad .
\end{equation}
The astrophysical  units are:  $t_4$  and $t_{0,4}$ which  are $t$
and  $t_0$ expressed  in $10^4$ \mbox{yr} units, $R_{pc}$ and
$R_{0,pc}$ which are $R$ and  $R_0$  expressed in  $pc$, $\dot
{R}_{kms}$ and $\dot {R}_{0,kms}$ which are
 $\dot{R} $ and  $\dot{R}_0$   expressed
in $\frac{km}{s}$. Therefore the previous formula becomes
\begin{equation}
\dot {R}_{kms} =24.49 \frac{1}{(t_4-t_{0,4} )}
\frac{R_{pc}^4-R_{0,pc}^4} {R_{0,pc} ^3} \left  (
1+\frac{R_{pc}^4-R_{0,pc}^4}{R_{0,pc}^4} \right )^{-\frac{3}{4}}
\label{velocitym2astro} \quad .
\end{equation}
On introducing   $R_{0,pc}=0.1$ , $R_{pc}=0.78$ , $\dot {R}_{kms}
=  34.5  \frac{km}{s}$ , the approximated age of A39  is  found to
be $t_4-t_{0,4}=50 $ and $\dot {R}_{0,kms}  =  181.2 $.

\subsection{Asymmetry - Exponential medium}

Given the Cartesian   coordinate system $(x,y,z)$ , the plane
$z=0$ will be called equatorial plane and in  polar coordinates
$z= R \sin ( \theta) $,
 where
$\theta$ is the polar angle and $R$ the distance from the origin .
The presence of a non homogeneous medium in which the expansion
takes place can be modeled assuming an exponential behavior for
the number of particles of the type
\begin{equation}
n (z) = n_0 \exp {- \frac {z}{h} } \quad = n_0 \exp {- \frac
{R\times \sin (\theta) }{h} } \quad  ,
\end{equation}
where  $R$ is the radius of the shell , $n_0$ is the number of
particles at $R=R_0$ and $h$ the scale. The 3D expansion will be
characterized by the following properties
\begin {itemize}
\item Dependence of the momentary radius of the shell
      on  the polar angle $\theta$ that has a range
      $[-90 ^{\circ}  \leftrightarrow  +90 ^{\circ} ]$.

\item Independence of the momentary radius of the shell
      from  $\phi$ , the azimuthal  angle  in the x-y  plane,
      that has a range
      $[0 ^{\circ}  \leftrightarrow  360 ^{\circ} ]$.
\end {itemize}
The mass swept, $M$,  along the solid angle $ \Delta\;\Omega $,
between 0 and $R$ is
\begin{equation}
M(R)= \frac { \Delta\;\Omega } {3}  m_H n_0 I_m(R) + \frac{4}{3}
\pi R_0^3 n_0 m_H \quad  ,
\end {equation}
where
\begin{equation}
I_m(R)  = \int_{R_0} ^R r^2 \exp { - \frac {r \sin (\theta) }{ h}
} dr \quad ,
\end{equation}
where $R_0$ is the initial radius and $m_H$ the mass of the
hydrogen  . The integral is
\begin{eqnarray}
I_m(R)  = \frac { h \left( 2\,{h}^{2}+2\,R_0h\sin \left( \theta
\right) +{R_0}^{2} \left( \sin \left( \theta \right)  \right) ^{2}
\right) {{\rm e}^{-{\frac {R_0 \sin \left( \theta \right) }{h}}}}
} { \left( \sin \left( \theta \right)  \right) ^{3} }
           \nonumber\\
- \frac { h \left( 2\,{h}^{2}+2\,Rh\sin \left( \theta \right)
+{R}^{2} \left( \sin \left( \theta \right)  \right) ^{2} \right)
{{\rm e}^{-{\frac {R \sin \left( \theta \right) }{h}}}} } { \left(
\sin \left( \theta \right)  \right) ^{3} } \quad .
\end{eqnarray}
The conservation of the momentum gives
\begin{equation}
M(R)   \dot {R}= M(R_0) \dot {R_0} \quad  ,
\end{equation}
where $\dot {R}$  is the  velocity at $R$ and $\dot {R_0}$  the
initial velocity at $R=R_0$.

In this differential equation of the first order in $R$ the
variable can be separated and the integration term by term gives
\begin{equation}
\int_{R_0}^{R}  M(r)  dr = M(R_0) \dot {R_0} \times ( t-t_0) \quad
,
\end{equation}
where  $t$ is the time and $t_0$ the time at $R_0$. The resulting
non linear equation ${\mathcal{F}}_{NL}$ expressed in
astrophysical units is
\begin{eqnarray}
{\mathcal{F}}_{NL} = - 6\,{{\rm e}^{- \,{\frac {{\it
R_{0,pc}}\,\sin \left( \theta
 \right) }{{\it h_{pc}}}}}}{{\it h_{pc}}}^{4}
- \,{\it h_{pc}}\,{{\rm e}^{-
 \,{\frac {{\it R_{0,pc}}\,\sin \left( \theta \right) }{{\it h_{pc}}}}}}
 \left( \sin \left( \theta \right)  \right) ^{3}{{\it R_{0,pc}}}^{3}
\nonumber \\
- 6 \,{{\it h_{pc}}}^{3}{{\rm e}^{- \,{\frac {{\it R_{0,pc}}\,\sin
\left( \theta \right) }{{\it h_{pc}}}}}}\sin \left( \theta \right)
{\it R_{0,pc}}-
 3\,{{\it h_{pc}}}^{2}{{\rm e}^{- \,{\frac {{\it R_{0,pc}}\,\sin \left(
\theta \right) }{{\it h_{pc}}}}}} \left( \sin \left( \theta
\right)
 \right) ^{2}{{\it R_{0,pc}}}^{2}
\nonumber \\
- \,{{\it R_{0,pc}}}^{4} \left( \sin
 \left( \theta \right)  \right) ^{4}+ 6\,{{\rm e}^{- \,{\frac {{
\it R_{pc}}\,\sin \left( \theta \right) }{{\it h_{pc}}}}}}{{\it
h_{pc}}}^{4}+
 4\,{{\rm e}^{- \,{\frac {{\it R_{pc}}\,\sin \left( \theta \right) }
{{\it h_{pc}}}}}}{{\it h_{pc}}}^{3}{\it R_{pc}}\,\sin \left(
\theta \right)
\nonumber  \\
+{ {\rm e}^{- \,{\frac {{\it R_{pc}}\,\sin \left( \theta \right)
}{{\it h_{pc}}}}}}{{\it h_{pc}}}^{2}{{\it R_{pc}}}^{2} \left( \sin
\left( \theta
 \right)  \right) ^{2}
\nonumber  \\
+ 2\,{{\rm e}^{- \,{\frac {{\it R_{0,pc}}\,\sin
 \left( \theta \right) }{{\it h_{pc}}}}}}{{\it h_{pc}}}^{3}{\it R_{pc}}\,\sin
 \left( \theta \right) + 2\,{{\rm e}^{- \,{\frac {{\it R_{0,pc}}\,
\sin \left( \theta \right) }{{\it h_{pc}}}}}}{{\it
h_{pc}}}^{2}{\it R_{pc}}\,
 \left( \sin \left( \theta \right)  \right) ^{2}{\it R_{0,pc}}
\nonumber \\
+{{\rm e}^{-
 \,{\frac {{\it R_{0,pc}}\,\sin \left( \theta \right) }{{\it h_{pc}}}}}}{
\it h_{pc}}\,{\it R_{pc}}\, \left( \sin \left( \theta \right)
\right) ^{3}{ {\it R_{0,pc}}}^{2}
\nonumber \\
+ \left( \sin \left( \theta \right)  \right) ^{4}{{\it
R_{0,pc}}}^{3}{\it R_{pc}}- 0.01 \left( \sin \left( \theta
 \right)  \right) ^{4}{{\it R_{0,pc}}}^{3}{ \dot{R}_{0,kms}}\,
\left( t_4-t_{0,4}
 \right)
=0 \quad  ,
\end{eqnarray}
where   $t_4$  and $t_{0,4}$
 are $t$ and  $t_0$
expressed  in $10^4$ \mbox{yr} units, $R_{pc}$ and $R_{0,pc}$  are
$R$ and  $R_0$  expressed in  $pc$, $\dot {R}_{kms}$ and $\dot
{R}_{0,kms}$ are $\dot{R} $ and  $\dot{R}_0$   expressed in
$\frac{km}{s}$, $\theta$ is expressed in radians and  $h_{pc}$ is
the  the scale , $h$  , expressed in $pc$. It is not possible to
find  $R_{pc}$   analytically  and a numerical method   should be
implemented. In  our case in order to find  the root of
${\mathcal{F}}_{NL}$, the FORTRAN SUBROUTINE  ZRIDDR from
\cite{press} has been used.

The unknown parameter $t_4-t_{0,4}$  can be found from different
runs  of the code once
 $R_{0,pc}$ is fixed as $\approx$ 1/10 of the observed equatorial
radius , $\dot {R}_{0,kms}$ is  200  or less and $h_{pc} \approx
2\times R_{0,pc}$.

\subsection{Asymmetry - Power law  medium}

A  possible form for a power law profile of the medium surrounding
the Homunculus nebula is
\begin{equation}
n (z) = n_0  \bigl ( \frac{z}{R_0} \bigr )^{-\alpha}
\quad  ,
\label{powerlaw}
\end{equation}
where  $z=R\times \sin ( \theta) $ is the distance from the equatorial plane,
$R$ is the instantaneous radius of expansion,
$n_0$ is the number
of particles at $R=R_0$, $R_0$ is the scale
and $\alpha$ is a coefficient $>0$.

The  swept   mass, $M$,  along the solid angle $ \Delta\;\Omega $
between 0 and $R$ is
\begin{equation}
M(R)=
\frac { \Delta\;\Omega } {3} 1.4 \,  m_H n_0 I_m(R)
+ \frac{4}{3} \pi R_0^3 n_0 m_H \,1.4
\quad  ,
\end {equation}
where
\begin{equation}
I_m(R)  = \int_{R_0} ^R r^2
\bigl ( \frac {r \sin (\theta) }{ R_0} \bigr )^{-\alpha}   dr
\quad ,
\end{equation}
where $R_0$ is the initial radius
and $m_H$ is the mass of hydrogen.
The integral is
\begin{eqnarray}
I_m(R)  =
\frac
{
{R}^{3} \left( {\frac {R\sin \left( \theta \right) }{{\it R0}}}
 \right) ^{-\alpha}
} { 3-\alpha } \quad .
\end{eqnarray}
Conservation of momentum gives
\begin{equation}
M(R)   \dot {R}=
M(R_0) \dot {R_0}
\quad,
\end{equation}
where $\dot {R}$  is the  velocity
at $R$ and
$\dot {R_0}$ is the initial velocity at $R=R_0$
,
$M(R)$ and $M(R_0)$
are the swept masses at $R$ and $R_0$ respectively

In this first-order differential equation in $R$, the
variables can be separated. Integration term-by-term gives
\begin{equation}
\int_{R_0}^{R}  M(r)  dr =
M(R_0) \dot {R_0} \times ( t-t_0)
\quad  ,
\end{equation}
where  $t$ is the time and $t_0$ is the time at $R_0$.
The resulting non-linear equation ${\mathcal{F}}_{NL}$
expressed in astrophysical units
is
\begin{eqnarray}
{\mathcal{F}}_{NL} =
- \,{{\it R_{0,pc}}}^{4}{\alpha}^{2}+{\it R_{pc}}\,{{\it R_{0,pc}}}^{3}{\alpha
}^{2}+{\it R_{pc}}\,{{\it R_{0,pc}}}^{3} \left( \sin \left( \theta \right)
 \right) ^{- \,\alpha}\alpha
\nonumber \\
+ 7.0\,{{\it R_{0,pc}}}^{4}\alpha- \,{{
\it R_{0,pc}}}^{4} \left( \sin \left( \theta \right)  \right) ^{- \,
\alpha}\alpha- 7.0\,{\it R_{pc}}\,{{\it R_{0,pc}}}^{3}\alpha
\nonumber  \\
+ 3\,{{\it R_{0,pc}
}}^{4} \left( \sin \left( \theta \right)  \right) ^{- \,\alpha}-
 12\,{{\it R_{0,pc}}}^{4}- 4.0\,{\it R_{pc}}\,{{\it R_{0,pc}}}^{3} \left( \sin
 \left( \theta \right)  \right) ^{- \,\alpha}
\nonumber \\
+ \,{{\it R_{pc}}}^{-
 \,\alpha+ 4.0}{{\it R_{0,pc}}}^{\alpha} \left( \sin \left( \theta
 \right)  \right) ^{- \,\alpha}+ 12\,{\it R_{pc}}\,{{\it R_{0,pc}}}^{3}
\nonumber \\ -
 0.122\,{{\it R_{0,pc}}}^{3}{\it { \dot{R}_{0,kms}} }\, \left
( t_4-{\it t_{0,4}}
 \right)
- 0.01\,{{\it R_{0,pc}}}^{3}{\it { \dot{R}_{0,kms}} }\,
\left( t_4-{
\it t_{0,4}} \right) {\alpha}^{2}
\nonumber  \\
+ 0.0714\,{{\it R_{0,pc}}}^{3}{\it
{ \dot{R}_{0,kms}} }\, \left( t_4-{\it t_{0,4}} \right) \alpha
=0
\quad ,
\label{nl_power}
\end{eqnarray}
where $t_4$
and
$t_{0,4}$
 are $t$ and  $t_0$
expressed in $10^4$ \mbox{yr} units,
$R_{pc}$ and $R_{0,pc}$  are
$R$ and  $R_0$  expressed in  $pc$,
$\dot {R}_{kms}$
and
$\dot {R}_{0,kms}$
are
$\dot{R} $ and  $\dot{R}_0$   expressed
in $\frac{km}{s}$ and
$\theta$ is expressed in radians.

It is not possible to find $R_{pc}$ analytically and
a numerical method must be implemented.
In  our case, in order
to find  the root of  ${\mathcal{F}}_{NL}$,
the FORTRAN SUBROUTINE ZRIDDR from \cite{press} has been used.
The unknown parameters,
$R_{0,pc}$  and
$\dot {R}_{0,kms}$, are found from different runs
of the code, $t_4-t_{0,4}$ is an input parameter.

\section{Applications of the law of motion}
\label{sec_application_motion}
From a practical point  of view,
$\epsilon$ ,
the percentage  of
reliability  of our code can also be  introduced,
\begin{equation}
\epsilon  =(1- \frac{\vert( R_{\mathrm {pc,obs}}- R_{pc,\mathrm {num}}) \vert}
{R_{pc,\mathrm {obs}}}) \cdot 100
\,,
\label{efficiency}
\end{equation}
where $R_{pc,\mathrm {obs}}$ is the   radius as given
by the astronomical observations in parsec ,
and  $R_{pc,\mathrm {num}}$ the radius  obtained from our  simulation
in parsec.

In order to  test the  simulation over different angles,  an
observational percentage  of
reliability
,$\epsilon_{\mathrm {obs}}$,
is  introduced which  uses
both the size and the shape,
\begin{equation}
\epsilon_{\mathrm {obs}}  =100(1-\frac{\sum_j |R_{pc,\mathrm {obs}}-R_{pc,\mathrm {num}}|_j}{\sum_j
{R_{pc,\mathrm {obs}}}_{,j}})
,
\label{efficiencymany}
\end{equation}
where
the  index $j$  varies  from 1 to the number of
available observations.

\subsection{Simulation of a  PN , Ring nebula}

A typical set of parameters that allows
us  to simulate
the Ring nebula is reported in
Table~\ref{parameters}.

\begin{table}
      \caption{Data of the simulation of the PN Ring nebula }
         \label{parameters}
      \[
         \begin{array}{cc}
            \hline
            \noalign{\smallskip}
\mbox {Initial ~expansion~velocity~,${\dot R}_{{0,kms}}$  } & 200             \\
\mbox {Age~($t_4-t_{0,4}$)  }                     & 0.12 \\
\mbox {Initial~radius~ $R_{0,pc}$  }                        & 0.035 \\
\mbox {scaling~ $h_{pc}$    }                                & \mbox {$2\times R_{0,pc}$}\\
            \noalign{\smallskip}
            \hline
         \end{array}
      \]
   \end{table}

\begin{figure}
  \begin{center}
\includegraphics[width=10cm]{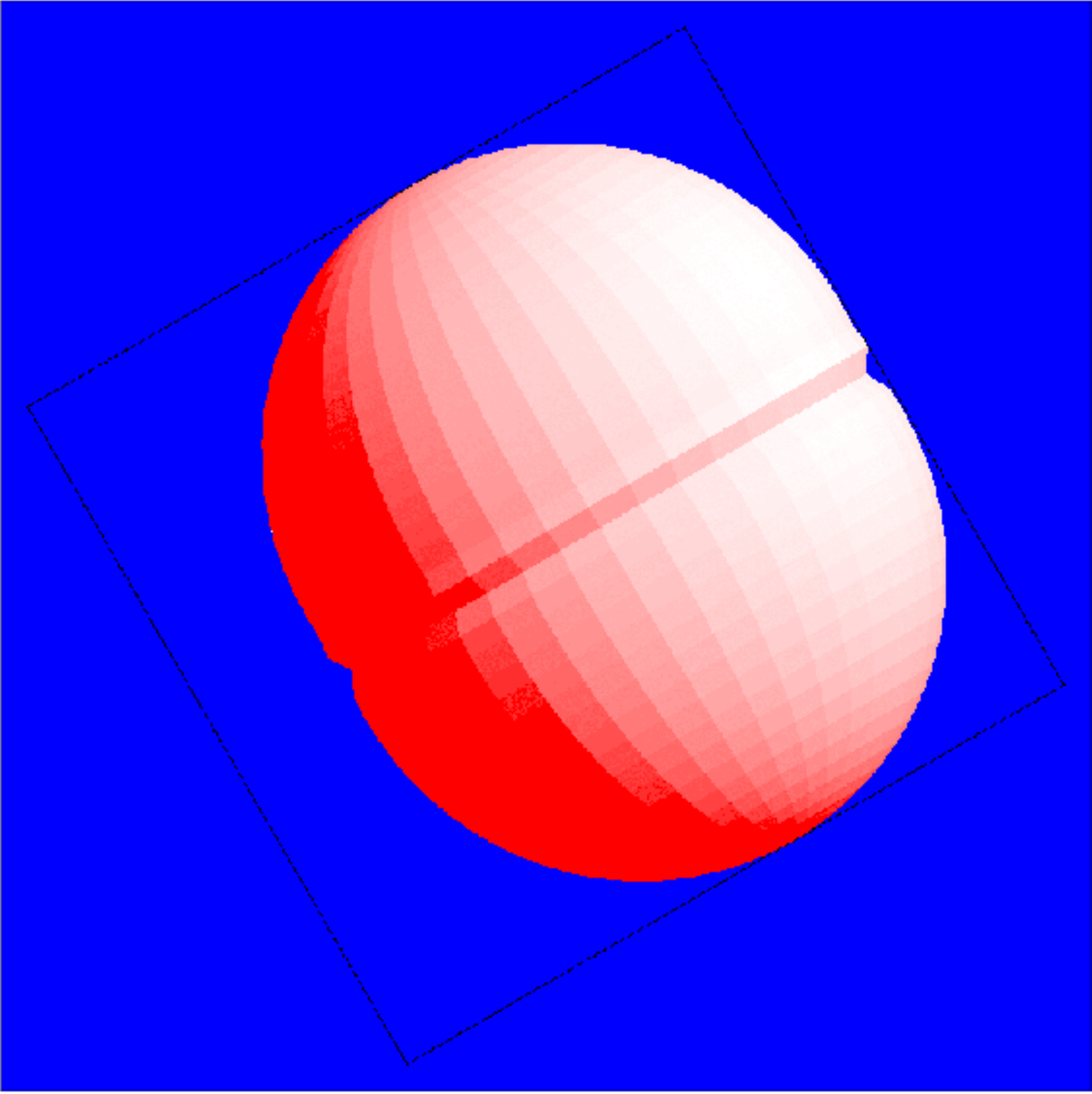}
  \end {center}
\caption { Continuous  three-dimensional surface of the PN Ring
nebula : the three Eulerian angles characterizing the point of
view are
     $ \Phi $=180    $^{\circ }  $,
     $ \Theta $=90   $^{\circ }$
and  $ \Psi $=-30    $^{\circ }   $.
Physical parameters as in Table~\ref{parameters}.
          }%
    \label{ring_faces}
    \end{figure}

\begin{figure}
  \begin{center}
\includegraphics[width=10cm]{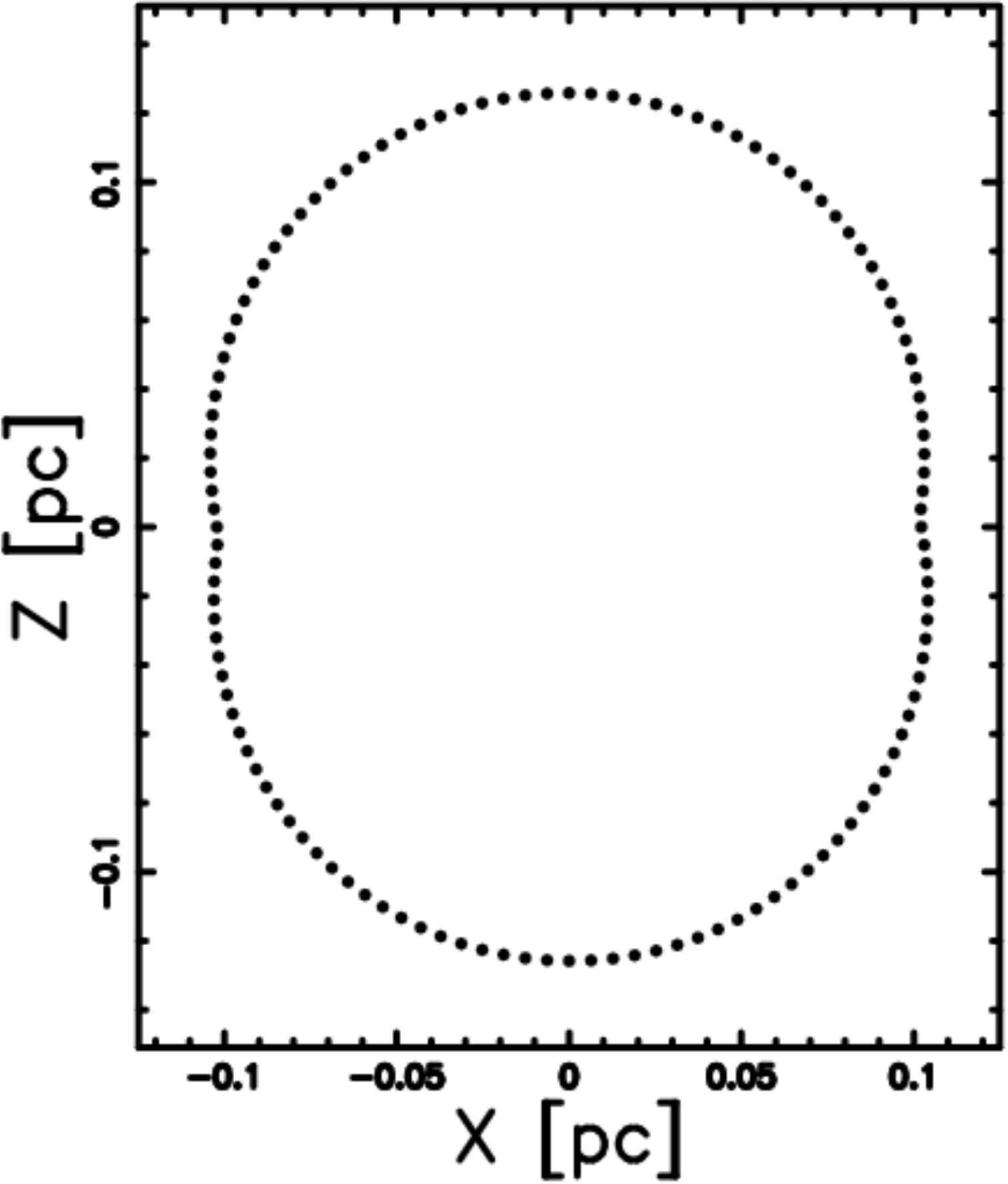}
  \end {center}
\caption { Section of the PN Ring nebula  on the {\it x-z}  plane.
The horizontal and vertical axis are in $pc$. Physical parameters
as in Table~\ref{parameters}.
          }%
    \label{ring_cut}
    \end{figure}

\begin{figure}
  \begin{center}
\includegraphics[width=10cm]{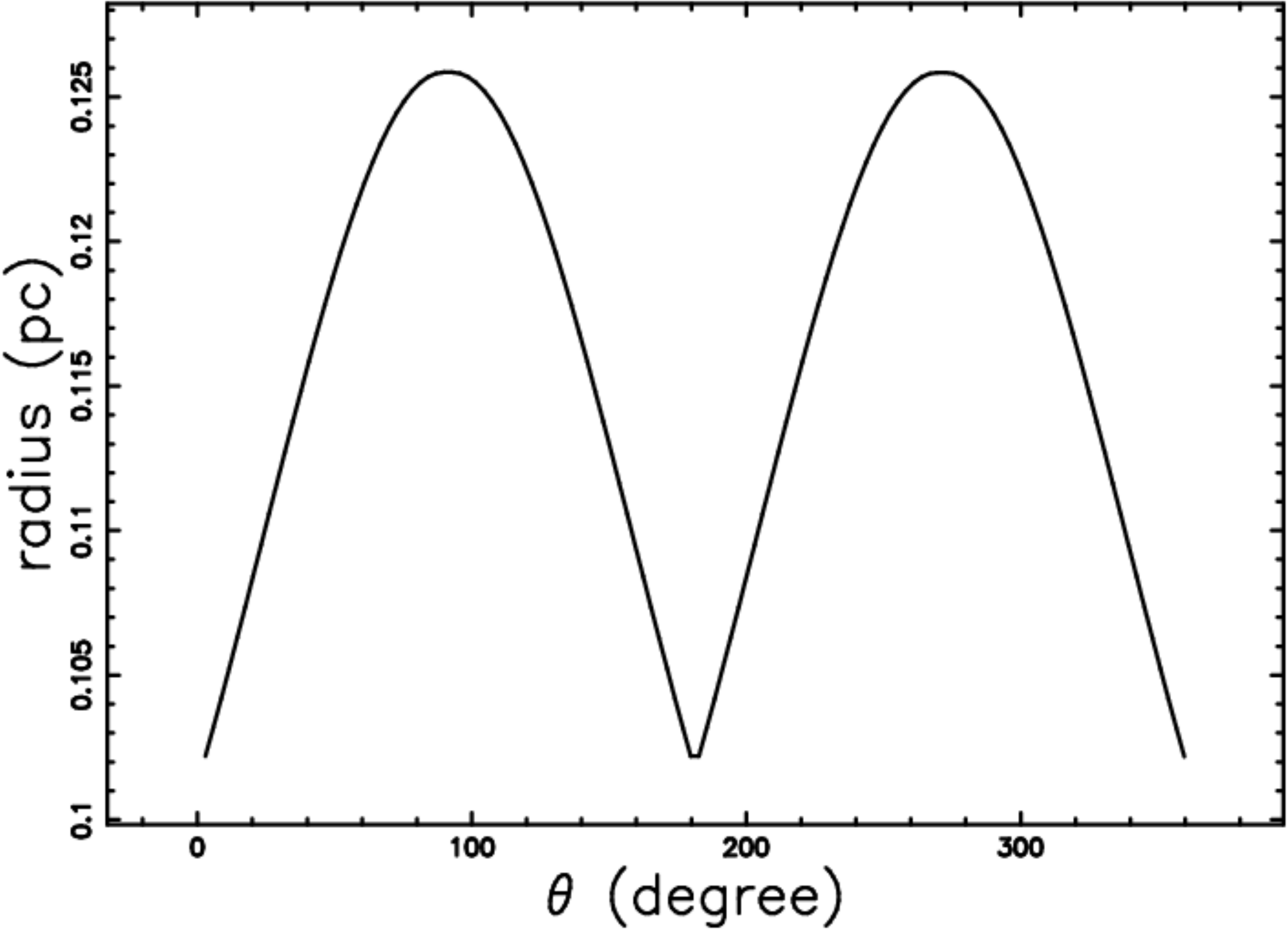}
  \end {center}
\caption { Radius in $pc$  of the PN Ring nebula as a function of
the position angle in degrees. Physical parameters as in
Table~\ref{parameters}.
          }%
    \label{ring_radius}
    \end{figure}

\begin{figure}
  \begin{center}
\includegraphics[width=10cm]{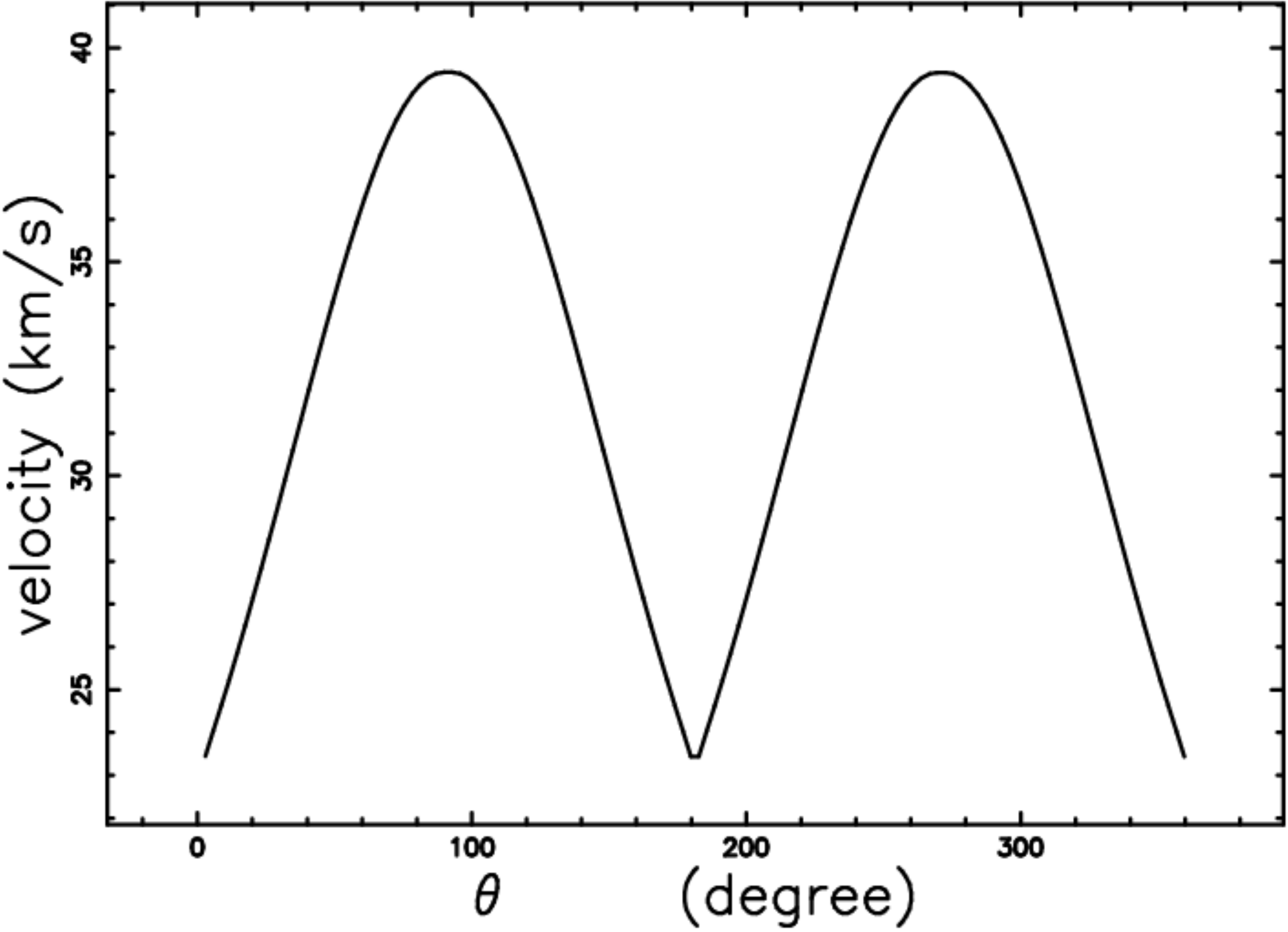}
  \end {center}
\caption { Velocity  in $\frac{km}{s}$  of the PN Ring nebula as a
function of the position angle in degrees. Physical parameters as
in Table~\ref{parameters}.
          }%
    \label{ring_velocity}
    \end{figure}

\begin{figure}
  \begin{center}
\includegraphics[width=10cm]{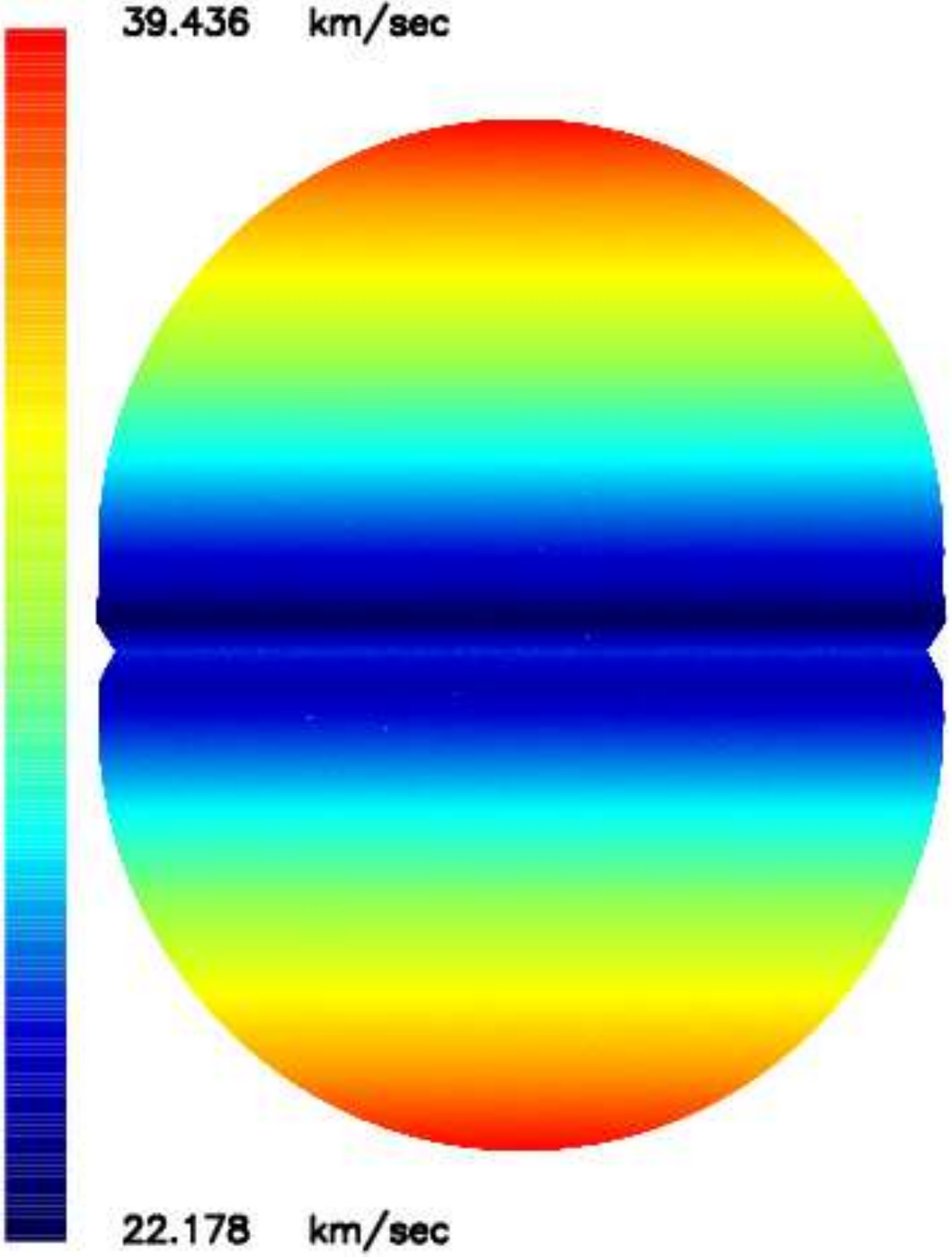}
  \end {center}
\caption {
  Map of the expansion velocity   in $\frac{km}{s}$
  relative to the simulation of the PN Ring nebula
  when 300000 random points are selected on the surface.
  Physical parameters as in Table~\ref{parameters}.
          }%
    \label{ring_velocity_field}
    \end{figure}

The complex 3D behavior of the advancing Ring nebula  is reported
in Figure~\ref{ring_faces} and Figure~\ref{ring_cut}
reports the asymmetric expansion in  a  section crossing the center.
In order to better visualize the asymmetries
Figure~\ref{ring_radius} and  Figure~\ref{ring_velocity}
report the radius  and the velocity
as a function of the position angle $\theta$.
The combined effect of spatial asymmetry and field of velocity
are reported in Figure \ref{ring_velocity_field}.

The efficiency  of our code in  reproducing
  the observed radii
as given by formula~(\ref{efficiency} )
and  the efficiency when the age is five time greater
are  reported in
Table~\ref{tab:rad}.

An analogous formula  allows us to compute the efficiency
in the computation of the maximum velocity ,
see Table~\ref{tab:vel}.

\begin{table}
      \caption{Reliability of the radii of the PN Ring nebula.}
         \label{tab:rad}
      \[
         \begin{array}{lccc}
            \hline
            \noalign{\smallskip}
~~~~     & R_{\mathrm{up}}(\mathrm{pc})  ~polar~direction  &
           R_{\mathrm{eq}}  (\mathrm{pc})~equatorial~plane  \\
            \noalign{\smallskip}
            \hline
            \noalign{\smallskip}
R_{\mathrm {obs}}                       &  0.14    & 0.1        \\
R_{\mathrm {num}}   (\mbox{our~code})   &  0.125   & 0.102
        \\
\mbox {$\epsilon$} (\%)      &  89   &   97       \\
\mbox {$\epsilon$}~(\%)~for~a~time~5~times~greater       &  27   &   41       \\
            \noalign{\smallskip}
            \hline
         \end{array}
      \]
   \end{table}

\begin{table}
      \caption{Reliability of the velocity
          of the PN Ring nebula
        }
         \label{tab:vel}
      \[
         \begin{array}{lc}
            \hline
            \noalign{\smallskip}
~~~~     & V  ({\frac {km}{s}  })  ~maximum~velocity\\
            \noalign{\smallskip}
            \hline
            \noalign{\smallskip}
V_{\mathrm {obs}}          &  48.79       \\
V_{\mathrm {num}}          &  39.43       \\
\mbox {$\epsilon$} (\%)    &  80.81       \\
\mbox {$\epsilon$} (\%) ~(\%)~for~a~time~5~times~greater   &  35.67      \\
            \noalign{\smallskip}
            \hline
         \end{array}
      \]
   \end{table}

\subsection{Simulation of PN ,  MyCn 18  }

A typical set of parameters that allows us  to simulate
  MyCn 18    is reported in
Table~\ref{parametersmycn18}.

\begin{table}
      \caption{Data of the simulation of the PN  MyCn 18 }
         \label{parametersmycn18}
      \[
         \begin{array}{cc}
            \hline
            \noalign{\smallskip}
\mbox {Initial ~expansion~velocity~,${\dot R}_{{0,kms}}$
 [km~s$^{-1}$}] &  200             \\
\mbox {Age~($t_4-t_{0,4}$) [10$^4$~yr]}                    & 0.2   \\
\mbox {Initial~radius~ $R_{0,pc}$ ~[pc] }                       & 0.001 \\
\mbox {scaling~ h     [pc] }                               & \mbox{$1.0\times R_0$} \\
            \noalign{\smallskip}
            \hline
         \end{array}
      \]
   \end{table}

\begin{figure}
  \begin{center}
\includegraphics[width=10cm]{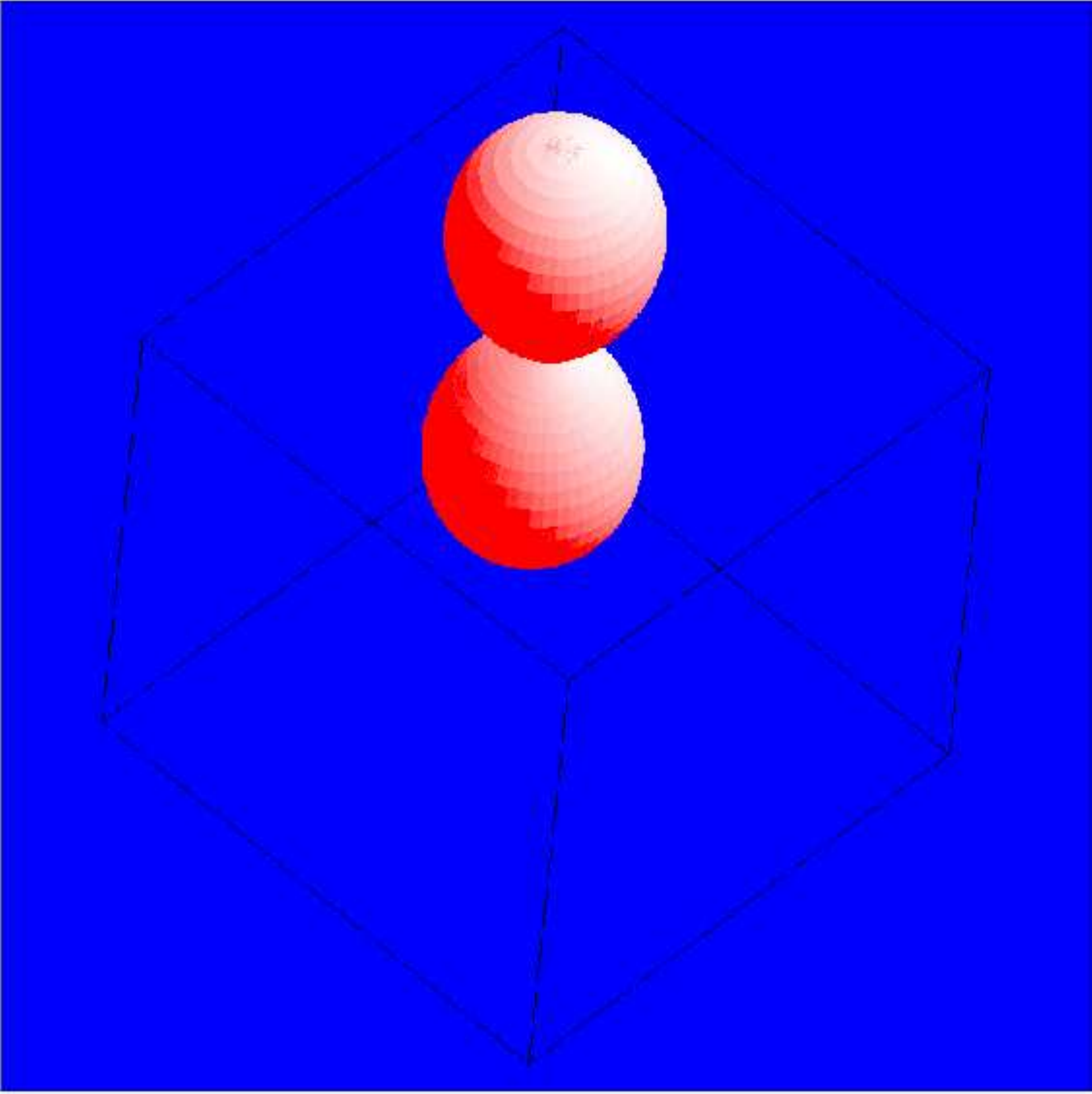}
  \end {center}
\caption { Continuous  three-dimensional surface of the PN MyCn 18
: the three Eulerian angles characterizing the point of view are
     $ \Phi   $=130     $^{\circ }  $,
     $ \Theta $=40   $^{\circ }$
and  $ \Psi   $=5     $^{\circ }   $.
Physical parameters as in Table~\ref{parametersmycn18}.
          }%
    \label{mycn18_faces}
    \end{figure}

\begin{figure}
  \begin{center}
\includegraphics[width=4cm]{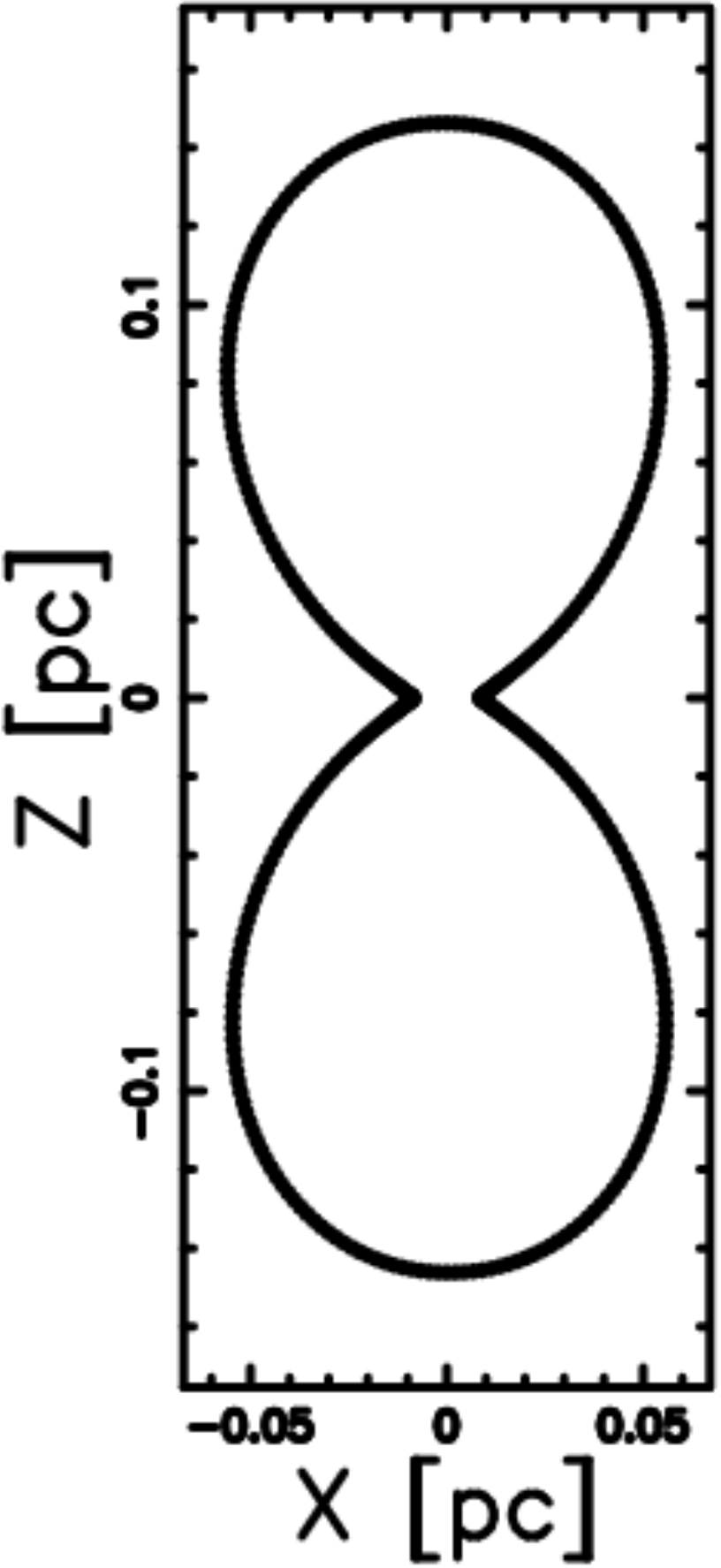}
  \end {center}
\caption { Section of the PN MyCn 18  on the {\it x-z}  plane.
Physical parameters as in Table~\ref{parametersmycn18}.
          }%
    \label{mycn18_cut}
    \end{figure}

\begin{figure}
  \begin{center}
\includegraphics[width=10cm]{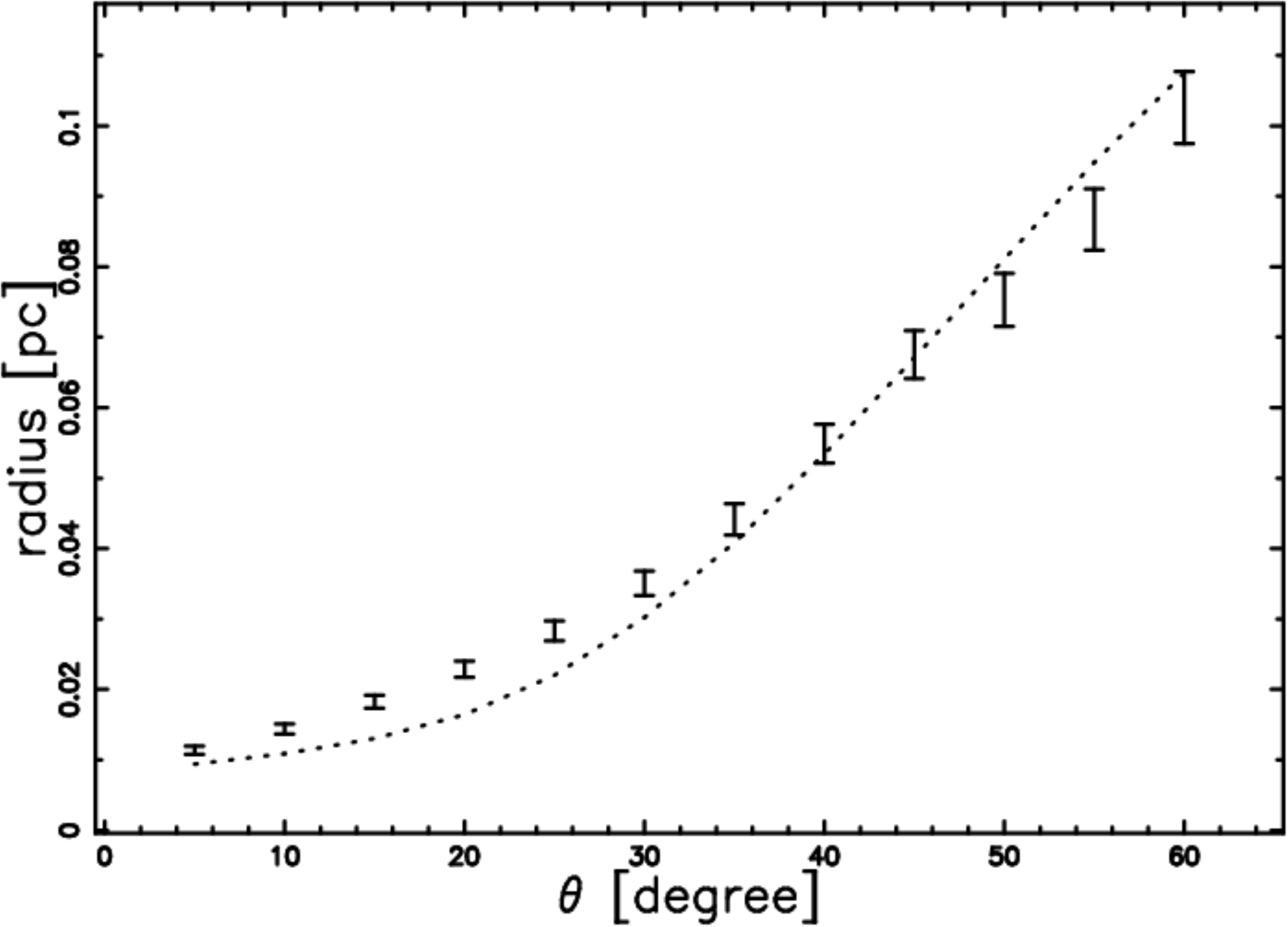}
  \end {center}
\caption { Radius in pc  of  the PN MyCn 18 as a function of
latitude from  $0^{\circ}$  to $60^{\circ}$ ( dotted line) when
the physical parameters are those of   Table~\ref{parametersmycn18}.
The points with error bar (1/10 of the value) represent the data
of Table 1 in  Dayal et al. 2000.
          }%
    \label{mycn18_radius}
    \end{figure}

The bipolar  behavior of the advancing
MyCn 18
 is reported
in Figure~\ref{mycn18_faces} and Figure~\ref{mycn18_cut}
reports the expansion in  a section crossing the center.
It is   interesting to point out  the  similarities
between our Figure~\ref{mycn18_cut} of
MyCn 18
and Figure~1 in  \cite{Morisset2008}
which define the parameters $a$ and $h$
of  the Atlas of synthetic line profiles.
In order to better visualize the two lobes
Figure~\ref{mycn18_radius}
reports  the radius
as a function of the position angle $\theta$.

\begin{figure}
  \begin{center}
\includegraphics[width=4cm]{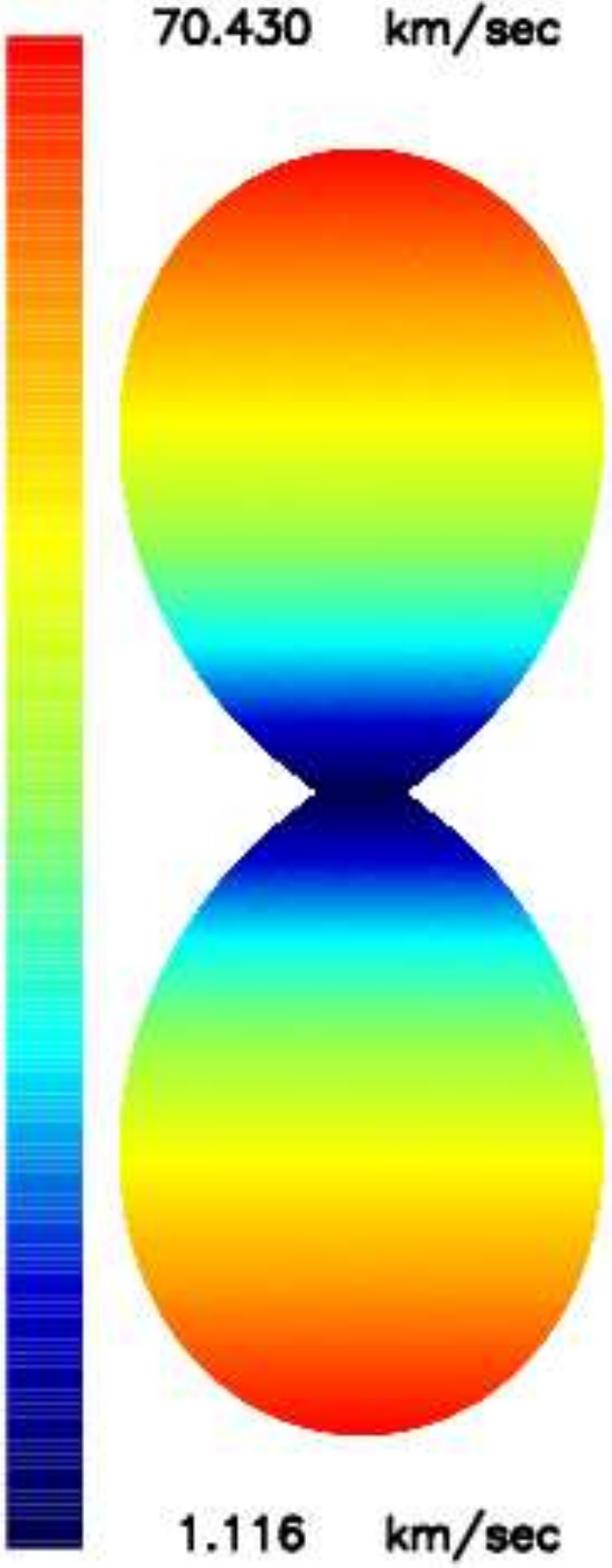}
  \end {center}
\caption {
  Map of the expansion velocity   in $\frac{km}{s}$
  relative to the simulation of the PN MyCn 18
  when 300000 random points are selected on the surface.
  Physical parameters as in Table~\ref{parametersmycn18}.
          }%
    \label{mycn18_velocity_field}
    \end{figure}

The combined effect of spatial asymmetry and field of velocity
are reported in Figure \ref{mycn18_velocity_field}.

The efficiency  of our code in  reproducing   the spatial
shape  over 12 directions of  MyCn 18
as given by formula~(\ref{efficiencymany} ) is reported in
Table~\ref{tab:radmycn18}.
This Table also reports the efficiency in simulating the
shape of the velocity.

\begin{figure}
  \begin{center}
\includegraphics[width=10cm]{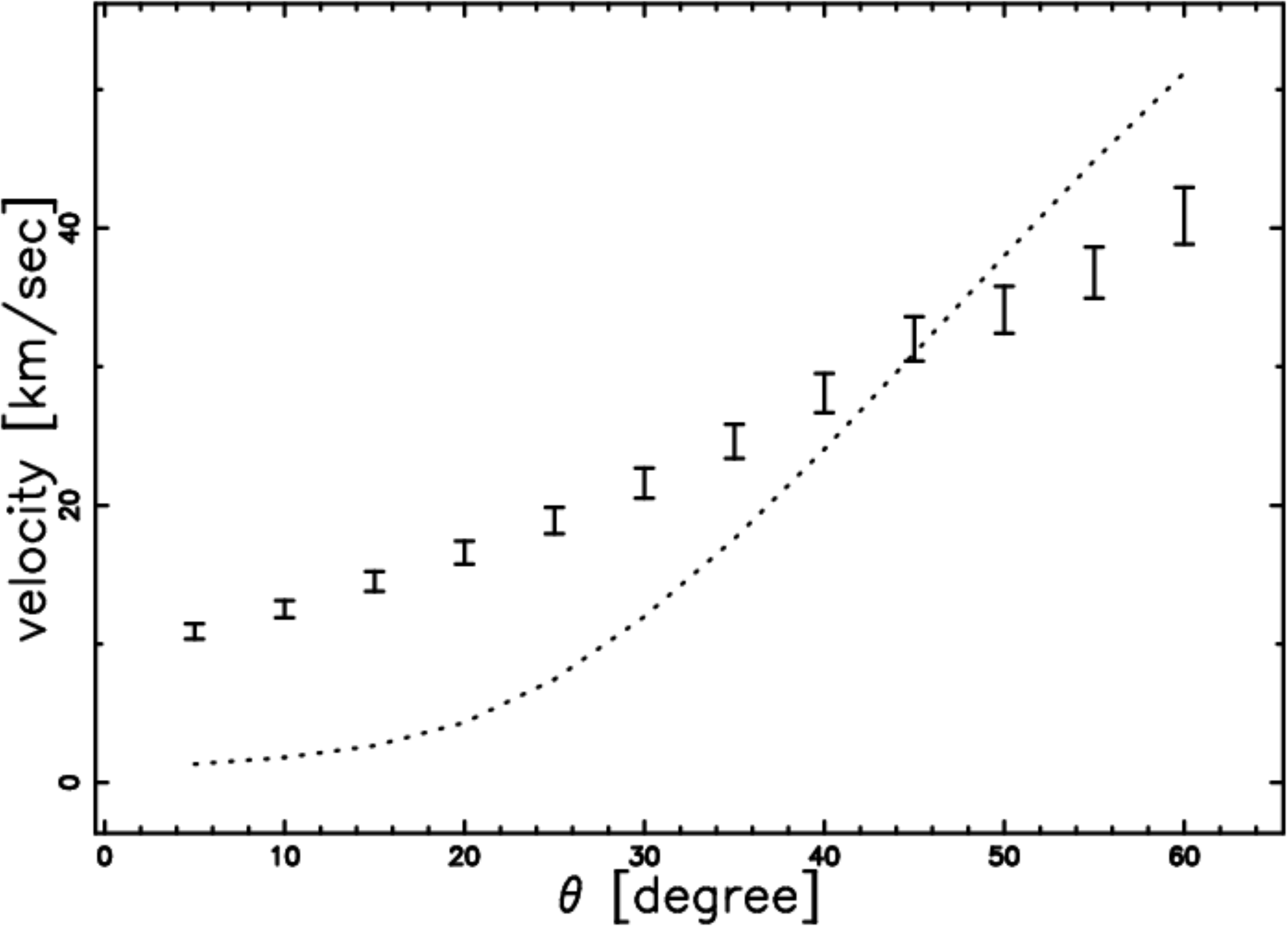}
  \end {center}
\caption { Velocity  in $\frac{km}{s}$  of the PN  MyCn 18 as a
function of the latitude in degrees when  the physical parameters
are those of
 Table~\ref{parametersmycn18}, dotted  line.
The points with error bar (1/10 of the value)
represent the data of
Table 1 in  Dayal et al. 2000.
          }%
    \label{mycn18_velocity}
    \end{figure}

\begin{table}
      \caption{Reliability of the spatial and velocity
              shape of the PN MyCn 18.}
         \label{tab:radmycn18}
      \[
         \begin{array}{lcc}
            \hline
            \noalign{\smallskip}
~~~                                & radius   &  velocity    \\
\mbox {$\epsilon_{obs}$} (\%)      &  90.66   &  57.68        \\
            \noalign{\smallskip}
            \hline
         \end{array}
      \]
   \end{table}
Figure ~ \ref{mycn18_velocity} reports our results
as well those of Table~1 in \cite{Dayal2000}.

\subsection{Simulation of \etacar in an exponentially varying medium  }

A typical set of parameters  which  allows the Homunculus nebula
around \etacar  to be simulated in the presence of a medium whose
density decreases exponentially is reported in Table
\ref{parametershom}. Table \ref{efficiency_exp} presents numbers
concerning the quality of fit.

\begin{figure}
  \begin{center}
\includegraphics[width=10cm]{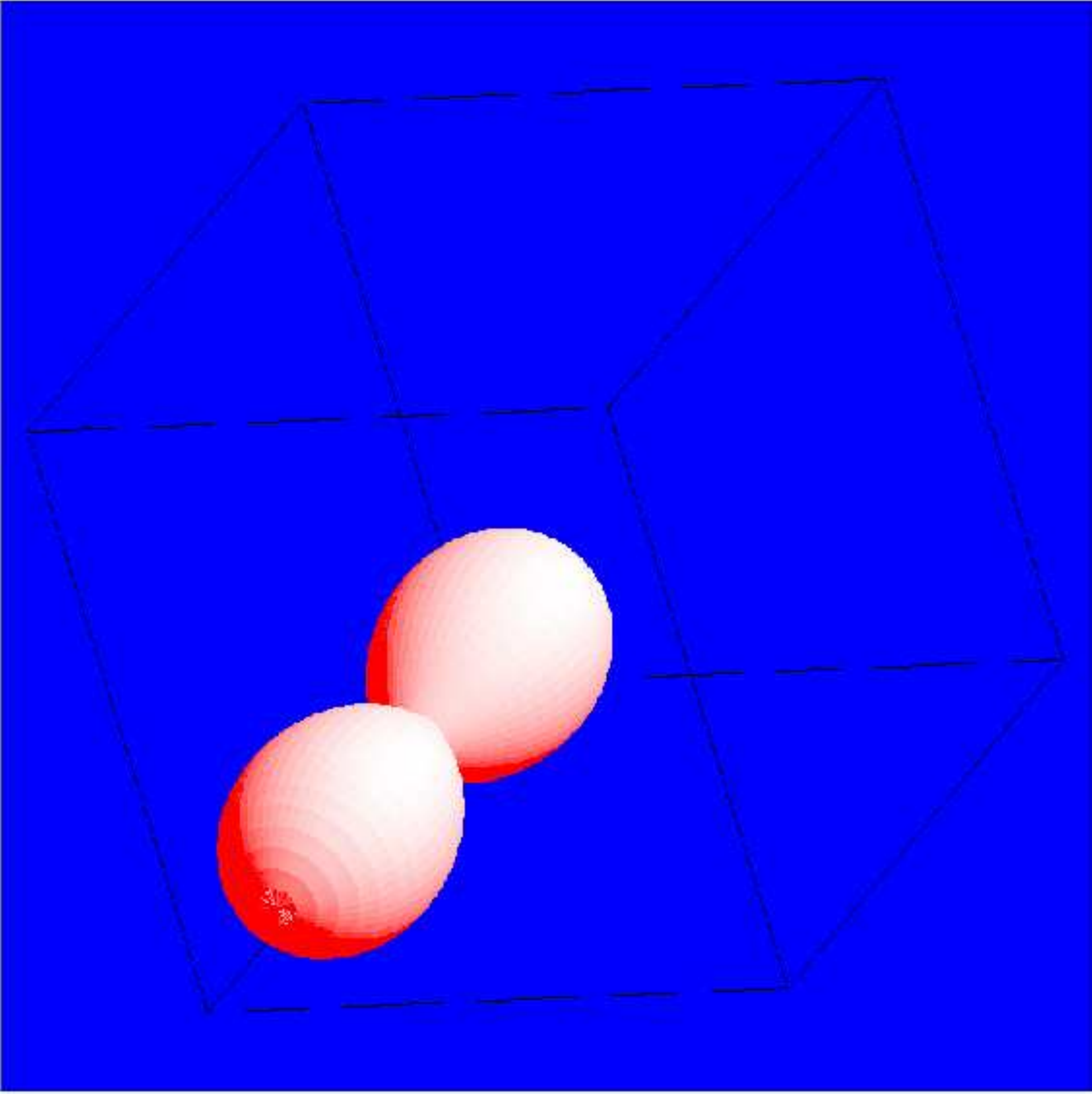}
  \end {center}
\caption { Simulations lead to this picture of the hybrid
Homunculus/\etacar nebula  for an exponentially varying medium.
The orientation of the figure is characterized by the Euler angles
, which are
     $ \Phi   $=130$^{\circ }$,
     $ \Theta $=40$^{\circ }$
and  $ \Psi   $=-140$^{\circ }$. Physical parameters as in
Table~\ref{parametershom}.
          }%
    \label{eta_faces}
    \end{figure}

\begin{figure}
  \begin{center}
\includegraphics[width=10cm]{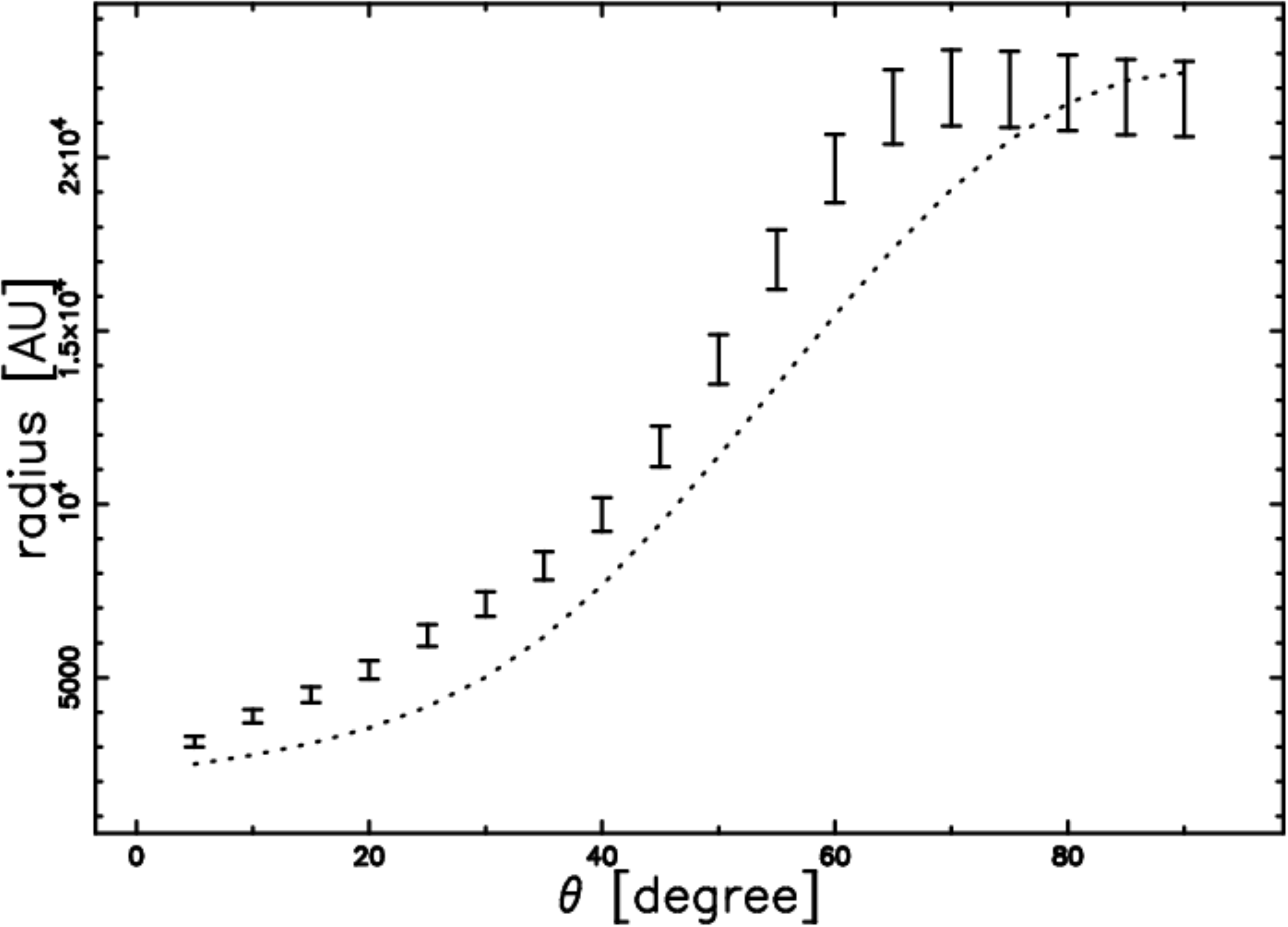}
  \end {center}
\caption {
 Radius of the  hybrid
Homunculus/\etacar nebula  as a function of latitude for an
 exponentially varying medium (dotted line)
and  astronomical data with error bar. Physical parameters as in
Table~\ref{parametershom}.
          }%
    \label{eta_radius}
    \end{figure}

\begin{figure}
  \begin{center}
\includegraphics[width=10cm]{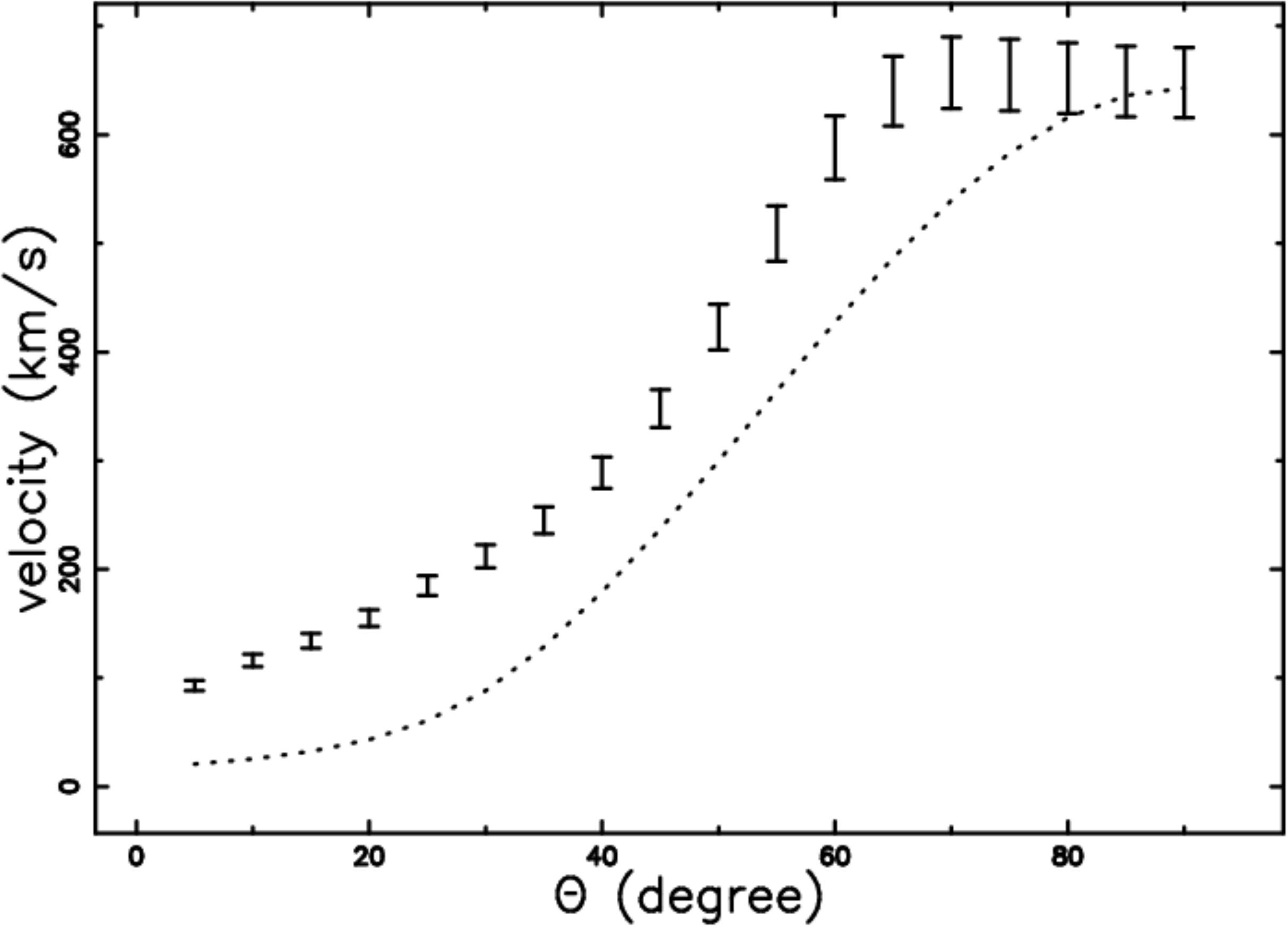}
  \end {center}
\caption { Velocity of the hybrid Homunculus/\etacar nebula as a
function of latitude for an exponentially
 varying medium
(dotted line) and  astronomical data with error bar Physical
parameters as in Table~\ref{parametershom}.
          }%
    \label{eta_velocity}
    \end{figure}

\begin{table}
      \caption{
 Parameter values used to simulate
the observations of the hybrid  Homunculus/\etacar nebula for a
medium varying exponentially (first 4 values) or a power law (2nd
set of 4 values)}
         \label{parametershom}
      \[
         \begin{array}{cc}
            \hline
            \noalign{\smallskip}
 \mbox {Initial ~expansion~velocity, ${\dot R}_{{0,1}}$
 [km~s$^{-1}$}]    & 8 000              \\
 \mbox {Age~($t_4-t_{0,4}$) [10$^4$~yr]} & 0.0158  \\
 \mbox {Scaling~h     [pc] }             & 0.0018 \\
 \mbox {Initial~radius~ $R_0$ ~[pc] }    & 0.001 \\
            \noalign{\smallskip}
            \hline
            \noalign{\smallskip}
 \mbox {Initial ~expansion~velocity, ${\dot R}_{{0,1}}$
 [km~s$^{-1}$}]    & 40,000              \\
\mbox {Age~($t_4-t_{0,4}$) [10$^4$~yr]} & 0.0158 \\
\mbox {Initial~radius~ $R_0$ ~[pc] }    & 0.0002 \\
\mbox {Power~law~coefficient $\alpha$ } & 2.4    \\
            \noalign{\smallskip}
            \hline
            \noalign{\smallskip}
         \end{array}
      \]
   \end{table}
\begin{table}
      \caption{
Agreement between observations and simulations of the hybrid
Homunculus/\etacar nebula, for an exponentially varying medium. }
         \label{efficiency_exp}
      \[
         \begin{array}{lcc}
            \hline
            \noalign{\smallskip}
~~~                                     & radius   &  velocity    \\
\mbox {$\epsilon$} (\%)-polar~ direction & 97       &  99          \\
\mbox {$\epsilon$} (\%)-equatorial~ direction & 87       &  19          \\
\mbox {$\epsilon_{obs}$} (\%)      &  85   &  75           \\
            \noalign{\smallskip}
            \hline
         \end{array}
      \]
   \end{table}
The bipolar character of the Homunculus is
 shown
in Figure~\ref{eta_faces}. In order to better visualize the two
lobes, Figure~\ref{eta_radius} and  Figure~\ref{eta_velocity} show
the radius and velocity as a function of the angular position
$\theta$. The  orientation of the observer is characterized by the
three Euler   angles $(\Phi, \Theta, \Psi)$, see
\cite{Goldstein2002};  different  Euler angles produce different
observed shapes.

\begin{figure}
  \begin{center}
\includegraphics[width=4cm]{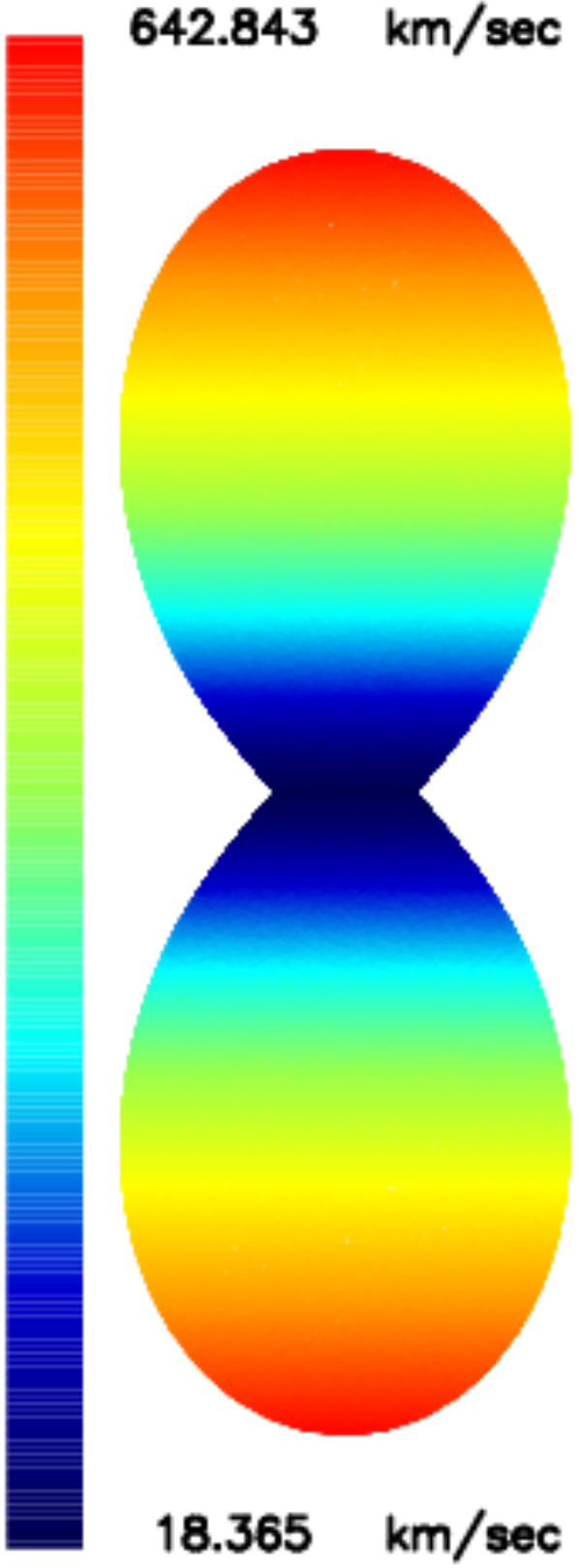}
  \end {center}
\caption { Map of the expansion velocity for an exponentially
varying medium  relative to the hybrid Homunculus/\etacar nebula.
  Physical parameters as in Table~\ref{parametershom}.
          }%
    \label{eta_velocity_field}
    \end{figure}

The velocity field is  shown in Figure \ref{eta_velocity_field}.

The accuracy with which our code reproduces the spatial shape and
the velocity field over 18 directions of the Homunculus nebula as
given by formula~(\ref{efficiencymany}) is reported in
Table~\ref{efficiency_exp}. From a careful analysis of
Table~\ref{efficiency_exp} we can conclude that the spatial shape
over 18 directions is well modeled by an exponential medium ,
$\epsilon_{obs} = 85 \%$. The overall efficiency of the field is
smaller $\epsilon_{obs} = 75 \%$. We can therefore conclude that
formula (\ref{efficiencymany}) which gives the efficiency  over
all the range of polar  angles represents a better  way  to
describe  the results in  respect to the efficiency  in a single
direction as given by formula (\ref{efficiency}).

\subsection{Simulation of \etacar for a power law medium}

For assumed parameters see Table \ref{parametershom},
Table~\ref{efficiency_power} reports the accuracy of radius and
velocity in two directions.

\begin{table}
      \caption{
 Agreement between model for a power
law medium and observations for the hybrid  Homunculus/\etacar
nebula. }
         \label{efficiency_power}
      \[
         \begin{array}{lcc}
            \hline
            \noalign{\smallskip}
~~~                                     & radius   &  velocity    \\
\mbox {$\epsilon$} (\%)-polar~ direction & 78       &  76          \\
\mbox {$\epsilon$} (\%)-equatorial~ direction & 6   &  2          \\
\mbox {$\epsilon_{obs}$} (\%)                 & 79   &  73          \\
            \noalign{\smallskip}
            \hline
         \end{array}
      \]
   \end{table}
\subsection{Simulation of a spherical  SNR , \snr}

According to the power solution as given by (\ref{rpower}) and
 to the data used in Section \ref{sphericalpowerlaw},
$\epsilon$=98.54 \%.

\subsection{Simulation of a asymmetric  SNR , \sn1006}

\begin{figure*}
\begin{center}
\includegraphics[width=10cm,angle=-90]{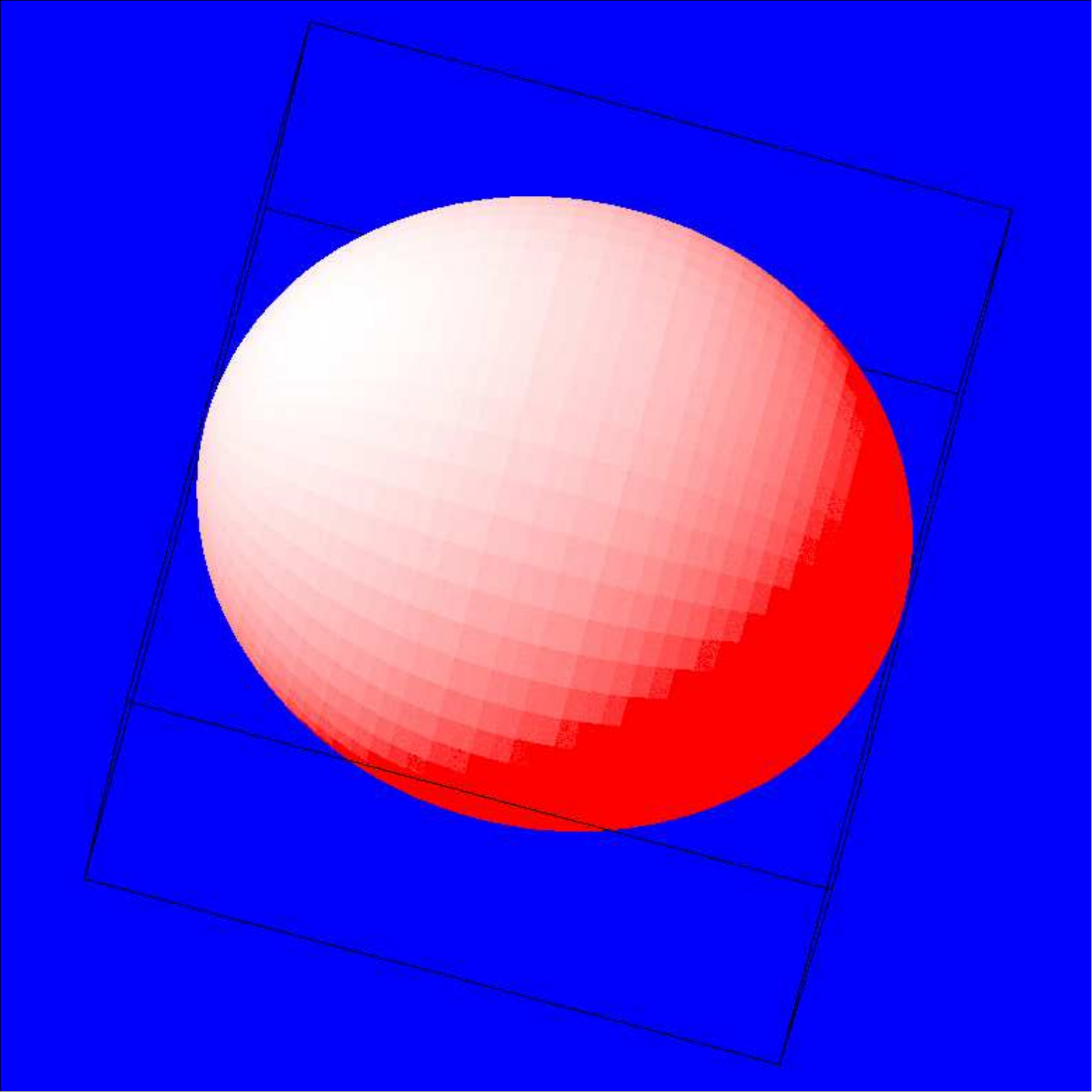}
\end {center}
\caption 
{
The shape of the expanding envelope  modeled 
by 2500 faces particularized for \sn1006.
The three Eulerian angles characterizing the point of
view are
     $ \Phi $=75    $^{\circ }  $,
     $ \Theta $=90   $^{\circ }$
and  $ \Psi $=75    $^{\circ }   $. 
} \label{sn10063d}
    \end{figure*}

According to the  numerical code developed in \cite{Zaninetti2000}
in the case  of a Gaussian profile  we have
 $\epsilon= 94.9 \%$  in the polar direction
 and $\epsilon= 92.5 \%$  in the equatorial direction.
From a practical point of view,  the  
range   of the polar angle $\theta$ ($180 ^{\circ}$ ) 
is  divided 
into  $n_{\theta}$ steps and the range of the azimuthal angle 
$\phi$ ($360^{\circ }$ ) into  $n_{\phi}$   steps.
This  yields  ($n_{\theta}$ +1) ($n_{\phi}$ +1) directions of 
motion which  can  also  be identified with  the number of vertexes
of the polyhedron representing the volume occupied by the explosion~;
this   polyhedron  varies  from  a sphere  
to an  irregular  shape  on the basis of the  swept--up  material 
in each direction~.
In the plots showing the  expansion surface  of the explosion,
the number  of vertexes 
($n_{\theta}$+1)$\cdot$($n_{\phi}$ +1)
and  the number  
of the faces 
$n_{\theta}\cdot\;n_{\phi}$; 
are specified,
 for example in Figure ~\ref{sn10063d}   $n_{\theta}$=50  
and $n_{\phi}$=50.

\section{Diffusion}

\label{sec_diffusion}

The mathematical diffusion  allows us  to  follow the number density
of particles from high values
(injection) to low values (absorption).
We recall that the number density is expressed
in $\frac{particles}{unit~ volume}$ and  the symbol $C$ is
used  in the mathematical diffusion and
the symbol $n$ in an astrophysical context.
The density $\rho$ is obtained
by
multiplying $n$
by the mass of  hydrogen , $m_H$ ,
 and by a multiplicative factor , $f$,
which varies from 1.27 in   \cite{Kim2000} to 1.4
in \cite{Dalgarno1987}
\begin{equation}
\rho  = f  m_H n
\quad .
\end{equation}
The physical process that allows  the particles
to diffuse  is hidden in the mathematical diffusion.
In our case the physical process can be the random walk
with a time step equal  to the Larmor gyroradius.
In the Monte Carlo diffusion
the step-length  of the random walk
is generally taken as  a fraction of the side of the
considered box.
Both mathematical diffusion
and Monte Carlo diffusion use the concept of absorbing-boundary  which is  the spatial coordinate where
the diffusion path  terminates.

In the following,  3D mathematical diffusion from a
sphere and 1D  mathematical as well  Monte Carlo diffusion
in presence of drift and  are considered.

\subsection{3D diffusion from a spherical source}
\label{mathematical}
Once the number density , $C$, and the diffusion coefficient ,$D$,
are introduced ,  Fick'~s first equation
changes   expression on the basis  of the adopted
environment  ,  see for example equation~(2.5) in  \cite{berg}.
In three dimensions  it is
\begin{equation}
\frac {\partial C }{\partial t} =
D \nabla^2 C
\quad,
\label{eqfick}
\end {equation}
where $t$ is the time  and $ \nabla^2$ is
the Laplacian differential operator.

In presence of the  steady state condition:
\begin{equation}
D \nabla^2 C   = 0
\quad .
\label{eqfick_steady}
\end {equation}

The  number density rises from 0 at {\it r=a}  to a
maximum value $C_m$ at {\it r=b} and then  falls again
to 0 at {\it  r=c}~.
The  solution to  equation~(\ref{eqfick_steady})
 is
\begin{equation}
C(r) = A +\frac {B}{r}
\quad,
\label{solution}
\end {equation}
where $A$ and $B$  are determined by  the boundary conditions~,
\begin{equation}
C_{ab}(r) =
C_{{m}} \left( 1-{\frac {a}{r}} \right)  \left( 1-{\frac {a}{b}}
 \right) ^{-1}
\quad a \leq r \leq b
\quad,
\label{cab}
\end{equation}
and
\begin{equation}
C_{bc}(r)=
C_{{m}} \left( {\frac {c}{r}}-1 \right)  \left( {\frac {c}{b}}-1
 \right) ^{-1}
\quad b \leq r \leq c
\quad.
\label{cbc}
\end{equation}
These solutions can be found in
\cite{berg}
or in
\cite{crank} .

\subsection{1D diffusion with drift, mathematical diffusion}

In one dimension  and in the presence of a drift velocity
,$u$,
along the  radial direction
the diffusion is governed by  Fick's second equation ,
see equation~(4.5) in \cite{berg} ,
\begin{equation}
\frac {\partial C }{\partial t} =
D  \frac {\partial^2C}{\partial r^2} -  {\vec{u}} \frac {\partial C}{\partial r}
\quad ,
\label{eqfick_1_drift}
\end {equation}
where ${\vec{u}}$  can take  two directions.
The  number density rises from 0 at {\it r=a}  to a
maximum value $C_m$ at {\it r=b} and then  falls again
to 0 at {\it  r=c}~.
The general solution to  equation~(\ref{eqfick_1_drift})
in presence of a steady state is
\begin{equation}
C(r) = A + B e^{{\frac{\vec {u}}{D}}r }
\quad.
\end{equation}
We now assume  that   {\it u}  and {\it r}
do not  have  the same  direction  and therefore
{\it u } is negative ;  the solution is
\begin{equation}
C(r) = A + B e^{-{\frac{u}{D}}r }
\quad ,
\label{solution_1D_drift}
\end{equation}
and now the velocity $u$ is a scalar.

\begin{figure}
  \begin{center}
\includegraphics[width=10cm]{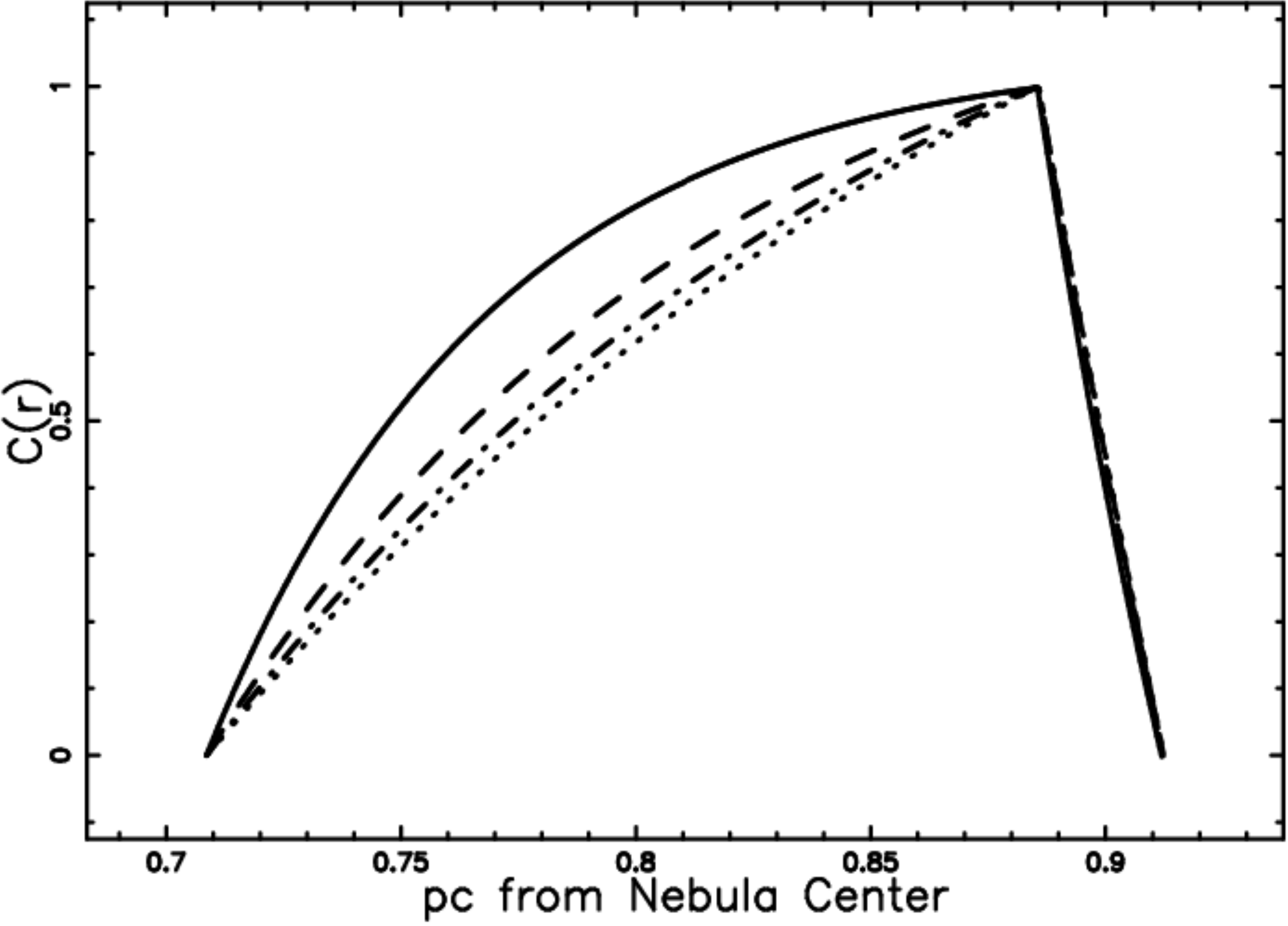}
  \end {center}
\caption {
 Number density of the PN A39  as a
 function of the distance in pc from
 the injection
 when  $u=1$   , $C_m=1 $, $a=69.6~arcsec $, $b=87~arcsec$ , $c=89.6~arcsec$
 and
 $D=2 $       (full line        ),
 $D=7 $       (dashed           ),
 $D=12 $       (dot-dash-dot-dash)
    and
 $D=17 $       (dotted ).
The conversion from $arcsec$ to   $pc$ is done assuming
a distance of 2100 $pc$ for A39.
          }%
    \label{diffusion}
    \end{figure}

The boundary-conditions  give
\begin{equation}
C_{a,b,drift}(r) =
C_m
\frac
{
e ^{-\frac{u}{D} a} -   e ^{-\frac{u}{D} r}
}
{
e ^{-\frac{u}{D} a} -   e ^{-\frac{u}{D} b}
}
\quad a \leq r \leq b ~\quad downstream~side
\quad ,
\label{cab_drift}
\end{equation}
and
\begin{equation}
C_{b,c,drift}(r) =
C_m
\frac
{
e ^{-\frac{u}{D} c} -   e ^{-\frac{u}{D} r}
}
{
e ^{-\frac{u}{D} c} -   e ^{-\frac{u}{D} b}
}
\quad b \leq r \leq c  ~\quad upstream~side
\quad .
\label{cbc_drift}
\end{equation}
A typical plot of the number density for different values
of the diffusion coefficient is reported in
Figure~\ref{diffusion}.

\subsection{1D diffusion with drift, random walk  }

Given a 1D segment of length $side$ we can implement
the random walk with
step-length  $\lambda$
by introducing the numerical parameter $NDIM=\frac{side}{\lambda}$~.
We now report the adopted  rules
when the  injection is  in the middle of the grid   :
\begin {enumerate}
\item    The first of the  $NPART$  particles  is chosen.
\item    The random  walk of a particle starts in the middle of
         the grid.
         The probabilities of  having  one step
         are $p_1$ in the
         negative direction  (downstream)
         ,$ p_1 =  \frac{1}{2} - \mu \times \frac{1}{2}$,
         and $p_2$ in the positive direction  (upstream)
         , $ p_2 =  \frac{1}{2}+ \mu \times \frac{1}{2}$,
         where $\mu$ is a  parameter that characterizes
         the asymmetry ($0 \leq \mu \leq 1 $).

\item    When the  particle reaches    one of the two
         absorbing points ,  the motion starts
         another  time from (ii) with
         a different diffusing  pattern.

\item    The number of visits is  recorded on ${\mathcal M}$ ,
         a one--dimensional grid.
\item    The random walk terminates when all  the $NPART$
         particles  are processed.
\item    For the sake  of normalization the
         one--dimensional visitation or number density grid
         ${\mathcal M}$ is divided by  $NPART$.

\end  {enumerate}

\begin{figure*}
\begin{center}
\includegraphics[width=10cm]{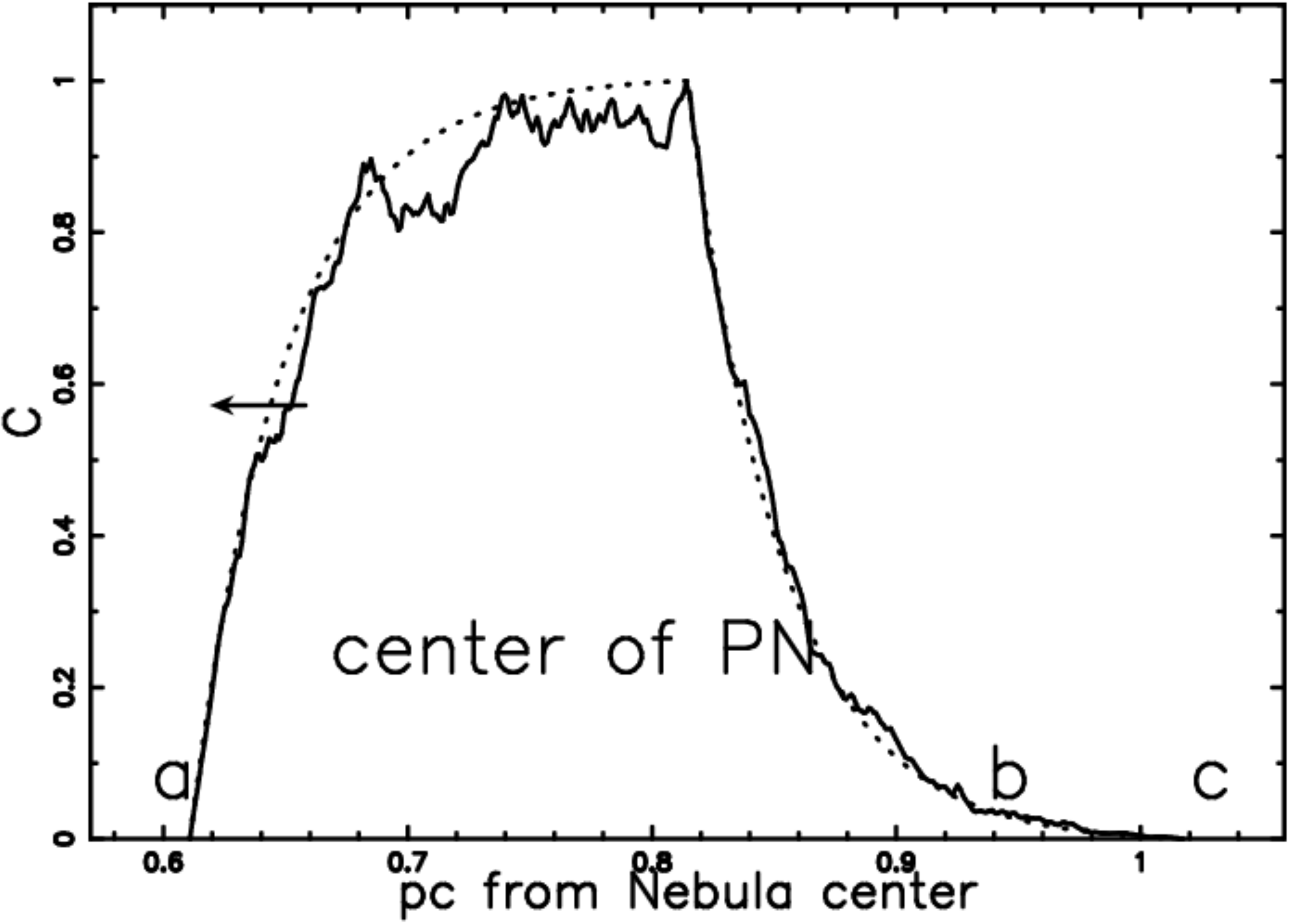}
\end {center}
\caption { Number density relative to PN  A39  of the 1D
asymmetric random walk (full line), NDIM=401 ,NPART=200
,$side=40~arcsec$ , $\lambda=0.1~arcsec$ and  $\mu$ =- 0.013. For
astrophysical purposes  $\mu$ is negative. The theoretical number
density as represented by formulas~(\ref{cab_drift}) and
(\ref{cbc_drift}) is reported
 when  $u=1$   , $C_m=1 $, $a=60~arcsec$, $b=80~arcsec$ , $c=100~arcsec$
 and
 $D=3.84 $       (dotted line ).
The conversion from $arcsec$ to   $pc$ is done assuming
a distance  of 2100 $pc$ for A39.
}
\label{montec}
    \end{figure*}

There is a systematic change of the average particle
position along the
$x$-direction:
\begin{equation}
\langle dx \rangle =    \mu~\lambda \quad,
\label{formula1}
\end {equation}
for each time step.
If the time step is $dt=\frac{\lambda}{v_{tr}}$
where   ${v_{tr}}$ is the transport velocity,
the asymmetry  ,$\mu$ ,
that characterizes the random walk is
\begin{equation}
\mu =  \frac{u}{v_{tr}}
\quad  .
\end{equation}
Figure~\ref{montec} reports ${\mathcal M}(x)$,
the  number  of  visits
generated by  the Monte Carlo simulation
as well as  the mathematical solution represented
by formulas~(\ref{cab_drift}) and (\ref{cbc_drift}).

The solutions of the mathematical   diffusion
equations~(\ref{cab_drift})   and  (\ref{cbc_drift})
can be rewritten at the light of the
random walk  and are
\begin{equation}
C_{a,b,MC}(r) =
C_m
\frac
{
e ^{-\frac{2\mu}{\lambda} a} -   e ^{-\frac{2\mu}{\lambda} r}
}
{
e ^{-\frac{2\mu}{\lambda} a} -   e ^{-\frac{2\mu}{\lambda} b}
}
\quad a \leq r \leq b ~\quad downstream~side
\quad ,
\label{cab_MC}
\end{equation}
and
\begin{equation}
C_{b,c,MC}(r) =
C_m
\frac
{
e ^{-\frac{2\mu}{\lambda} c} -   e ^{- \frac{2\mu}{\lambda} r}
}
{
e ^{-\frac{2\mu}{\lambda} c} -   e ^{-\frac{2\mu}{\lambda} b}
}
\quad b \leq r \leq c  ~\quad upstream~side
\quad .
\label{cbc_MC}
\end{equation}

\section{Radiative transfer equation}
\label{sec_transfer}
The transfer equation in the presence of emission only , see for
example \cite{rybicki}
 or
\cite{Hjellming1988}
 ,
 is
 \begin{equation}
\frac {dI_{\nu}}{ds} =  -k_{\nu} \zeta I_{\nu}  + j_{\nu} \zeta
\label{equazionetrasfer} \quad ,
\end {equation}
where  $I_{\nu}$ is the specific intensity , $s$  is the line of
sight , $j_{\nu}$ the emission coefficient, $k_{\nu}$   a mass
absorption coefficient, $\zeta$ the  mass density at position s
and the index $\nu$ denotes the interested frequency of emission.
The solution to  equation~(\ref{equazionetrasfer})
 is
\begin{equation}
 I_{\nu} (\tau_{\nu}) =
\frac {j_{\nu}}{k_{\nu}} ( 1 - e ^{-\tau_{\nu}(s)} ) \quad  ,
\label{eqn_transfer}
\end {equation}
where $\tau_{\nu}$ is the optical depth at frequency $\nu$
\begin{equation}
d \tau_{\nu} = k_{\nu} \zeta ds \quad.
\end {equation}
We now continue analyzing the case of an
 optically thin layer
in which $\tau_{\nu}$ is very small ( or $k_{\nu}$  very small )
and the density  $\zeta$ is substituted with our number density
C(s) of  particles. Two cases are taken into account :   the
emissivity is proportional to the number density and the
emissivity is proportional to the square of the number density .
In the  linear case
\begin{equation}
j_{\nu} \zeta =K  C(s) \quad  ,
\end{equation}
where $K$ is a  constant function. This can be the case of
synchrotron radiation in presence of a  isotropic distribution of
electrons with a power law distribution in energy, $N(E)$,
\begin{equation}
N(E)dE = K_s E^{-\gamma_f} \label{spectrum} \quad  ,
\end{equation}
where $K_s$ is a constant. In this case  the emissivity is
\begin{eqnarray}
j_{\nu} \rho    \approx 0.933 \times 10^{-23} \alpha (\gamma_f)
K_s
 H_{\perp}
^{\frac{\gamma_f +1}{2} }   \bigl (
 \frac{6.26 \times 10^{18} }{\nu}
\bigr )^{\frac{\gamma_f -1}{2} } \frac {erg} {s\, cm^3 \,Hz
\,rad^2} ,
\end{eqnarray}
where $\nu$ is the frequency and   $\alpha (\gamma_f)$  is a
slowly varying function of $\gamma_f$ which is of the order of
unity and is given by
\begin{eqnarray}
\alpha(\gamma_f) =   2^{(\gamma_f -3)/2}
\frac{\gamma_f+7/3}{\gamma_f +1} \Gamma \bigl ( \frac {3\gamma_f
-1 }{12} \bigr ) \Gamma \bigl ( \frac {3\gamma_f +7 }{12} \bigr )
\quad ,
\end{eqnarray}
for  $\gamma_f \ge \frac{1}{2}$,
 see formula
(1.175 ) in  \cite{lang} . The synchrotron emission is widely used
to explain  the radiation observed in SNR, see
\cite{Martinell2004,Berezhko2004,Bamba2005,Decourchelle2005,Bykov2008,Katsuda2010}.
This non thermal radiation continuum emission was also  detected
in a PN associated with a very long-period OH/IR variable star
(V1018 Sco), see~\cite{Cohen2006}.

In the  quadratic  case
\begin{equation}
j_{\nu} \zeta =K_2  C(s)^2 \quad  , \label{eqn_transfer_square}
\end{equation}
where $K_2$ is a  constant function. This is true for
\begin{itemize}
\item Free-free radiation from a thermal plasma, see formula
(1.219) in  \cite{lang}  . This  radiation process was  adopted by
\cite{Gonzales_2006} in the little Homunculus.

\item Thermal bremsstrahlung and recombination radiation , see
formula (1.237) in  \cite{lang}  . This  radiation process was
adopted in PNs by~\cite{Blagrave2006,Schwarz2006,Gruenwald2007}.
\end{itemize}

The intensity is now
\begin{equation}
 I_{\nu} (s) = K
\int_{s_0}^s   C (s\prime) ds\prime \quad  \mbox {optically thin
layer} \quad linear~case \quad , \label{transport1}
\end {equation}
or
\begin{equation}
 I_{\nu} (s) = K_2
\int_{s_0}^s   C (s\prime)^2 ds\prime \quad  \mbox {optically thin
layer} \quad quadratic~case \quad . \label{transport2}
\end {equation}
In the Monte Carlo experiments the number density is memorized  on
the   grid ${\mathcal M}$ and the intensity is
\begin{equation}
{\it I}\/(i,j) = \sum_k  \triangle\,s \times  {\mathcal M}(i,j,k)
\quad  \mbox {optically thin layer} \quad linear~case \quad,
\label{thin1}
\end{equation}
or
\begin{equation}
{\it I}\/(i,j) = \sum_k  \triangle\,s \times  {\mathcal
M}(i,j,k)^2 \quad  \mbox {optically thin layer} \quad
quadratic~case \quad  , \label{thin2}
\end{equation}
where $\triangle$s is the spatial interval between the various
values and  the sum is performed over the   interval of existence
of the index $k$. The theoretical intensity is then obtained by
integrating the intensity at a given frequency  over the solid
angle of the source.

In order to deal with the transition to the optically thick case,
the intensity is given by
\begin{eqnarray}
\label{transition} {\it I}\/(i,j) = \frac {1}{K_a}  (1 - \exp (-
K_a\sum_k
\triangle\,s \times {\mathcal S}(i,j,k)))  \\
 \quad
\mbox {Thin $\longmapsto $ Thick }\quad, \nonumber
\end{eqnarray}
where  $K_a$ is a constant that represents the absorption.
Considering the  Taylor expansion of the last formula
(\ref{transition}), equation~(\ref{thin1}) is obtained.

\section{Images}

\label{sec_image} 
The image of a PN , \etacar and a SNR  can be
modeled once an analytical or numerical law for the intensity of
emission as a function of the radial distance from the center is
given. 
Simple analytical results for the radial intensity
can be deduced in the rim model when
the length of the layer and the number density are constants.
The integration of the solutions to the mathematical diffusion
along the line of sight allows us  to deduce analytical formulas
in the spherical case. The complexity of the intensity in the
aspherical case can be attached only from a numerical point of
view.
\subsection{3D Constant Number density in a rim model}

\label{rimsec} 
\begin{figure*}
\begin{center}
\includegraphics[width=10cm]{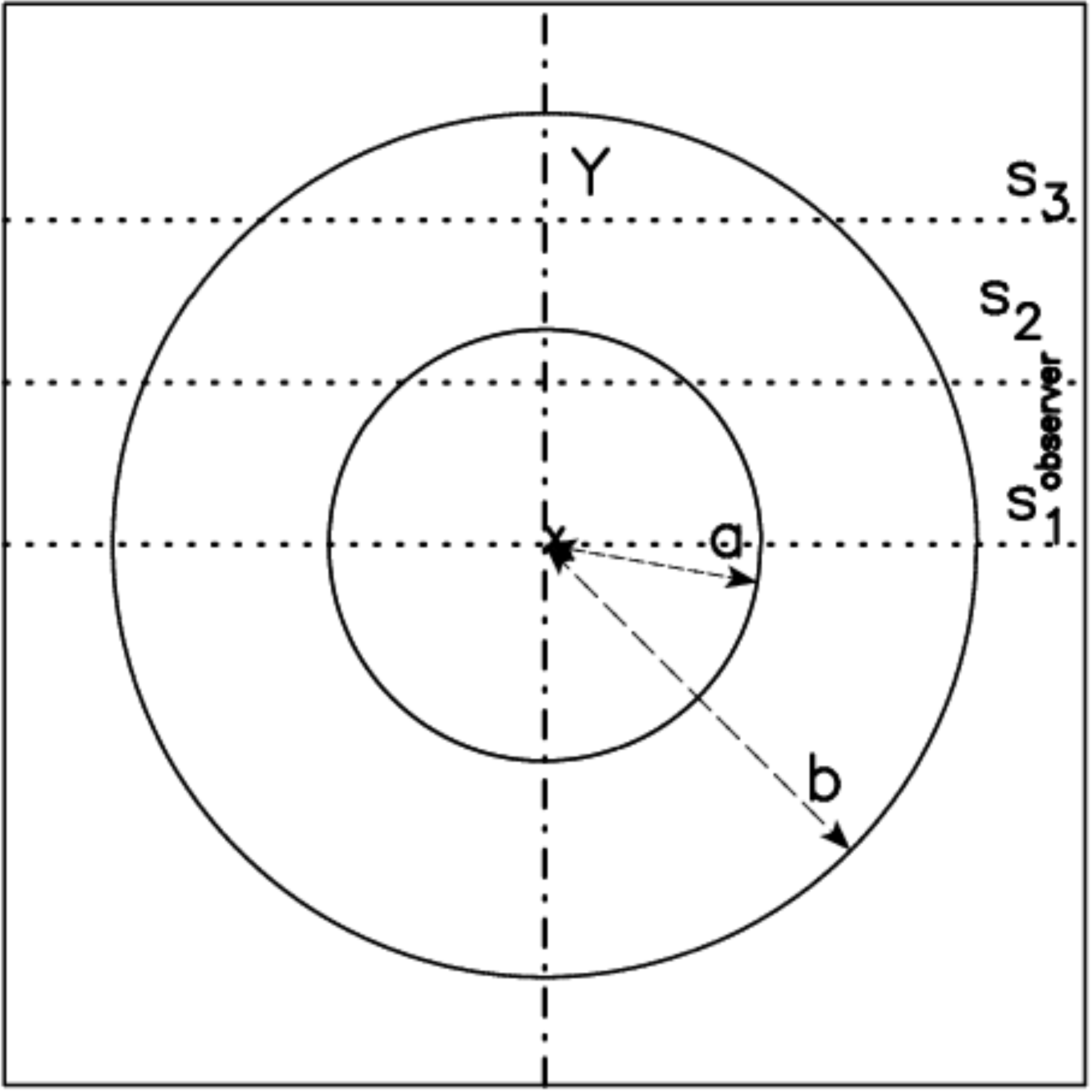}
\end {center}
\caption { The two circles (section of spheres)  which   include
the region with constant density are   represented through
 a full line.
The observer is situated along the x direction, and three lines of
sight are indicated. }
\label{plotab}
    \end{figure*}

\begin{figure*}
\begin{center}
\includegraphics[width=10cm]{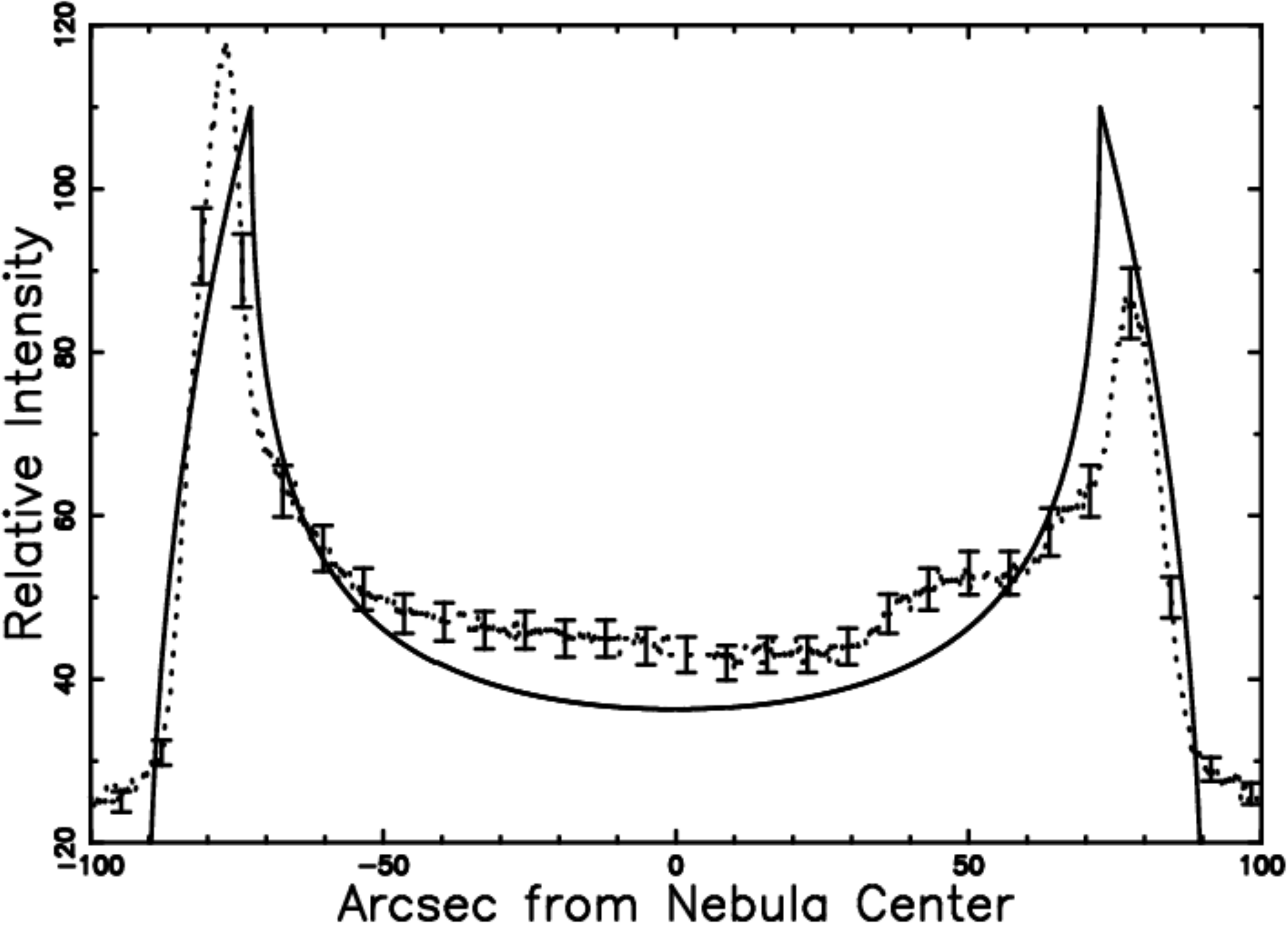}
\end {center}
\caption {
 Cut of the mathematical  intensity ${\it I}$
 of the rim model ( equation~(\ref{irim}))
 crossing the center    (full  line  ) of the PN A39
 and  real data         (dotted line with some error bar ) .
 The number of data is  801  and
for  this  model $\chi^2$ = 1.487 against $\chi^2$ = 0.862 of the
rim model fully described in ~Jacoby et al. (2001). }
\label{rim_cut_square}
    \end{figure*}

We assume that the number density $C$ is constant
and in particular rises from 0 at $r=a$ to a maximum value $C_m$ ,
remains constant up to $r=b$ and then falls again to 0. This
geometrical  description is reported in Figure~\ref{plotab}.
The length of sight , when the observer is situated at the
infinity of the $x$-axis , is the locus parallel to the $x$-axis
which  crosses  the position $y$ in a Cartesian $x-y$ plane and
terminates at the external circle of radius $b$. The locus length
is
\begin{eqnarray}
l_{0a} = 2 \times ( \sqrt { b^2 -y^2} - \sqrt {a^2 -y^2})
\quad  ;   0 \leq y < a  \nonumber  \\
l_{ab} = 2 \times ( \sqrt { b^2 -y^2})
 \quad  ;  a \leq y < b    \quad .
\label{length}
\end{eqnarray}
When the number density $C_m$ is constant between two spheres of
radius $a$ and $b$ the intensity of radiation is
\begin{eqnarray}
I_{0a} =C_m \times 2 \times ( \sqrt { b^2 -y^2} - \sqrt {a^2
-y^2})
\quad  ;   0 \leq y < a  \nonumber  \\
I_{ab} =C_m \times  2 \times ( \sqrt { b^2 -y^2})
 \quad  ;  a \leq y < b    \quad .
\label{irim}
\end{eqnarray}
The comparison   of observed data of A39 and the theoretical
intensity is reported in Figure~\ref{rim_cut_square} when  data
from  Table~\ref{dataab} are used.

The ratio between the theoretical intensity at the maximum ,
$(y=b)$ ,
 and at the minimum , ($y=0$) ,
is given by
\begin{equation}
\frac {I(y=b)} {I(y=0)} = \frac {\sqrt {b^2 -a^2}} {b-a} \quad .
\label{ratioteorrim}
\end{equation}

 \begin{table}
 \caption[]{Simulation of the PN A39 with the rim model }
 \label{dataab}
 \[
 \begin{array}{llc}
 \hline
 \hline
 \noalign{\smallskip}
 symbol  & meaning & value  \\
 \noalign{\smallskip}
 \hline
 \noalign{\smallskip}
a  & radius~of~the~internal~sphere    & 72.5^{\prime\prime} \\
\noalign{\smallskip} b                 &
radius~of~the~external~sphere    & 90.18^{\prime\prime} \\
\noalign{\smallskip} R_{shell}         &
observed~radius~of~the~shell       & 77^{\prime\prime} \\
\noalign{\smallskip} \delta\,r_{shell,t}  & theoretical~ thickness
~of~the~shell       & 17.6^{\prime\prime} \\ \noalign{\smallskip}
\delta\,r_{shell}  & observed~ thickness ~of~the~shell       &
10.1^{\prime\prime} \\ \noalign{\smallskip} \frac {I_{limb}}
{I_{center}} &  ratio~ of~ observed~ intensities  & (1.88-2.62)
\\ \noalign{\smallskip} \frac {I_{max}} {I(y=0)} & ratio~ of~
theoretical~ intensities  & 3.03      \\ \noalign{\smallskip}
 \hline
 \hline
 \end{array}
 \]
 \end {table}

\subsection{3D diffusion from a sphere, square dependence}

\begin{figure*}
\begin{center}
\includegraphics[width=10cm]{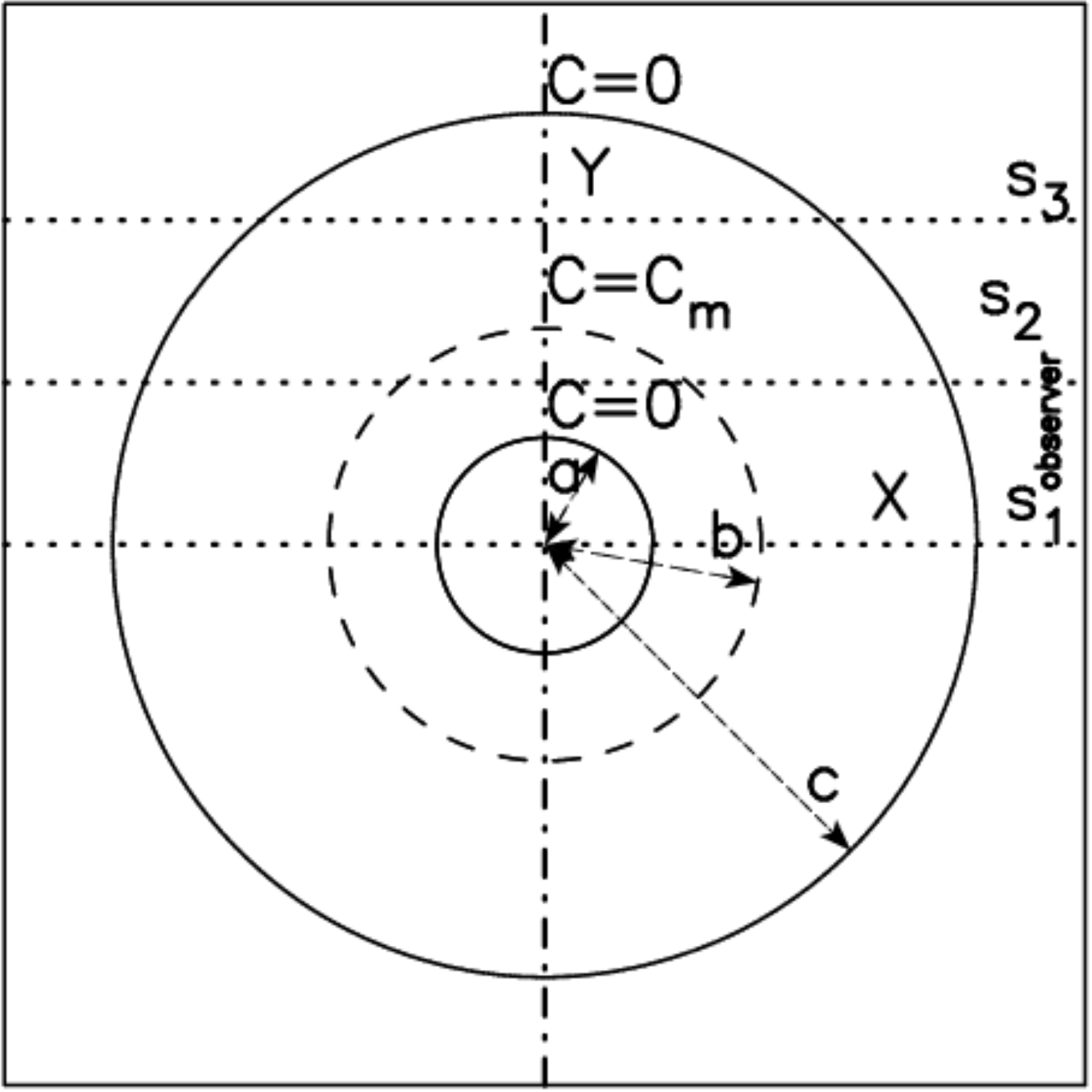}
\end {center}
\caption { The spherical source inserted in the great  box is
represented through a dashed line, and the two absorbing
boundaries with a full line. The observer is situated along the x
direction, and three lines of sight are indicated. Adapted from
Figure~3.1 by Berg (1993) . }
\label{plot}
    \end{figure*}

Figure~\ref{plot} shows a spherical shell  source of radius  $b$
between a spherical absorber of radius $a$ and a spherical
absorber of radius $c$.

The  number density rises from 0 at {\it r=a}  to a maximum value
$C_m$ at {\it r=b} and then  falls again to 0 at {\it  r=c}~.

The numbers density  to be used are formulas (\ref{cab}) and
(\ref{cbc})  once $r=\sqrt{x^2+y^2}$ is imposed ; these two
numbers density  are  inserted in
formula~(\ref{eqn_transfer_square})  which represents the transfer
equation with a quadratic dependence on the number density. An
analogous  case was   solved   in  \cite{Zaninetti2007_c} by
adopting a linear dependence on  the number density~.
 The     geometry of the phenomena fixes
 three different zones ($0-a,a-b,b-c$) in the variable $y$,
  ;
 the first piece , $I^I(y)$ , is
\begin{eqnarray}
 I^I(y)= \int _{\sqrt{a^2-y^2}} ^{\sqrt{b^2-y^2}} 2 C_{ab}^2 dx
 + \int _{\sqrt{b^2-y^2}} ^{\sqrt{c^2-y^2}}  2C_{bc}^2dx \nonumber\\
~= -2 \frac { {\it C_m}^{2}{b}^{2} } { y ( {b}^{2}-2\,ba+{a}^{2} )
( {c}^{2}-2\,cb+{b}^{2}
 )
} \bigl  [ -2\,{a}^{2}\arctan ( {\frac { \sqrt
{{a}^{2}-{y}^{2}}}{y}} ) cb-2\,\sqrt {{a}^{2}-{y}^{2}}ycb
\nonumber \\
- 2\,ay\ln  ( \sqrt {{a}^{2}-{y}^{2}}+a ) {b}^{2}+2\,{a}^{2}
\arctan ( {\frac {\sqrt {{b}^{2}-{y}^{2}}}{y}} ) cb
\nonumber \\
+2\,y \sqrt {{b}^{2}-{y}^{2}}cb+2\,a\ln  ( \sqrt
{{b}^{2}-{y}^{2}}+b
 ) y{b}^{2}+2\,a\ln  ( \sqrt {{b}^{2}-{y}^{2}}+b ) y{
c}^{2}
\nonumber  \\
-2\,cy\ln  ( \sqrt {{b}^{2}-{y}^{2}}+b ) {b}^{2}-2\,{
c}^{2}\arctan ( {\frac {\sqrt {{b}^{2}-{y}^{2}}}{y}} ) ba-2
\,y\sqrt {{b}^{2}-{y}^{2}}ba-
\nonumber \\
2\,cy\ln  ( \sqrt {{b}^{2}-{y}^{2}}+ b ) {a}^{2}+2\,c\ln  ( \sqrt
{{c}^{2}-{y}^{2}}+c ) y{ b}^{2}
\nonumber  \\
+2\,c\ln  ( \sqrt {{c}^{2}-{y}^{2}}+c ) y{a}^{2}+2\, \sqrt
{{c}^{2}-{y}^{2}}yba+2\,{c}^{2}\arctan ( {\frac {\sqrt {{c} ^{2}
-{y}^{2}}}{y}} ) ba
\nonumber \\
-2\,ay\ln  ( \sqrt {{a}^{2}-{y}^{2}} +a ) {c}^{2}+\sqrt
{{a}^{2}-{y}^{2}}y{c}^{2}
\nonumber  \\
-4\,c\ln  ( \sqrt {{c}^{2}-{y}^{2}}+c ) yba+4\,ay\ln  ( \sqrt
{{a}^{2}- {y}^{2}}+a ) cb
\nonumber \\
-\sqrt {{c}^{2}-{y}^{2}}y{a}^{2}+{a}^{2}\arctan
 ( {\frac {\sqrt {{a}^{2}-{y}^{2}}}{y}} ) {b}^{2}-y\sqrt {{
b}^{2}-{y}^{2}}{c}^{2}-{c}^{2}\arctan ( {\frac {\sqrt {{c}^{2}-{y
}^{2}}}{y}} ) {a}^{2}
\nonumber   \\
-\sqrt {{c}^{2}-{y}^{2}}y{b}^{2}+y\sqrt {{b
}^{2}-{y}^{2}}{a}^{2}+{c}^{2}\arctan ( {\frac {\sqrt {{b}^{2}-{y}
^{2}}}{y}} ) {b}^{2}
\nonumber  \\
-{a}^{2}\arctan ( {\frac {\sqrt {{b}^{2 }-{y}^{2}}}{y}} )
{b}^{2}+\sqrt {{a}^{2}-{y}^{2}}y{b}^{2}
\nonumber \\
-{c}^{2 }\arctan ( {\frac {\sqrt {{c}^{2}-{y}^{2}}}{y}} )
{b}^{2}+{ a}^{2}\arctan ( {\frac {\sqrt {{a}^{2}-{y}^{2}}}{y}} )
{c}^ {2} \bigr ]
\\
~ 0 \leq y < a \quad. \nonumber \label{I_1}
\end{eqnarray}
The second piece , $I^{II}(y)$ , is
 \begin{eqnarray}
 I^{II}(y)=  \int _0 ^{\sqrt{b^2-y^2}} 2 C_{ab}^2 dx
 + \int _{\sqrt{b^2-y^2}} ^{\sqrt{c^2-y^2}}  2C_{bc}^2dx \nonumber\\
~= 2 \frac { \,{b}^{2}{{\it C_m}}^{2} } { y (
{b}^{2}-2\,ba+{a}^{2} )  ( {c}^{2}-2\,cb+{b}^{2}
 )
} \bigl [ y\sqrt {{b}^{2}-{y}^{2}}{c}^{2}+{a}^{2 }\arctan ( {\frac
{\sqrt {{b}^{2}-{y}^{2}}}{y}} ) {b}^{2}
\nonumber \\
-{ c}^{2}\arctan ( {\frac {\sqrt {{b}^{2}-{y}^{2}}}{y}} ) {b}^ {2}
\nonumber  \\
-y\sqrt {{b}^{2}-{y}^{2}}{a}^{2}+{c}^{2}\arctan ( {\frac { \sqrt
{{c}^{2}-{y}^{2}}}{y}} ) {a}^{2}+{c}^{2}\arctan ( { \frac {\sqrt
{{c}^{2}-{y}^{2}}}{y}} ) {b}^{2}+\sqrt {{c}^{2}-{y} ^{2}}y{a}^{2}
\nonumber  \\
+\sqrt {{c}^{2}-{y}^{2}}y{b}^{2}+2\,a\ln  ( y
 ) y{b}^{2}+2\,a\ln  ( y ) y{c}^{2}+2\,cy\ln  (
\sqrt {{b}^{2}-{y}^{2}}+b ) {b}^{2}
\nonumber \\
-2\,{a}^{2}\arctan ( { \frac {\sqrt {{b}^{2}-{y}^{2}}}{y}} )
cb-2\,y\sqrt {{b}^{2}-{y}^ {2}}cb-2\,a\ln  ( \sqrt
{{b}^{2}-{y}^{2}}+b ) y{b}^{2}
\nonumber  \\
-2\,a \ln  ( \sqrt {{b}^{2}-{y}^{2}}+b ) y{c}^{2}-2\,{c}^{2}
\arctan ( {\frac {\sqrt {{c}^{2}-{y}^{2}}}{y}} ) ba-2\, \sqrt
{{c}^{2}-{y}^{2}}yba
\nonumber  \\
-2\,c\ln  ( \sqrt {{c}^{2}-{y}^{2}}+c
 ) y{a}^{2}-2\,c\ln  ( \sqrt {{c}^{2}-{y}^{2}}+c ) y{
b}^{2}+2\,{c}^{2}\arctan ( {\frac {\sqrt {{b}^{2}-{y}^{2}}}{y}}
 ) ba
\nonumber   \\
+2\,y\sqrt {{b}^{2}-{y}^{2}}ba+2\,cy\ln  ( \sqrt {{b}^
{2}-{y}^{2}}+b ) {a}^{2}-4\,a\ln  ( y ) ycb+4\,c\ln
 ( \sqrt {{c}^{2}-{y}^{2}}+c ) yba
\bigr ]
 \\
  a \leq y < b  \quad.
\nonumber \label{I_2}
\end{eqnarray}
The third  piece , $I^{III}(y)$ , is
 \begin{eqnarray}
 I^{III}(y)= \int_0 ^{\sqrt{c^2-y^2}}  2C_{bc}^2 dx \nonumber\\
~= 2 \frac { {b}^{2}{{\it C_m}}^{2} } { y \left(
{b}^{2}-2\,ba+{a}^{2} \right)  \left( {c}^{2}-2\,cb+{b}^{2}
 \right)
} \bigl [ y\sqrt {{c}^{2}-{y}^{2}}{b}^{2}+{c}^{2 }\arctan ( {\frac
{\sqrt {{c}^{2}-{y}^{2}}}{y}} ) {b}^{2}
\nonumber \\
-\sqrt {{a}^{2}-{y}^{2}}y{b}^{2} +{c}^{2}\arctan ( {\frac {\sqrt
{{ c}^{2}-{y}^{2}}}{y}} ) {a}^{2}+y\sqrt {{c}^{2}-{y}^{2}}{a}^{2}-
\sqrt {{a}^{2}-{y}^{2}}y{c}^{2}
\nonumber\\
+{a}^{2}\arctan ( {\frac {\sqrt {{ b}^{2}-{y}^{2}}}{y}} ) {b}^{2}
+\sqrt {{b}^{2}-{y}^{2}}y{c}^{2}- \sqrt
{{b}^{2}-{y}^{2}}y{a}^{2}-{c}^{2}\arctan ( {\frac {\sqrt {{
b}^{2}-{y}^{2}}}{y}} ) {b}^{2}
\nonumber  \\
-{a}^{2}\arctan ( {\frac { \sqrt {{a}^{2}-{y}^{2}}}{y}} ) {b}^{2}
-{a}^{2}\arctan ( { \frac {\sqrt {{a}^{2}-{y}^{2}}}{y}} )
{c}^{2}+2\,cy\ln  ( \sqrt {{b}^{2}-{y}^{2}}+b ) {a}^{2}
\nonumber\\
+2\,ay\ln  ( \sqrt {{a}^ {2}-{y}^{2}}+a ) {c}^{2}
+2\,{a}^{2}\arctan ( {\frac {\sqrt {{a}^{2}-{y}^{2}}}{y}} )
cb-2\,a\ln  ( \sqrt {{b}^{2}-{y}^{ 2}}+b ) y{c}^{2}
\nonumber \\
-2\,a\ln  ( \sqrt {{b}^{2}-{y}^{2}}+b
 ) y{b}^{2}
-2\,\sqrt {{b}^{2}-{y}^{2}}ycb-2\,{a}^{2}\arctan
 ( {\frac {\sqrt {{b}^{2}-{y}^{2}}}{y}} ) cb
\nonumber \\
+2\,cy\ln
 ( \sqrt {{b}^{2}-{y}^{2}}+b ) {b}^{2}
-2\,c\ln  ( \sqrt {{c}^{2}-{y}^{2}}+c ) y{b}^{2}-2\,c\ln  ( \sqrt
{{c}^ {2}-{y}^{2}}+c ) y{a}^{2}
\nonumber \\
-2\,y\sqrt {{c}^{2}-{y}^{2}}ba -2\,{c}^{ 2}\arctan ( {\frac {\sqrt
{{c}^{2}-{y}^{2}}}{y}} ) ba+2\,ay \ln  ( \sqrt {{a}^{2}-{y}^{2}}+a
) {b}^{2}
\nonumber \\
+2\,\sqrt {{a}^{2 }-{y}^{2}}ycb +2\,\sqrt
{{b}^{2}-{y}^{2}}yba+2\,{c}^{2}\arctan ( { \frac {\sqrt
{{b}^{2}-{y}^{2}}}{y}} ) ba
\nonumber\\
+4\,c\ln  ( \sqrt { {c}^{2}-{y}^{2}}+c ) yba -4\,ay\ln  ( \sqrt
{{a}^{2}-{y}^{2} }+a ) cb \bigr  ]
\\
 b \leq y < c  \quad.
\label{I_3} \nonumber
\end{eqnarray}

The profile  of ${\it I}$   made by the three pieces (~\ref{I_1}),
(~\ref{I_2}) and  (~\ref{I_3}), can be calibrated on the real data
of A39  and an acceptable match is   realized adopting the
parameters reported in Table~\ref{dataabcsquare}.
 \begin{table}
 \caption[]{Simulation of the PN A39 with 3D diffusion }
 \label{dataabcsquare}
 \[
 \begin{array}{llc}
 \hline
 \hline
 \noalign{\smallskip}
 symbol  & meaning & value  \\
 \noalign{\smallskip}
 \hline
 \noalign{\smallskip}
a  & radius~of~the~internal~absorbing~sphere    &
65.96^{\prime\prime} \\ \noalign{\smallskip} b  &
radius~of~the~shock                        & 80^{\prime\prime}  \\
\noalign{\smallskip} c                 &
radius~of~the~external~absorbing~sphere    & 103.5^{\prime\prime}
\\ \noalign{\smallskip} R_{shell}         &
observed~radius~of~the~shell       & 77^{\prime\prime} \\
\noalign{\smallskip} \delta\,r_{shell}  & observed~ thickness
~of~the~shell       & 10.1^{\prime\prime} \\ \noalign{\smallskip}
\frac {I_{limb}} {I_{center}} &  ratio~ of~ observed~ intensities
& (1.88-2.62)       \\ \noalign{\smallskip} \frac {I_{max}}
{I(y=0)} & ratio~ of~ theoretical~ intensities  & 2.84
\\ \noalign{\smallskip}
 \hline
 \hline
 \end{array}
 \]
 \end {table}

\begin{figure*}
\begin{center}
\includegraphics[width=10cm]{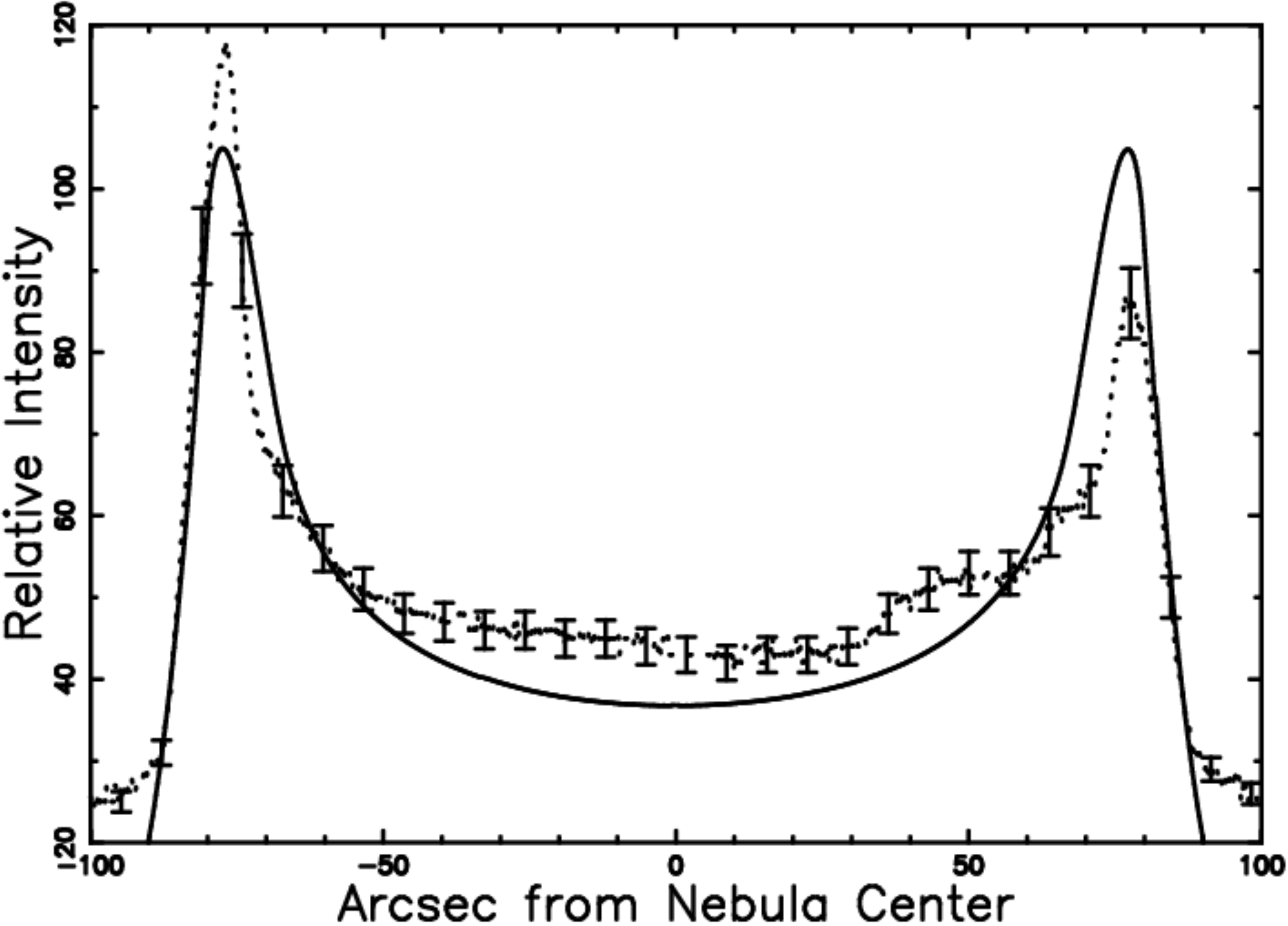}
\end {center}
\caption {
 Cut of the mathematical  intensity ${\it I}$
 (formulas (~\ref{I_1}), (~\ref{I_2}) and  (~\ref{I_3})) ,
 crossing the center    (full    line  ) of PN A39
 and  real data         (dotted  line with some error bar).
 The number of data is  801  and
for  this  model $\chi^2$ = 19.03 against $\chi^2$ = 12.60 of the
rim model fully described in ~Jacoby et al. (2001). }
\label{pn_cut_square}
    \end{figure*}

\begin{figure*}
\begin{center}
\includegraphics[width=10cm]{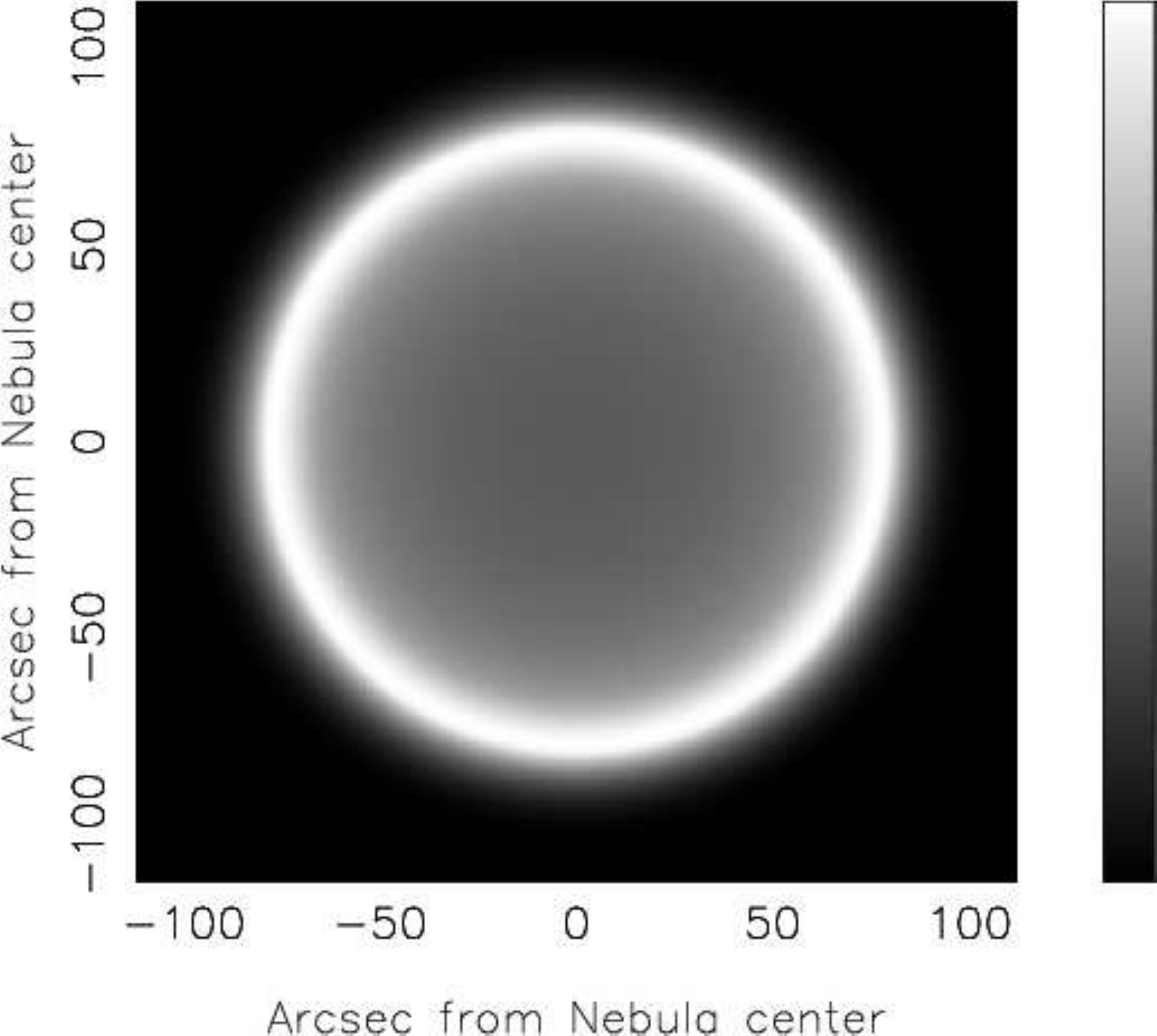}
\end {center}
\caption { Contour map  of  ${\it I}$ particularized to simulate
the PN A39. } \label{pnbri_square}
    \end{figure*}

The theoretical intensity can therefore be plotted as a function
of the distance from the center , see~Figure~\ref{pn_cut_square},
or as  an    image , see~Figure~\ref{pnbri_square}.

\begin{figure*}
\begin{center}
\includegraphics[width=10cm]{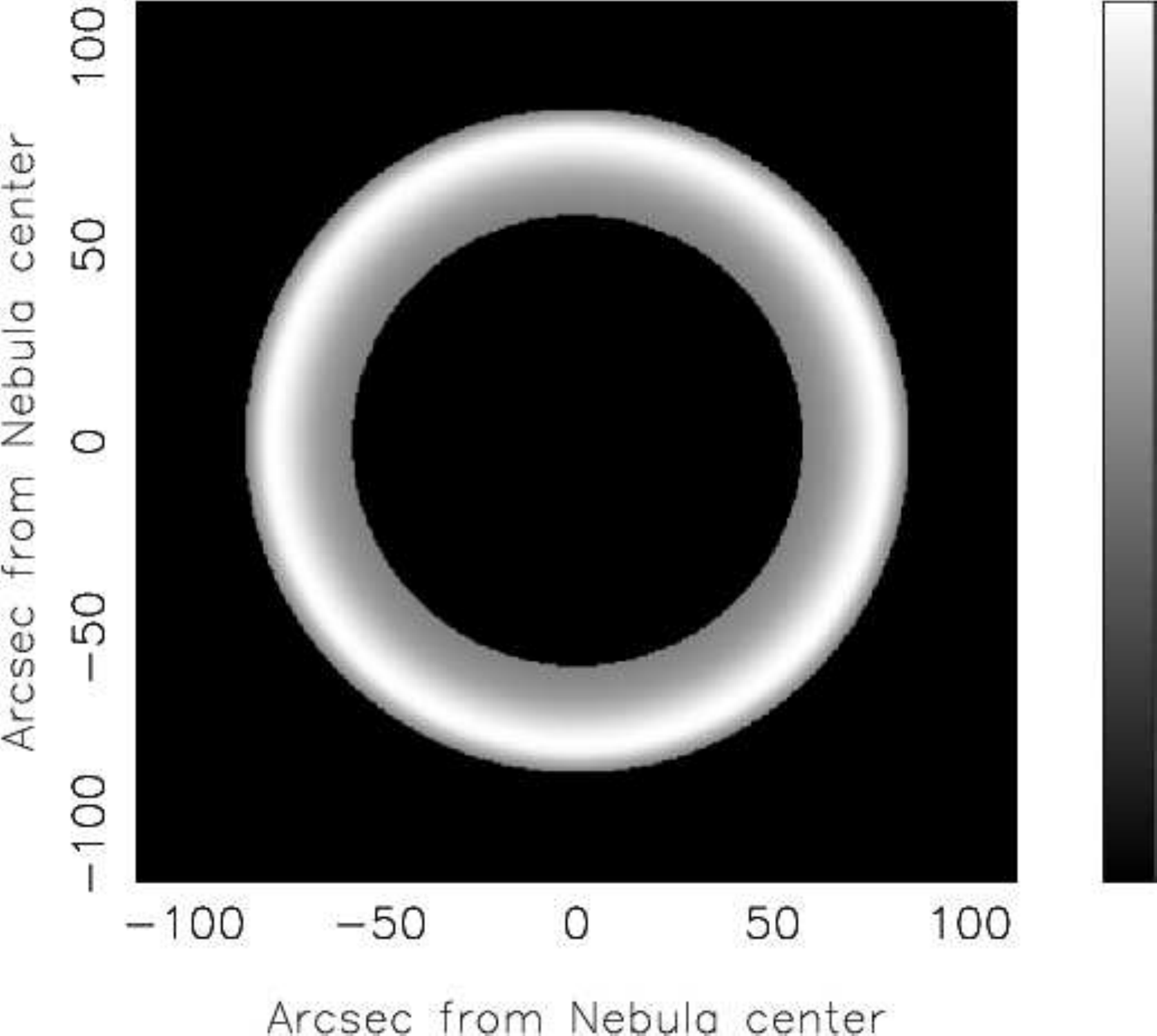}
\end {center}
\caption {
 The same  as  Figure~\ref{pnbri_square}  but with
 $I_{tr}= I_{max}/2$
} \label{hole_square}
    \end{figure*}

The effect of the  insertion of a threshold intensity , $I_{tr}$,
given by the observational techniques , is now analyzed. The
threshold intensity can be parametrized  to  $I_{max}$, the
maximum  value  of intensity characterizing the ring: a typical
image with  a hole  is visible in  Figure~\ref{hole_square} when
$I_{tr}= I_{max}/2$.

The position of  the minimum of ${\it I}$  is at $y=0$ and the
position of the maximum is situated at $y=b$.

The ratio between the theoretical intensity at maximum , $I_{max}$
at $y=b$  ,
 and at the minimum ($y=0$)
is given by
\begin{equation}
\frac {I_{max}} {I(y=0)} = \frac{Numerator}{Denominator} \quad,
\label{ratioteorsquare}
\end {equation}
where
\begin{eqnarray}
 Numerator=
( {b}^{2}-2\,ba+{a}^{2} )
\nonumber\\
\times  ( 2\,cb\ln  ( b
 ) -2\,c\ln  ( \sqrt {{c}^{2}-{b}^{2}}+c ) b+b\sqrt {
{c}^{2}-{b}^{2}}+{c}^{2}\arctan ( {\frac {\sqrt {{c}^{2}-{b}^{2}}
}{b}} )  )
 \quad,
\end{eqnarray}
and
\begin{eqnarray}
 Denominator= \\
2\,b ( {a}^{2}c-{c}^{2}a-2\,bca\ln  ( a ) +2\,bca\ln
 ( c ) -b{a}^{2}+b{c}^{2}-{b}^{2}c+{b}^{2}a+{b}^{2}a\ln
 ( a )
\nonumber \\
-{c}^{2}a\ln  ( b ) +{b}^{2}c\ln  ( b ) -{b}^{2}a\ln  ( b )
-{b}^{2}c\ln  ( c
 ) +{a}^{2}c\ln  ( b ) +{c}^{2}a\ln  ( a
 ) -{a}^{2}c\ln  ( c )  ) \nonumber
\quad.
\end{eqnarray}

The  ratio rim(maximum) /center(minimum)  of the observed
intensities  as well as the theoretical one  are reported in
Table~\ref{dataabcsquare} for A39 \cite{Zaninetti2009a}.

\subsection{3D diffusion from a sphere, linear dependence}

The  concentration rises from 0 at {\it r=a}  to a maximum value
$C_m$ at {\it r=b} and then  falls again to 0 at {\it  r=c}. The
concentrations to be used are formulas (\ref{cab}) and (\ref{cbc})
once $r=\sqrt{x^2+y^2}$ is imposed; these two concentrations are
inserted in formula~(\ref{eqn_transfer}) which  represents the
transfer equation.
 The geometry of the phenomenon fixes
 three different zones ($0-a,a-b,b-c$) for the variable $y$,
 see~\cite{Zaninetti2007_c,Zaninetti2009a};
 the first segment, $I^I(y)$, is
\begin{eqnarray}
I^I(y)=
\nonumber  \\
2 {\frac {b{\it C_m} \sqrt {{a}^{2}-{y}^{2}}}{-b+a}}-2 {\frac {b{
\it C_m} a\ln  \left( \sqrt {{a}^{2}-{y}^{2}}+a \right) }{-b+a}}
-2 { \frac {b{\it C_m} \sqrt {{b}^{2}-{y}^{2}}}{-b+a}} \nonumber
\\
+2 {\frac {b{\it C_m}
 a\ln  \left( \sqrt {{b}^{2}-{y}^{2}}+b \right) }{-b+a}}
 +2 {\frac {b {\it C_m} c\ln  \left( \sqrt {{b}^{2}-{y}^{2}}+b
\right) }{-c+b}} \nonumber \\ -2 { \frac {b{\it C_m} \sqrt
{{b}^{2}-{y}^{2}}}{-c+b}}  -2 {\frac {b{\it C_m}
 c\ln  \left( \sqrt {{c}^{2}-{y}^{2}}+c \right) }{-c+b}}+2 {\frac {b
{\it C_m} \sqrt {{c}^{2}-{y}^{2}}}{-c+b}}
\label{I_1l} \\
~ 0 \leq y < a \quad.  \nonumber
\end{eqnarray}

The second segment, $I^{II}(y)$, is
 \begin{eqnarray}
 I^{II}(y)=
-{\frac {b{\it C_m} a\ln  \left( {y}^{2} \right) }{-b+a}}-2 {\frac
{b {\it C_m} \sqrt {{b}^{2}-{y}^{2}}}{-b+a}}
\nonumber \\
+2 {\frac {b{\it C_m} a\ln
 \left( \sqrt {{b}^{2}-{y}^{2}}+b \right) }{-b+a}}
 +2 {\frac {b{\it C_m } c\ln  \left( \sqrt {{b}^{2}-{y}^{2}}+b
\right) }{-c+b}}
\nonumber \\
-2 {\frac { b{\it C_m} \sqrt {{b}^{2}-{y}^{2}}}{-c+b}}-2 {\frac
{b{\it C_m} c\ln
 \left( \sqrt {{c}^{2}-{y}^{2}}+c \right) }{-c+b}}
+2 {\frac {b{\it C_m } \sqrt {{c}^{2}-{y}^{2}}}{-c+b}}
\label{I_2l} \\
  a \leq y < b  \quad. \nonumber
\end{eqnarray}
The third segment, $I^{III}(y)$, is
 \begin{eqnarray}
 I^{III}(y)=
 {\frac {b{\it C_m} c\ln  \left( {y}^{2} \right) }{-c+b}}-2 {\frac
{b{ \it C_m} c\ln  \left( \sqrt {{c}^{2}-{y}^{2}}+c \right)
}{-c+b}} +2 { \frac {b{\it C_m} \sqrt {{c}^{2}-{y}^{2}}}{-c+b}}
 \\
 b \leq y < c  \quad.  \nonumber
\label{I_3l}
\end{eqnarray}

\begin{figure*}
\begin{center}
\includegraphics[width=10cm]{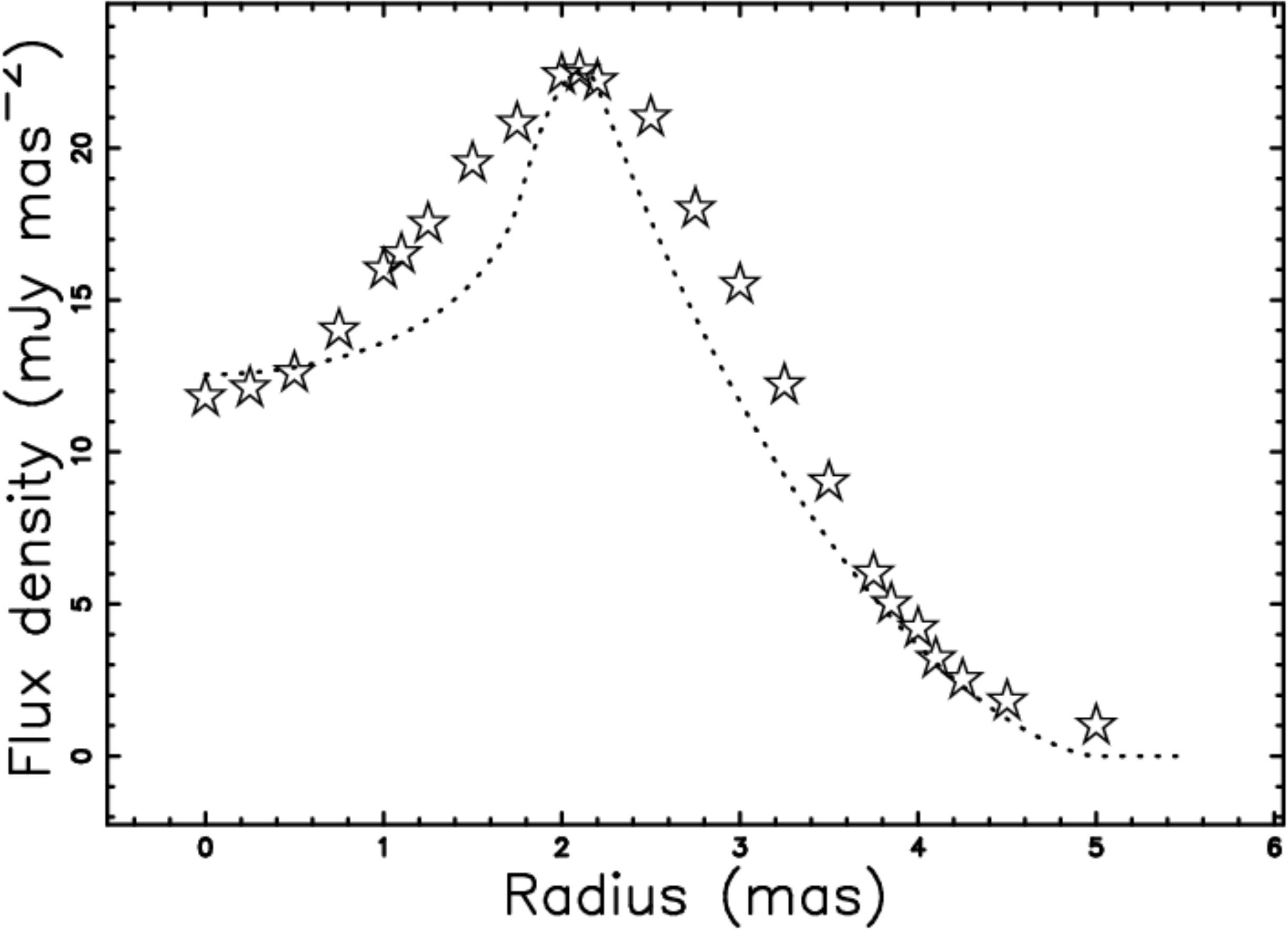}
\end {center}
\caption {
 Cross-section of the mathematical  intensity ${\it I}$
 (formulas (\ref{I_1l}), (\ref{I_2l}) and  (\ref{I_3l})),
 through the center  (dotted  line) of \snr
 and  real data  (empty stars),
 $\chi^2$ = 100.49
The real data made on  day  1889 of SNR \snr after  the explosion
have been extracted  by the author from Figure   3 of Marcaide et al.
(2009). Parameters as in Table~\ref{dataabc}. } \label{pn_cut}
    \end{figure*}

\begin{figure*}
\begin{center}
\includegraphics[width=10cm]{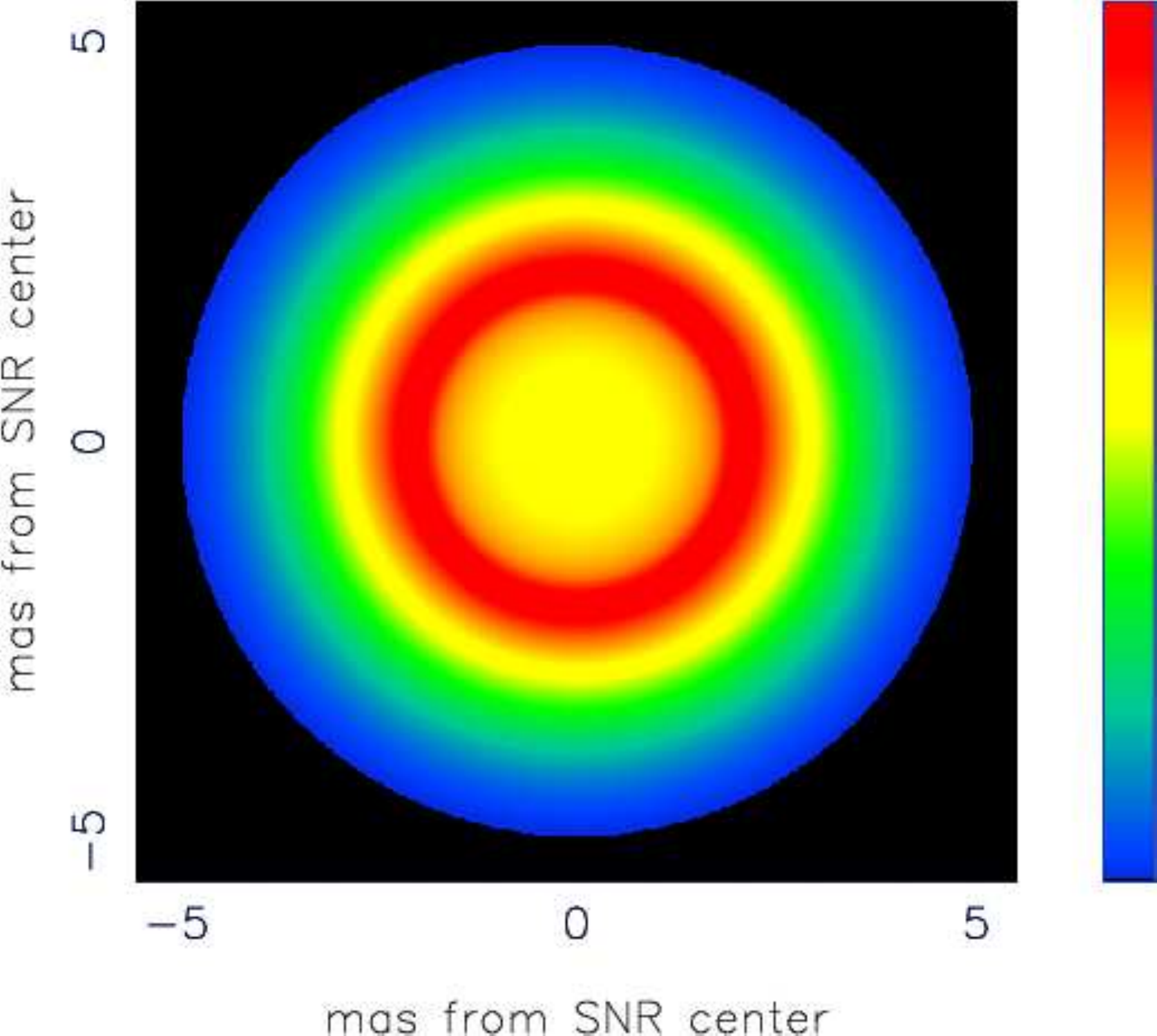}
\end {center}
\caption { Contour map  of  ${\it I}$ adjusted to simulate the SNR
\snr. Parameters as in Table~\ref{dataabc}. } \label{pnbri}
    \end{figure*}

The profile of ${\it I}$ made up of the three segments
(\ref{I_1l}), (\ref{I_2l}) and  (\ref{I_3l}), can be calibrated
against the real data of \snr  and an acceptable match can be
achieved by adopting the parameters reported in
Table~\ref{dataabc}.
 \begin{table}
 \caption[]{Simulation of the SNR  \snr by 3D diffusion,
  optically thin case}
 \label{dataabc}
 \[
 \begin{array}{lll}
 \hline
 \hline
 \noalign{\smallskip}
 symbol  & meaning & value  \\
 \noalign{\smallskip}
 \hline
 \noalign{\smallskip}
a  &  radius ~internal~sphere    & 1.76 (mas)  \\
\noalign{\smallskip} b  & radius~shock                        &
2.2  (mas)  \\ \noalign{\smallskip} c  & radius~external~sphere
& 5.0  (mas)  \\ \noalign{\smallskip} \frac {I_{limb}}
{I_{center}} &  ratio~observed~ intensities  & 1.7926       \\
\noalign{\smallskip} \frac {I_{max}} {I(y=0)} & ratio~
theoretical~ intensities  & 1.7927       \\ \noalign{\smallskip}
 \hline
 \hline
 \end{array}
 \]
 \end {table}
The theoretical intensity can therefore be plotted as a function
of the distance from the center, see Figure  \ref{pn_cut}, or as  a
contour map, see Figure  \ref{pnbri}.

\begin{figure*}
\begin{center}
\includegraphics[width=10cm]{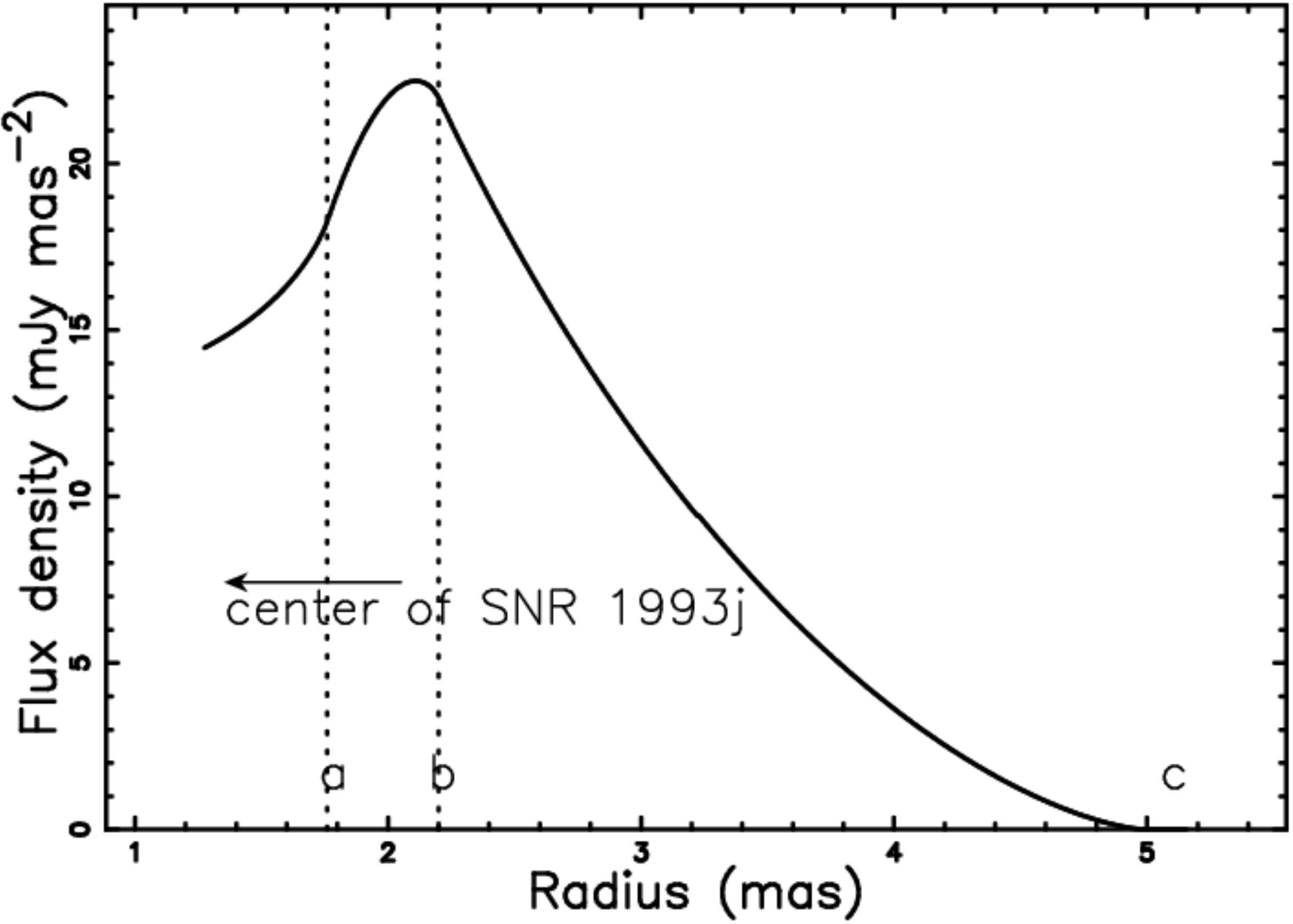}
\end {center}
\caption {
 Cross-section through the mathematical  intensity ${\it I}$
 towards the edge of the SNR \snr.
 The three  parameters which  characterize
 the expanding PN, {\it a  }, {\it b} and  {\it c},
 are reported.
Parameters as in Table~\ref{dataabc}. } \label{pn_cut_exp}
    \end{figure*}

The position of  the minimum of ${\it I}$  is at $y=0$ and the
position of the maximum is situated in the region $a \leq y < b $,
or more precisely at:
\begin{equation}
y={\frac {\sqrt {- \left( b-2\,a+c \right) a \left( ab-2\,bc+ac
\right) }}{b-2\,a+c}} \quad. \label{maximum}
\end{equation}
This  means that the maximum emission  is not at the position of
the shock, identified here as $b$, but shifted a little towards
the center; see Figure  \ref{pn_cut_exp}.

The ratio between the theoretical intensity at maximum , $I_{max}$
, as given by formula~(\ref{maximum}) and at minimum ($y=0$) is
given by

\begin{equation}
\frac {I_{max}} {I(y=0)} = \frac{Numerator}{Denominator} \quad,
\label{ratioteor}
\end {equation}
where
\begin{eqnarray}
 Numerator= \nonumber\\
 -a\ln  \left( -{\frac {\left( ac+ab-2\,cb \right) a}{b+c-2\,a}}
 \right) c+a\ln  \left( -{\frac {\left( ac+ab-2\,cb \right) a}{b+c-2
\,a}} \right) b-2\,\sqrt {{\frac {\left( c+b \right)  \left( b-a
 \right) ^{2}}{b+c-2\,a}}}c \nonumber\\
 -2\,a\ln  \left( \sqrt {{\frac {\left( c+b
 \right)  \left( b-a \right) ^{2}}{b+c-2\,a}}}+b \right) b+
\nonumber\\ 2\,c\ln
 \left( \sqrt {{\frac {\left( c+b \right)  \left( b-a \right) ^{2}}{b
+c-2\,a}}}+b \right) b+2\,\sqrt {{\frac {\left( c+b \right) \left(
b -a \right) ^{2}}{b+c-2\,a}}}a
\nonumber \\
 -2\,c\ln  \left( \sqrt
{{\frac {\left( a -c \right) ^{2} \left( c+b \right)
}{b+c-2\,a}}}+c \right) b+ \nonumber \\
2\,c\ln
 \left( \sqrt {{\frac {\left( a-c \right) ^{2} \left( c+b \right) }{b
+c-2\,a}}}+c \right) a+2\,\sqrt {{\frac {\left( a-c \right) ^{2}
 \left( c+b \right) }{b+c-2\,a}}}b-
 \nonumber \\
 2\,\sqrt {{\frac {\left( a-c
 \right) ^{2} \left( c+b \right) }{b+c-2\,a}}}a
 \quad,
\end{eqnarray}
and
\begin{eqnarray}
 Denominator= \nonumber\\
-2\,ac\ln  \left( a \right) +2\,ba\ln  \left( a \right) -2\,ba\ln
 \left( b \right) +2\,bc\ln  \left( b \right)
 -2\,bc\ln  \left( c
 \right) +2\,ac\ln  \left( c \right)
\quad.
\end{eqnarray}
 The observed ratio as well as the theoretical ratio are reported
in Table~\ref{dataabc}.

\begin{figure*}
\begin{center}
\includegraphics[width=10cm]{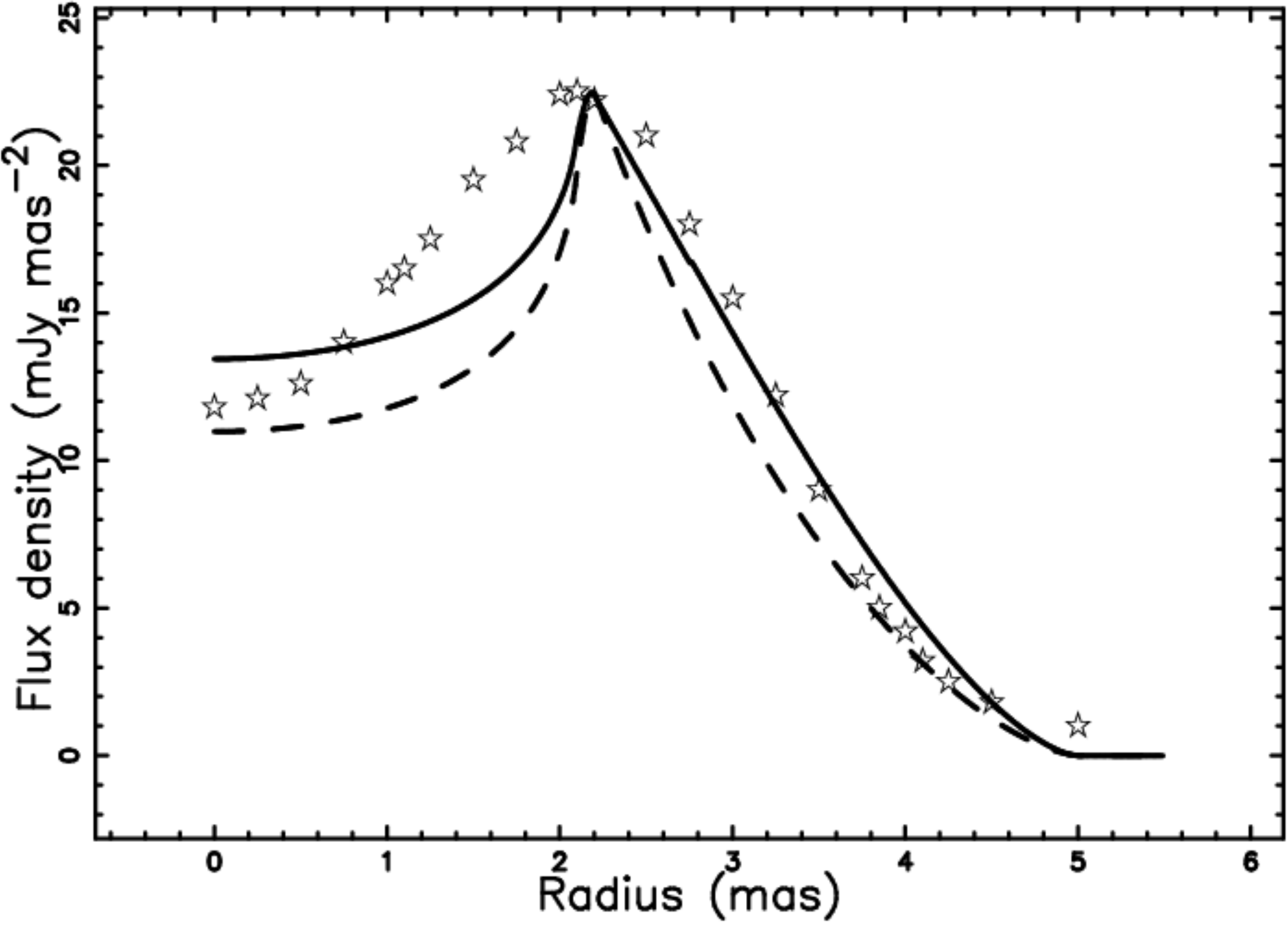}
\end {center}
\caption {
 Cross-section through the mathematical intensity ${\it I}$
 (formulas (\ref{I_1l}), (\ref{I_2l}) and  (\ref{I_3l})),
in the optically thin case (dashed line, $\chi^2=237.3$), and
optically thick case (full line, $\chi^2=84.7$)
 and  real data (empty stars) for SNR \snr.
Parameters as in Table~\ref{dataabcabsorption}. }
\label{absorption}
    \end{figure*}

The effect of  absorption is easily evaluated by applying
formula~(\ref{transition})  and fixing the value of $K_a$. The
result  is shown in Figure  \ref{absorption}.

 \begin{table}
 \caption[]{Simulation of the SNR  \snr with 3D diffusion,
  optically thick case with $K_a=0.2$.}
 \label{dataabcabsorption}
 \[
 \begin{array}{llc}
 \hline
 \hline
 \noalign{\smallskip}
 symbol  & meaning & value  \\
 \noalign{\smallskip}
 \hline
 \noalign{\smallskip}
a  & radius~internal~sphere    & 2.01 (mas)  \\
\noalign{\smallskip} b  & radius~of~shock                        &
2.2  (mas)  \\ \noalign{\smallskip}
c  & radius~external~sphere    & 5.0  (mas)  \\
\noalign{\smallskip} \frac {I_{limb}} {I_{center}} &
ratio~observed~ intensities  & 1.7926       \\
\noalign{\smallskip} \frac {I_{max}} {I(y=0)} & ratio~optically~
thin~case & 2.0491      \\ \noalign{\smallskip} \frac {I_{max}}
{I(y=0)} & ratio~optically~ thick~case& 1.6741      \\
\noalign{\smallskip}
 \hline
 \hline
 \end{array}
 \]
 \end {table}
\subsection{3D complex morphologies of PN }
\label{seccomplexpn}
\begin{figure}
  \begin{center}
\includegraphics[width=10cm]{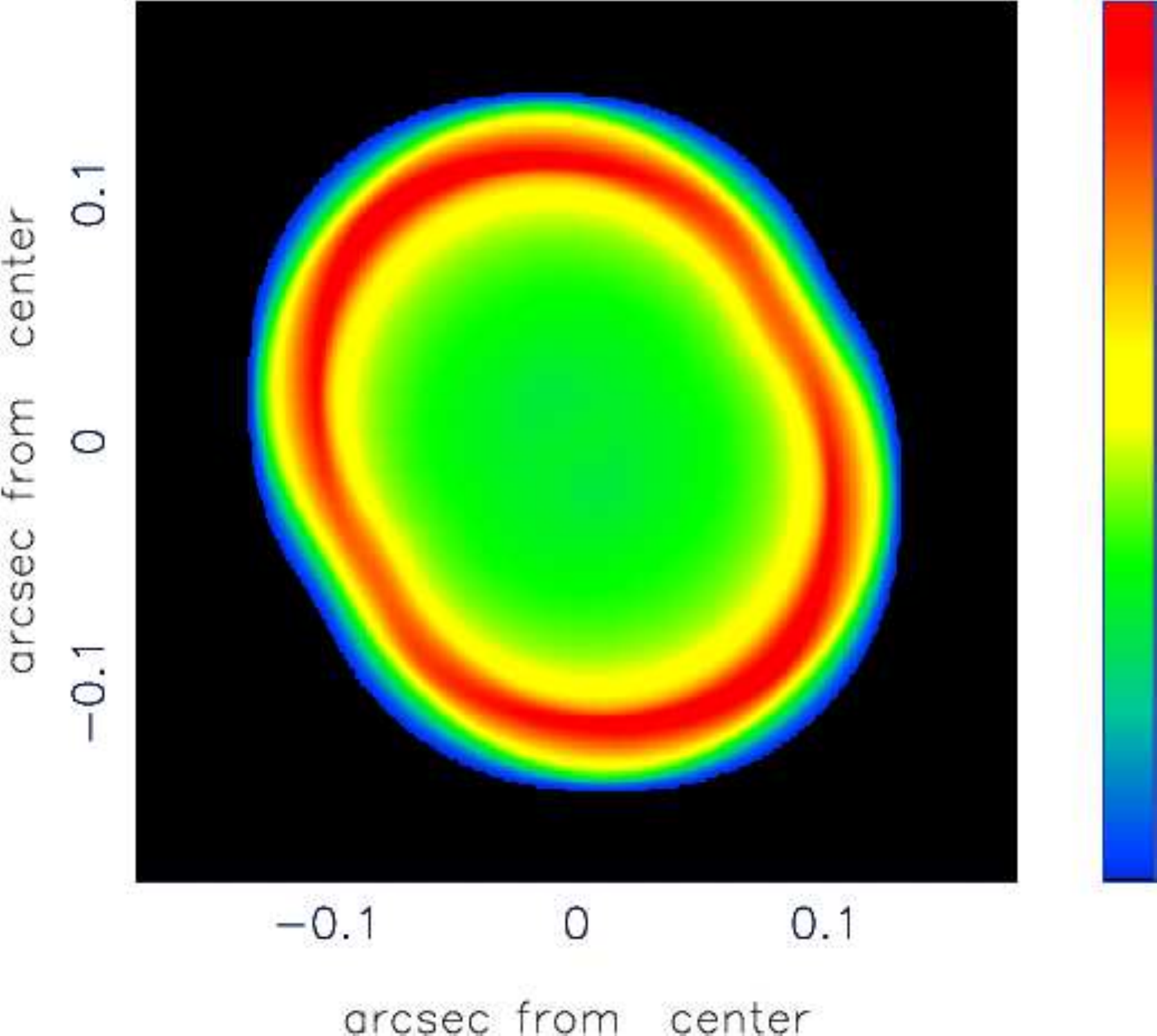}
  \end {center}
\caption {
Map of the theoretical intensity  of the PN  Ring nebula.
Physical parameters as in Table~\ref{parameters} and
$f$=12 .
 The three Eulerian angles 
 characterizing the point of view are 
     $ \Phi $=180    $^{\circ }  $, 
     $ \Theta $=90   $^{\circ }$
and  $ \Psi $=-30    $^{\circ }   $.
          }%
    \label{ring_heat}
    \end{figure}

\begin{figure*}
\begin{center}
\includegraphics[width=10cm]{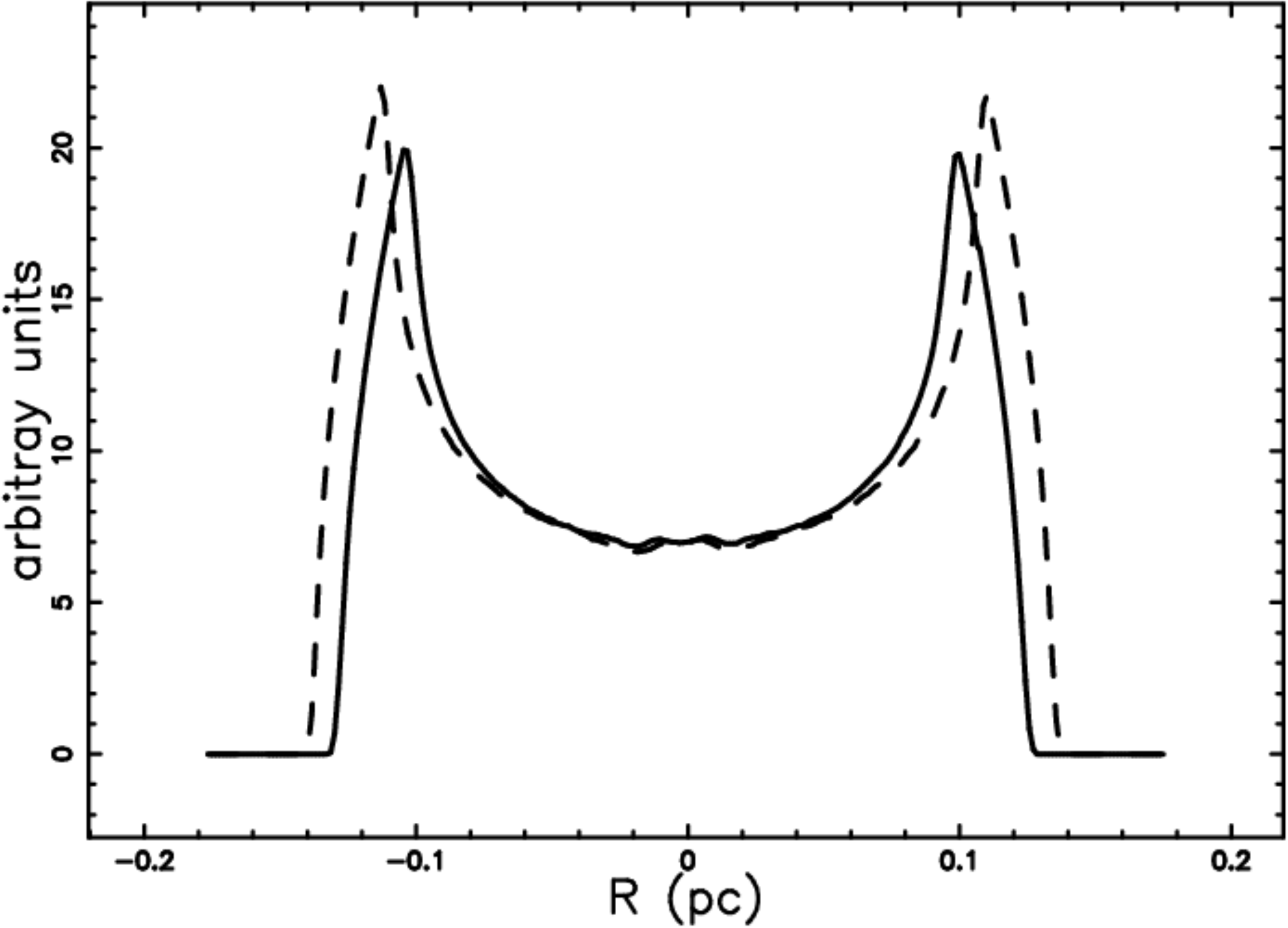}
\end {center}
\caption
{
 Two  cut of the mathematical  intensity ${\it I}$
 crossing the center    of the PN Ring nebula:
 equatorial cut          (full line)
 and  polar cut          (dotted line) .
}
\label{cut_xy_ring}
    \end{figure*}

The numerical approach to the intensity map can be implemented
when the ellipsoid that characterizes the expansion surface 
of the PN has a constant thickness  expressed , for example ,
as  $r_{min}/f $ where $r_{min}$ is the minimum radius of the
ellipsoid and $f$ an integer.
We remember that  $f=12$ has a physical basis 
in the symmetrical case  , see~\cite{Dalgarno1987}.
The numerical algorithm that allows us  to build the 
image  is now outlined 
\begin{itemize}
\item
A memory grid  ${\mathcal {M}} (i,j,k)$ that  contains 
$NDIM^3$ pixels is considered
\item
The points of the thick ellipsoid  are memorized
on  ${\mathcal {M}}$
\item
Each point of  ${\mathcal {M}} $  has spatial coordinates
$x,y,z$ 
which  can be  represented  by
the following $1 \times 3$  matrix ,$A$,
\begin{equation}
A=
 \left[ \begin {array}{c} x \\\noalign{\medskip}y\\\noalign{\medskip}{
\it z}\end {array} \right] 
\quad  .
\end{equation}
The point of view of the observer is characterized by the
Eulerian  angles   $(\Phi, \Theta, \Psi)$
and  therefore  by a total rotation 
 $3 \times 3$  matrix ,
$E$ , see \cite{Goldstein2002}. 
The matrix point  is now 
represented by the following $1 \times 3$  matrix , $B$,
\begin{equation}
B = E \cdot A 
\quad .
\end{equation}
\item 
The map in intensity is obtained by summing the points of the 
rotated images along a direction 
, for example along z , 
 ( sum over the range of one index, for example k ).
\end{itemize}
Figure~\ref{ring_heat} reports the rotated image 
of the Ring nebula and  Figure~\ref{cut_xy_ring} 
reports two cuts along the polar and equatorial 
directions.

\begin{figure*}
\begin{center}
\includegraphics[width=10cm]{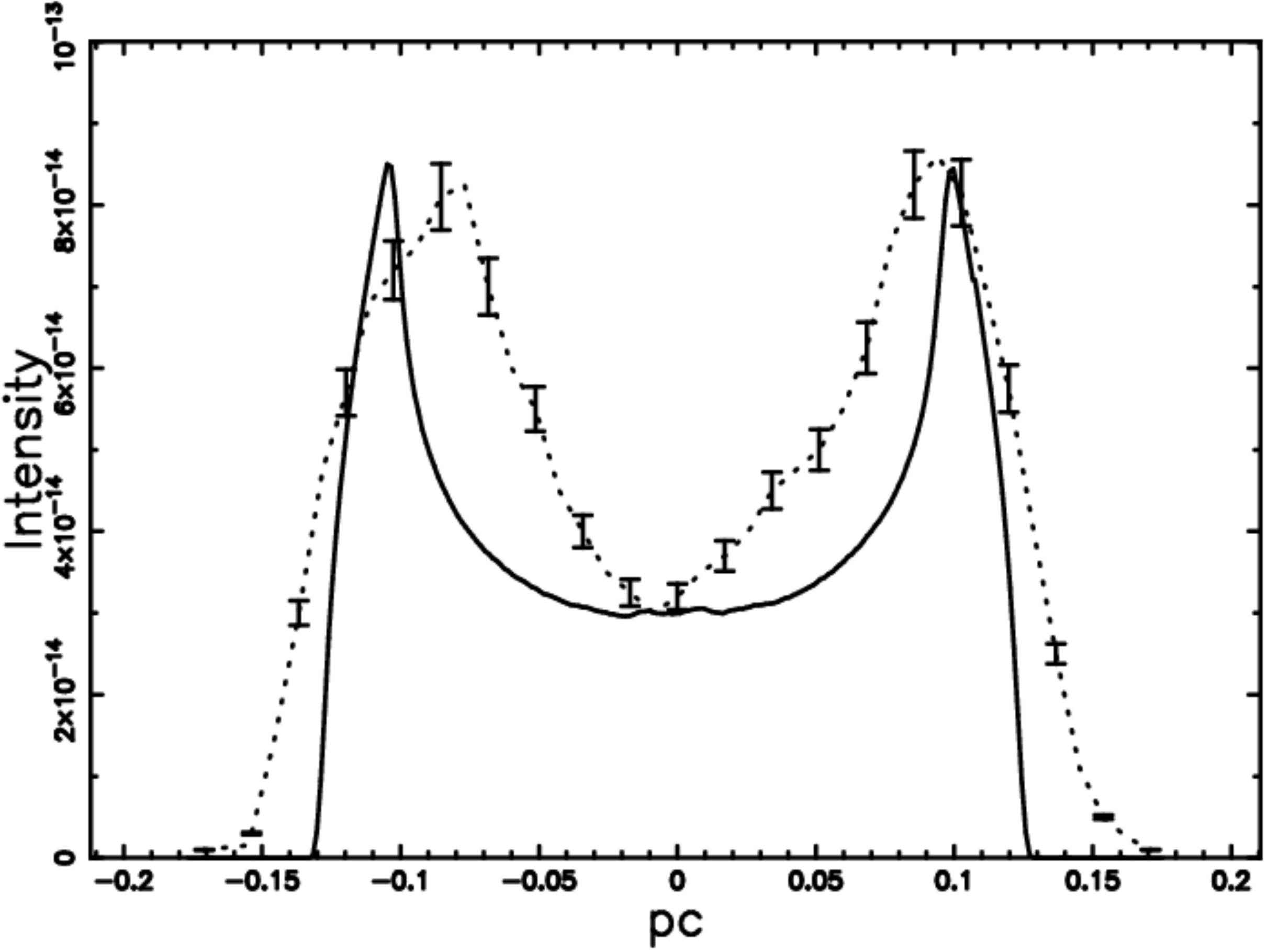}
\end {center}
\caption
{
 Cut of the mathematical  intensity ${\it I}$
 of the PN Ring Nebula      
 crossing the center    (full  line  ) 
 and  real data  of $H_{\beta}$ 
 (dotted line with some error bar ) .
 The number of data is  250  and 
for  this  model $\chi^2$ = 15.53~.
The real data are extracted  by the author 
from Figure 1 of Garnett and Dinerstein 2001.
}
\label{cut_confronto}
    \end{figure*}

Figure~\ref{cut_confronto} reports 
the comparison between  a theoretical and observed
east-west cut in $H_{\beta}$ 
that cross the center of the nebula, see Figure~1 in 
\cite{Garnett2001}.

A comparison can be made   with the color composite image
of Doppler-shifted $H_2$ emission as represented
in Figure~2 in \cite {Hiriart2004}.

\begin{figure}
  \begin{center}
\includegraphics[width=10cm]{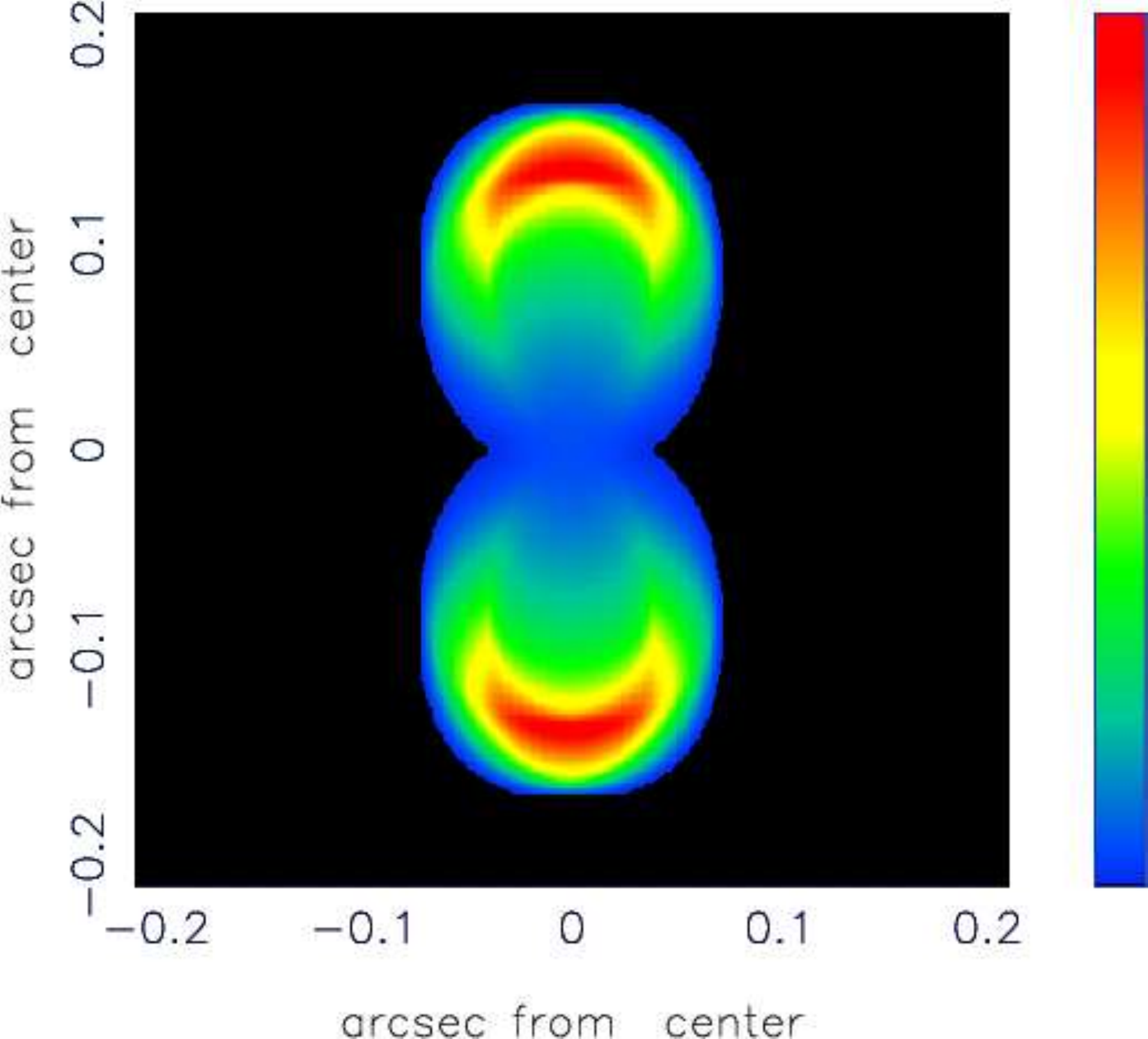}
  \end {center}
\caption {
Map of the theoretical intensity  of the PN  MyCn 18 .
Physical parameters as in Table~\ref{parametershom} and
$f$=12 .
The three Eulerian angles 
characterizing the point of view are $ \Phi $=180   $^{\circ }$
, $ \Theta $=90   $^{\circ }$
and  $ \Psi $=0  $^{\circ }$.
          }%
    \label{mycn18_heat}
    \end{figure}

\begin{figure}
  \begin{center}
\includegraphics[width=10cm]{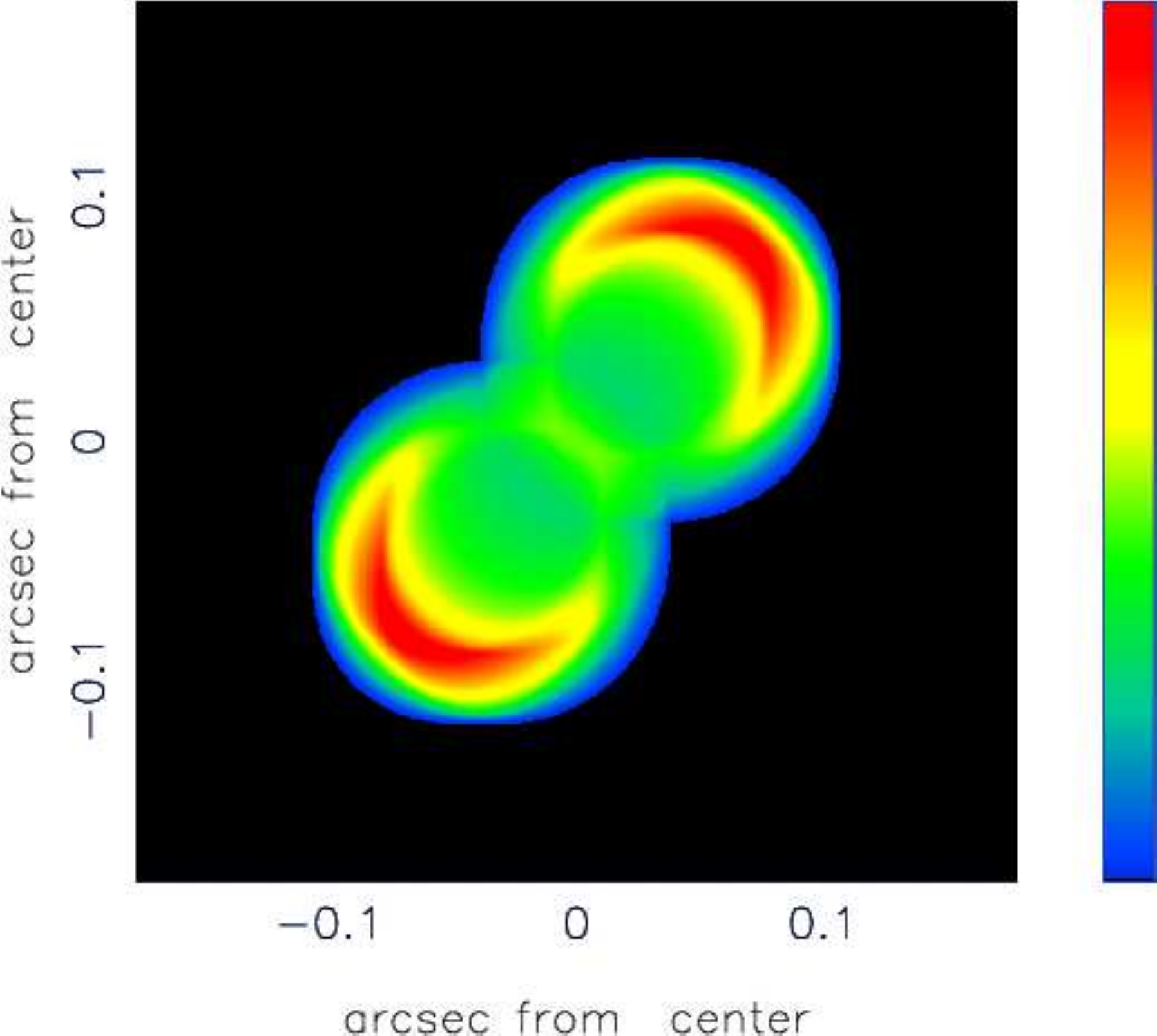}
  \end {center}
\caption {
Map of the theoretical intensity  of the 
rotated PN MyCn 18 .
Physical parameters as in Table~\ref{parametershom} and
$f$=12 .
The three Eulerian angles 
characterizing the point of view are 
     $ \Phi   $=130     $^{\circ }  $, 
     $ \Theta $=40   $^{\circ }$
and  $ \Psi   $=5     $^{\circ }   $.
          }%
    \label{mycn1840_heat}
    \end{figure}

\begin{figure*}
\begin{center}
\includegraphics[width=10cm]{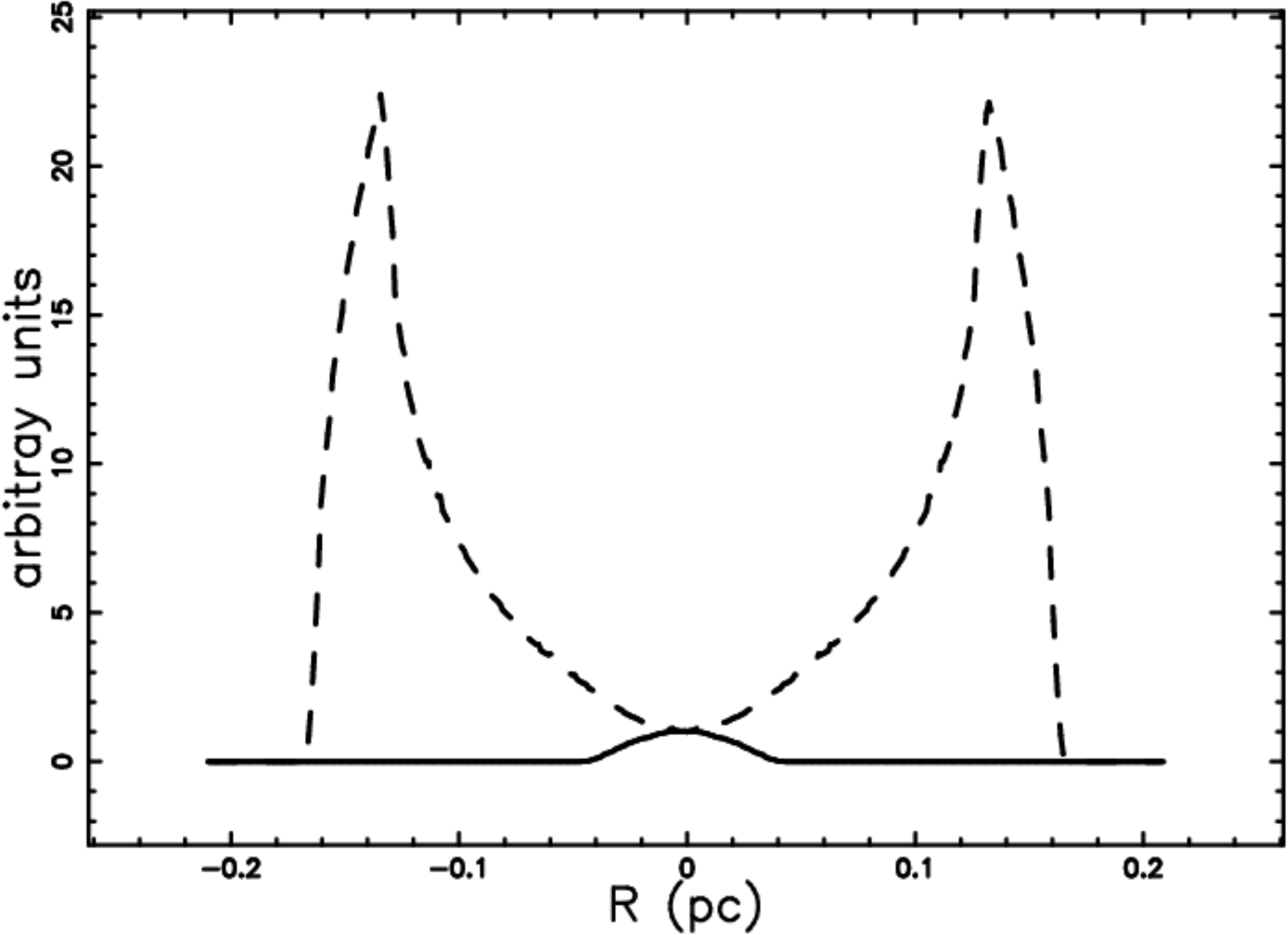}
\end {center}
\caption
{
 Two  cut of the mathematical  intensity ${\it I}$
 crossing the center    of the  PN  MyCn 18  :
 equatorial cut          (full line)
 and  polar cut          (dotted line) .
 Parameters as in Figure~\ref{mycn18_heat}.
}
\label{cut_xy_mycn18}
    \end{figure*}

\begin{figure*}
\begin{center}
\includegraphics[width=10cm]{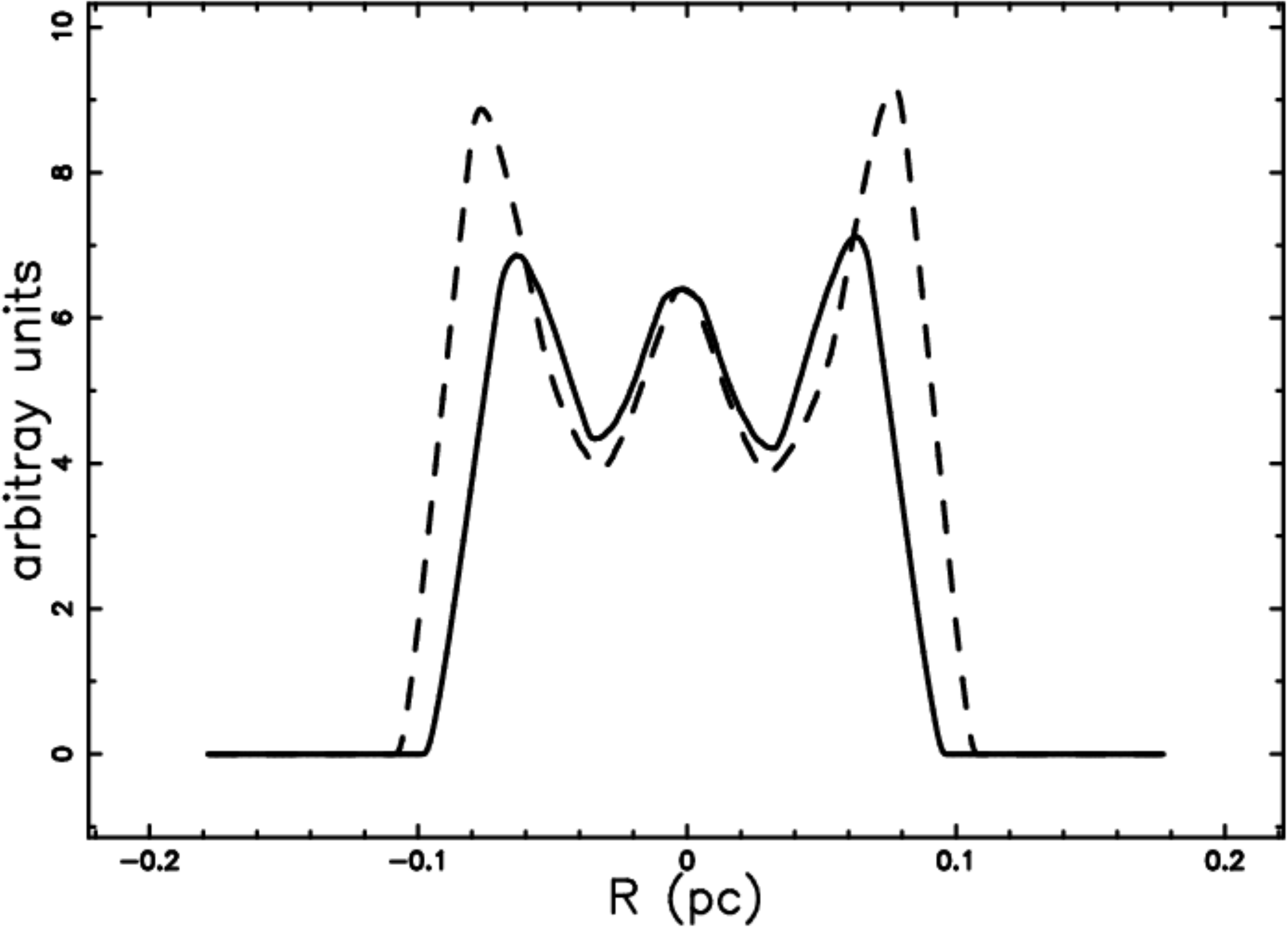}
\end {center}
\caption
{
 Two  cut of the mathematical  intensity ${\it I}$
 crossing the center    of the rotated  PN MyCn 18 nebula:
 equatorial cut          (full line)
 and  polar cut          (dotted line) .
 Parameters as in Figure~\ref{mycn1840_heat}.
}
\label{cut_xy_mycn1840}
    \end{figure*}

In order to explain some of the morphologies
which characterize the PN's we first map 
MyCn 18    with the polar axis in the vertical 
direction , see map in intensity in 
Figure~\ref{mycn18_heat}. 
The vertical  and horizontal cut in intensity
are reported in Figure~\ref{cut_xy_mycn18}.
The point of view of the observer as modeled 
by the Euler angles increases  the complexity
of the shapes : Figure~\ref{mycn1840_heat}  
reports the after rotation image 
and  Figure~\ref{cut_xy_mycn1840} the vertical and
horizontal rotated cut.
The after rotation image contains the double ring
and an enhancement in  intensity of  the central
region which  characterize
MyCn 18.

 
This central enhancement can be considered
one of the various morphologies that the PNs present
and is similar to  model $BL_1-F$ in Figure~3 
of  the Atlas of synthetic line profiles by \cite{Morisset2008}. 

\subsection{3D complex morphology of the hybrid \etacar }

\begin{figure}
  \begin{center}
\includegraphics[width=10cm]{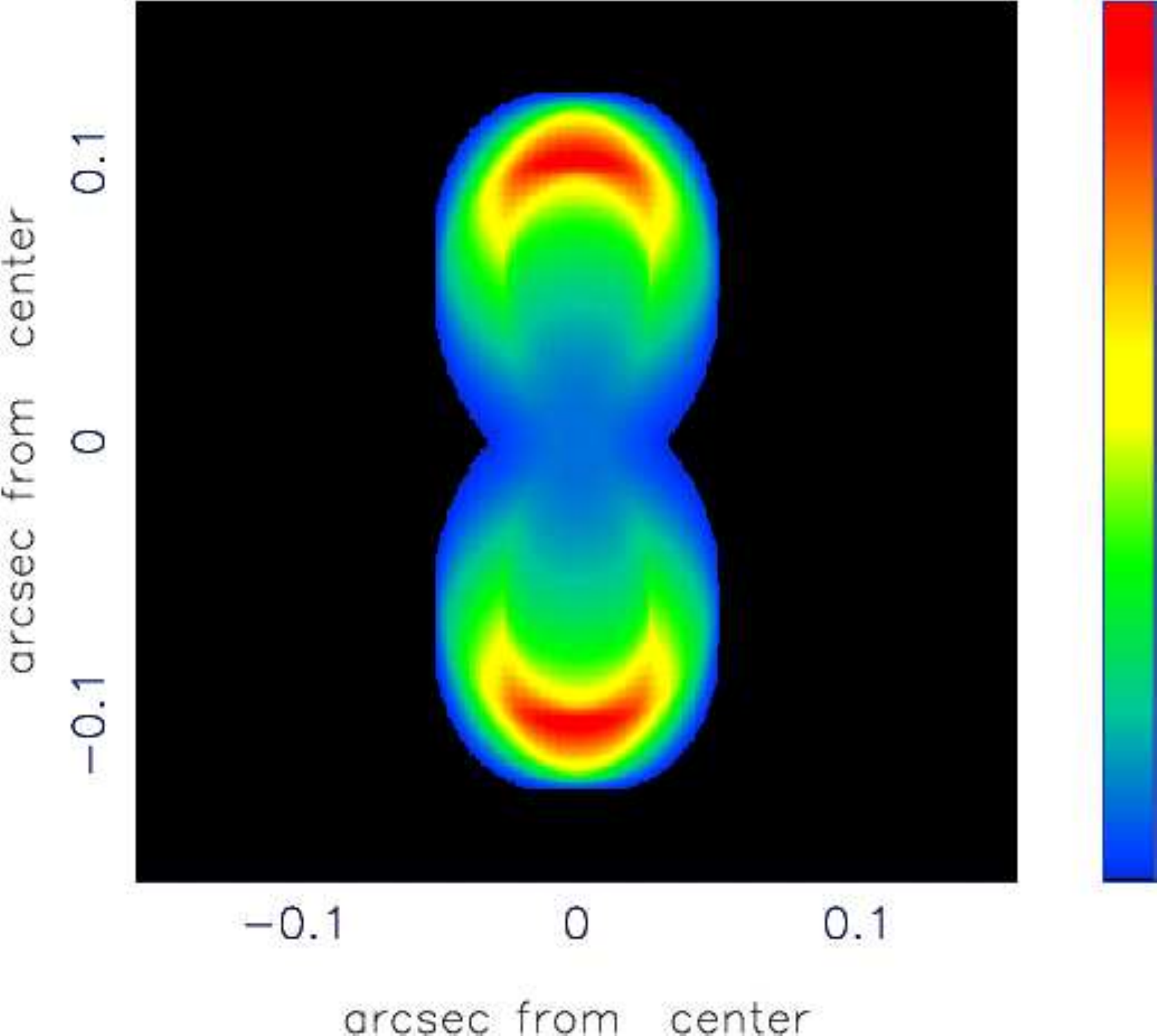}
  \end {center}
\caption {
Map of the theoretical intensity  of the hybrid 
Homunculus/\etacar  nebula
in the presence of an exponentially varying medium.
Physical parameters as in Table~\ref{parametershom}.
The three Euler angles
characterizing the   orientation
  are $ \Phi $=180$^{\circ }$,
$ \Theta $=90$^{\circ }$
and $ \Psi $=0$^{\circ }$.
This  combination of Euler angles corresponds
to the rotated image with the polar axis along the
z-axis.}%
    \label{eta_heat}
    \end{figure}

\begin{figure}
  \begin{center}
\includegraphics[width=10cm]{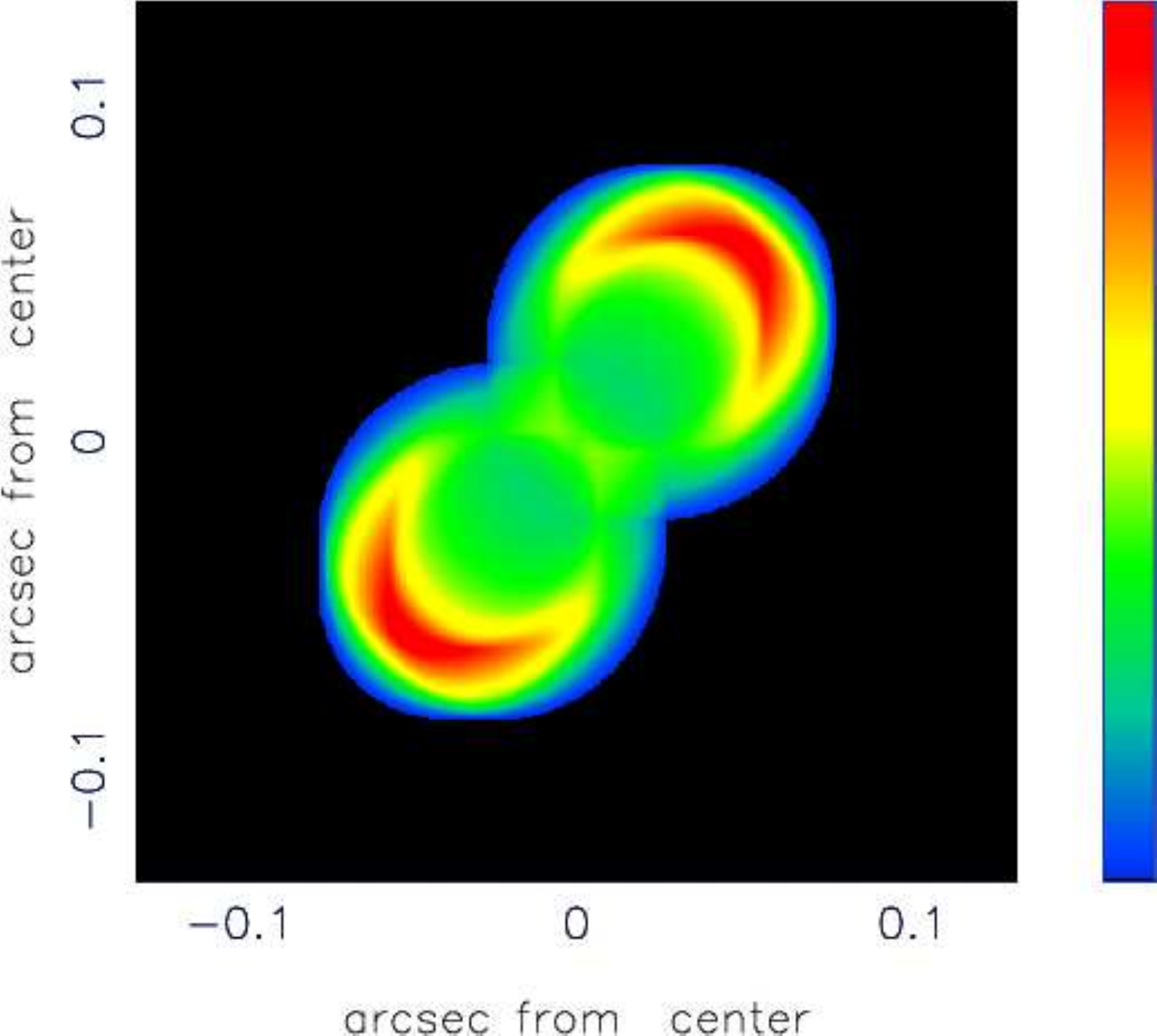}
  \end {center}
\caption {
Model map of the of the hybrid 
Homunculus/\etacar  nebula rotated in
accordance with the observations, for an exponentially
varying
medium.
Physical parameters as in Table~\ref{parametershom}.
The three Euler  angles characterizing
the orientation of the observer
are
     $ \Phi   $=130$^{\circ }$,
     $ \Theta $=40$^{\circ }$
and  $ \Psi   $=-140$^{\circ }$.
This  combination of Euler angles corresponds
to the observed image.
          }%
    \label{eta40_heat}
    \end{figure}

Here we adopt  the numerical algorithm developed 
in the previous Section \ref{seccomplexpn}.
An  ideal image of the
Homunculus
nebula having the polar axis aligned with the z-direction
which means  polar axis along the z-direction,
is shown in Figure~\ref{eta_heat}
and this should be compared
with the $H_2$ emission structure reported
in Figure~4 of \cite{Smith2006}.
A model for a realistically rotated Homunculus
is shown in Figure \ref{eta40_heat}.
This should be compared with
Figure 1 in \cite{Smith2000} or
Figure 1 in \cite{Smith2006}.

\begin{figure*}
\begin{center}
\includegraphics[width=10cm]{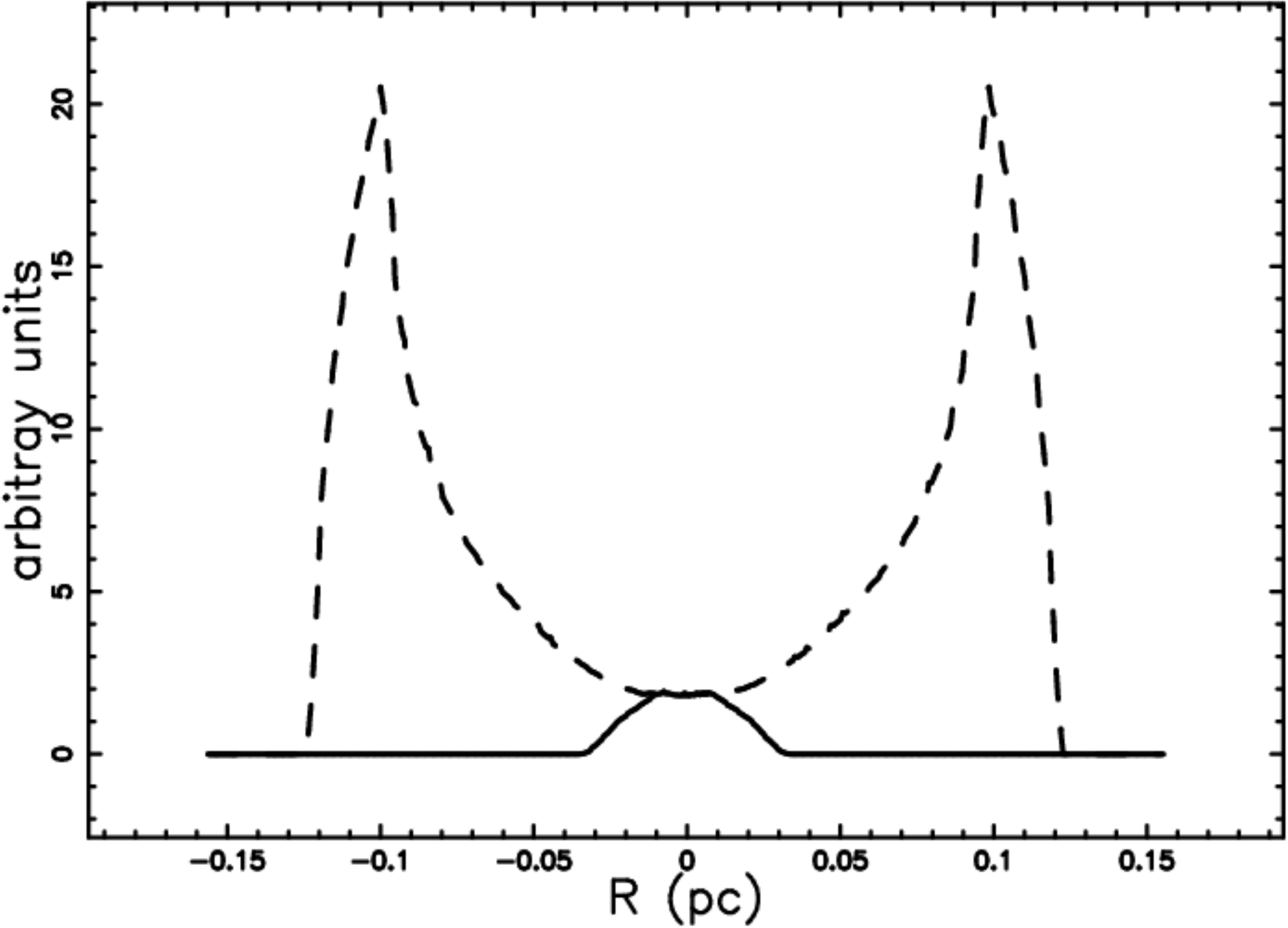}
\end {center}
\caption
{
 Two cuts of the model intensity
 across the center of the hybrid 
Homunculus/\etacar  nebula
 for  an exponentially varying medium:
 equatorial cut (full line)
 and polar cut  (dotted line).
 Parameters as in Figure~\ref{eta_heat}.
}
\label{cut_xy_eta}
    \end{figure*}

\begin{figure*}
\begin{center}
\includegraphics[width=10cm]{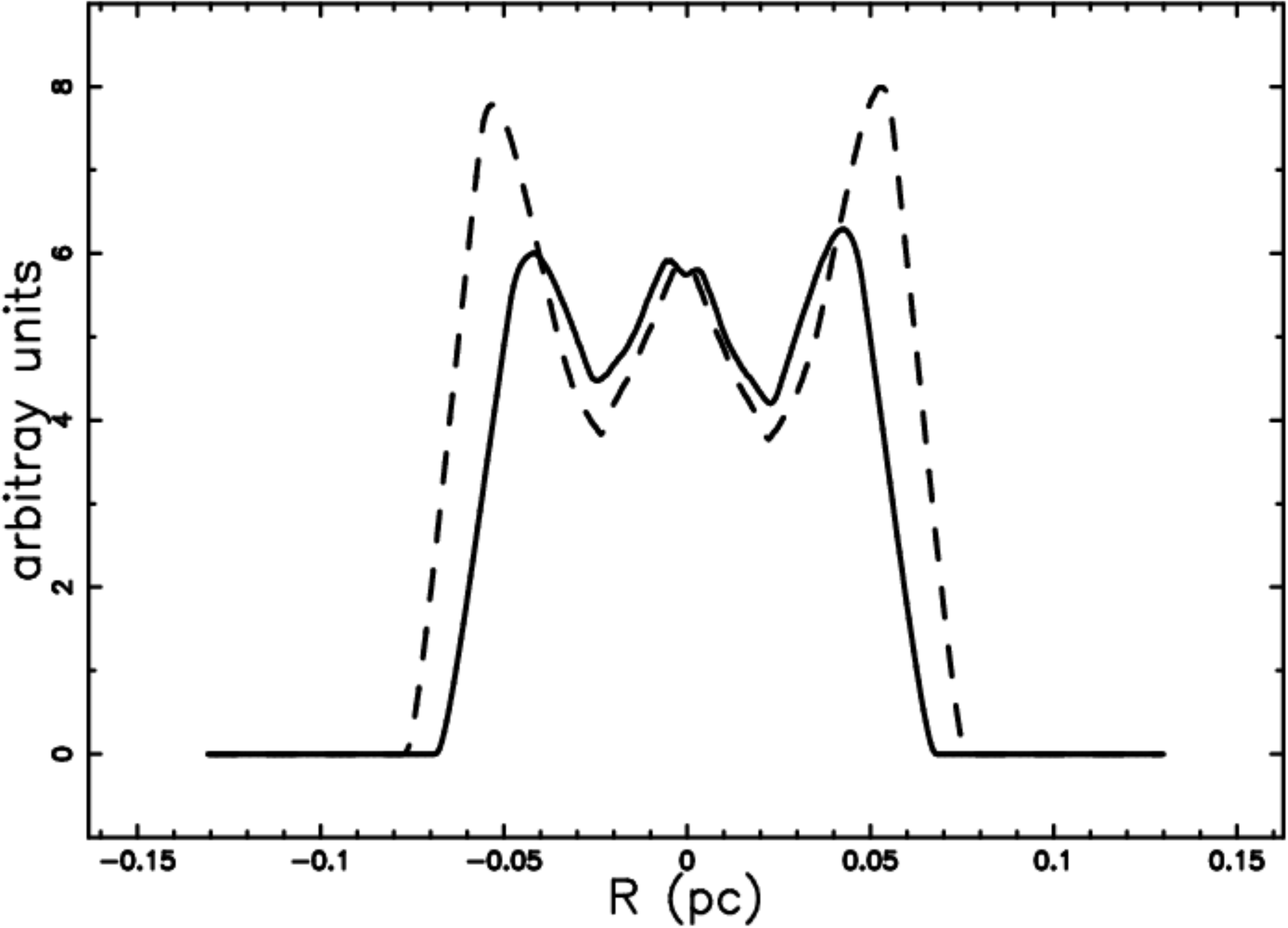}
\end {center}
\caption
{
 Two cuts of the model intensity
 across the center
 of the realistically rotated
 hybrid Homunculus/$\eta-car$  nebula
 for  an exponentially varying medium:
 equatorial cut (full line)
 and polar cut  (dotted line).
 Parameters as in Figure~\ref{eta40_heat}.
}
\label{cut_xy_eta40}
    \end{figure*}

The rotated image exhibits
a double ring
and an intensity enhancement in the central
region which characterizes the little
Homunculus,
see~\cite{Smith2002,Ishibashi2003,Smith_2005,Gonzales_2006}.
Figure~\ref{cut_xy_eta} and Figure~\ref{cut_xy_eta40} show two
cuts through the Homunculus nebula
without and with rotation.
The intensity enhancement is due to a projection effect
and is an alternative for the theory that associates
the little Homunculus
with  an eruption occurring some time
after the Great Eruption, see
\cite{Ishibashi2003,Smith_2005}. We briefly recall that  a central
enhancement is visible in one of the various
morphologies characterizing planetary nebulae.
This can be compared with   the model $BL_1-F$ in Figure~3 of  the Atlas of
synthetic line profiles by \cite{Morisset2008}.

Such cuts are common when analyzing planetary nebulae.
 As an example  Figure 4 in \cite{Jacoby2001} reports a
nearly symmetrical profile of the intensity
in the [OIII] image of A39, a nearly
spherical planetary  nebula.
Another example is the east-west cut in
$H{\beta}$
for the elliptical Ring nebula, crossing
the center of the nebula, see Figure~1
in \cite{Garnett2001}.
Such intensity
cuts  are not yet available
for $\eta$-Carinae and therefore can represent a new target for the
observers.

\subsection{3D complex morphology  of a SNR , \sn1006}

The theory of an asymmetric SNR  was developed in 
Sect.~4.1 of 
\cite{Zaninetti2000}  
in which an expansion surface
as a function of a non-homogeneous  ISM was computed:
in the same   paper Figure~8  models  SN1006.
The diffusing algorithm  adopted here is the 3D random walk 
from many injection points ( in the following IP)

\begin {enumerate}
\item    The first IP is chosen 
\item    The first of the  NPART  electrons is chosen.
\item    The random  walk of an electron starts where the
         selected IP is situated. 
         The electron moves in one of the six possible directions.

\item    After N steps the process restarts from (2)

\item    The number of visits is  recorded on ${\mathcal M^3}$ ,
         a three--dimensional grid.
\item    The random walk terminates when all  the NPART
         electrons are processed.
\item    The process restarts from (1) selecting another IP 
\item    For the sake  of normalization the
         one--dimensional visitation/concentration grid
         ${\mathcal M^3}$ is divided by  NPART.
\end  {enumerate}

The {\it IP}\/  are randomly  selected in 
space, and  the   radius is computed 
by using the method  of  bilinear interpolation on the four 
grid points that surround the selected latitude and longitude,
( 
\cite{press}
).
The  radius will be the selected value  
+ {\it R}/24  in order  to generate the  IP where 
the action of the shock is maximum. 

\begin{figure*}
\begin{center}
\includegraphics[width=4cm]{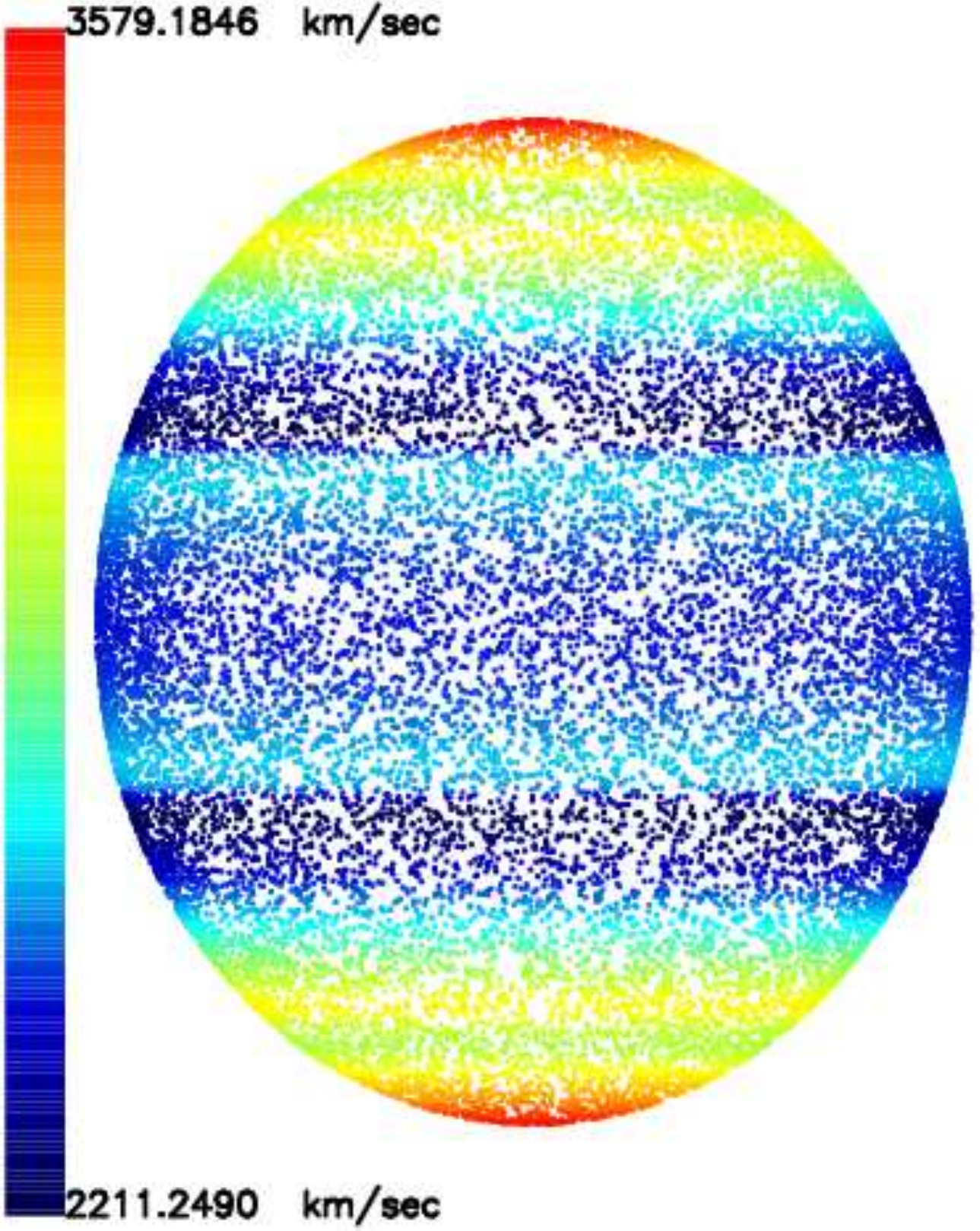}
\end {center}
\caption 
{
  Map of the expansion velocity  
  relative to the simulation of SNR  \sn1006 
  when 190000 random points are selected on the surface.
  The   physical parameters are the  same as in 
  Figure~8 of 
  \cite{Zaninetti2000}.
}
\label{sn1006_velo} 
    \end{figure*}

Our model gives radial velocities , $V_{theo}$ , 
2211~$km~s^{-1}$ $\leq~V_{theo}~\leq$ 3580~$km~s^{-1}$
and the map of the expansion velocity 
is reported in Figure~\ref{sn1006_velo}
from which it is possible to visualize the differences 
in the expansion velocities among the various regions
as well as the overall elliptical shape.

Before  continuing  
we should recall that 
in  the presence of discrete time steps on a 3D lattice the average
square radius ,$\langle R^2(N  )\rangle$,  after N steps
(see  
\cite {gould}, 
equation~(12.5~))  is
\begin {equation}
\langle R^2(N  )\rangle  \sim  6 DN \quad,
\label{rquadro}
\end   {equation}
from which the diffusion coefficient , $D$ , is derived
\begin {equation}
D= \frac {\langle R^2(N  )\rangle} {6 N} \quad.
\label{r2N}
\end   {equation}
The two boundaries in which the random walk is  taking 
place are now represented by two irregular surfaces.
It is  possible to simulate them by stopping the random walk  
after a number of iterations $N$ given by  
\begin {equation}
N = NINT (\frac {\overline R_{pc}}{24}\frac  {1}{\delta})^2 
\quad, 
\end {equation}
where $\overline R_{pc}$ represents the averaged radius in pc .
These are the iterations  after which according, 
to formula~(\ref{r2N}),  the walkers reach the boundaries at 
a radial distance given by  $\frac {\overline R_{pc}}{24}$
from the place of injection; 
in other words we are working
on an unbounded  lattice . 
The influence of velocity  
on the  flux  $F$ of radiation  can be inferred from 
the suspected  dependence  when  non-thermal emission
is considered, see equation~(9.29)  in
\cite{Mckee},  
\begin{equation}
 F= \chi_t \frac {1}{4} \mu_H n_0 v_s^3  
\quad, 
\end {equation}
where $\chi_T$ represents  the efficiency of 
conversion of the unitarian flux of kinetic energy,
$\mu_H$ the mass of the hydrogen nucleus,
$n_0$ the  particles/$cm^3$ and 
$v_s$ the velocity of the shock.
 
\begin{figure*}
\begin{center}
\includegraphics[width=10cm]{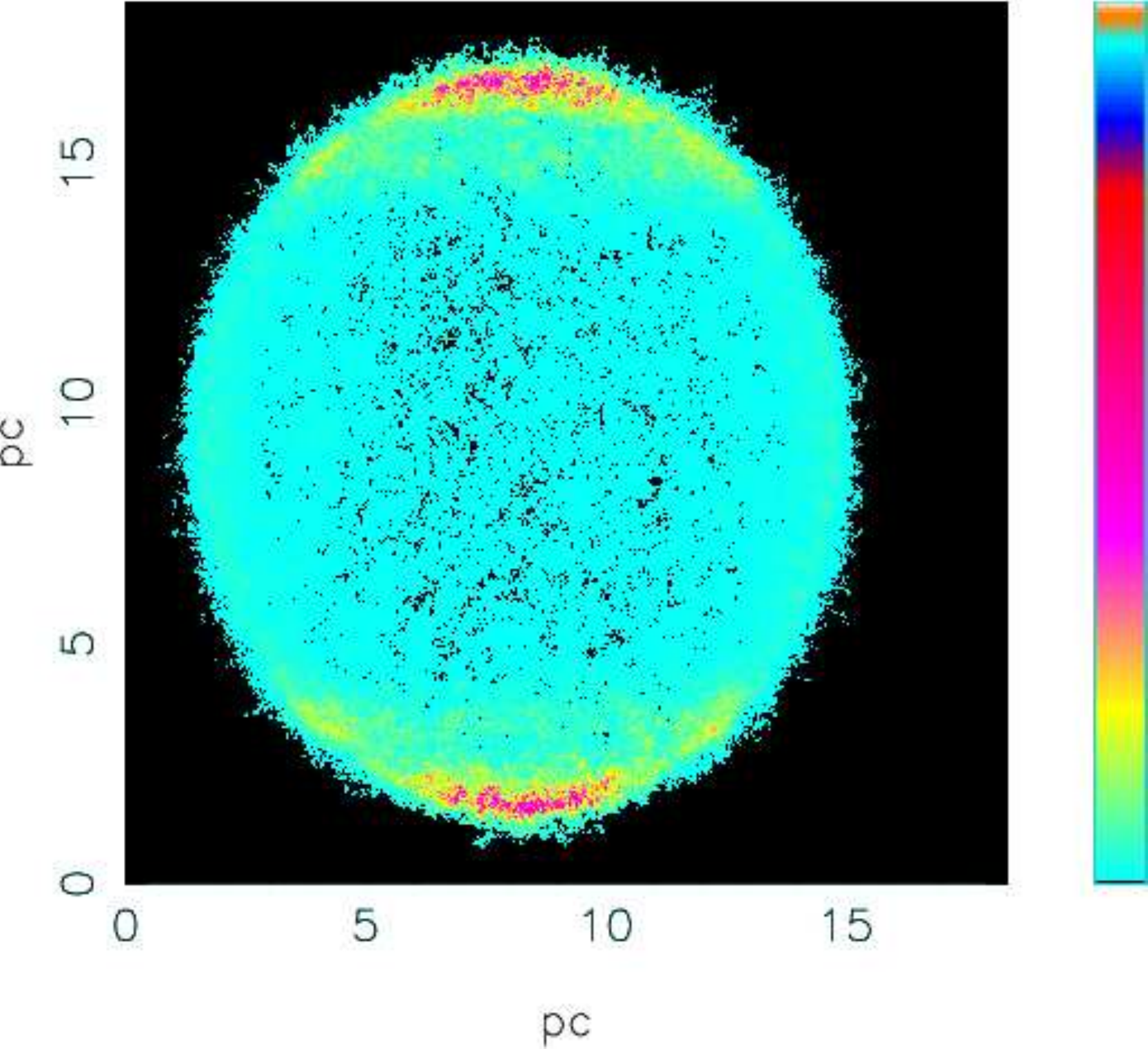}
\end {center}
\caption
{
 Contour  of the intensity {\it I} in the X-rays
of  SNR  \sn1006.  
 The parameters are 
 $ side_{SNR}$=18.37 pc ,
 $\delta=6.12~10^{-3}$~pc,
 $\rho  =2.8~10^{-3}$~pc,
 NDIM=3001,
 $IP=190000^2$ , 
 {\it NPART= 100}, 
 $NTIMES_{max}$=18,
 great box.
Optically thin layer.
}
\label{sn1006_colore}
    \end{figure*}

\begin{figure*}
\begin{center}
\includegraphics[width=10cm]{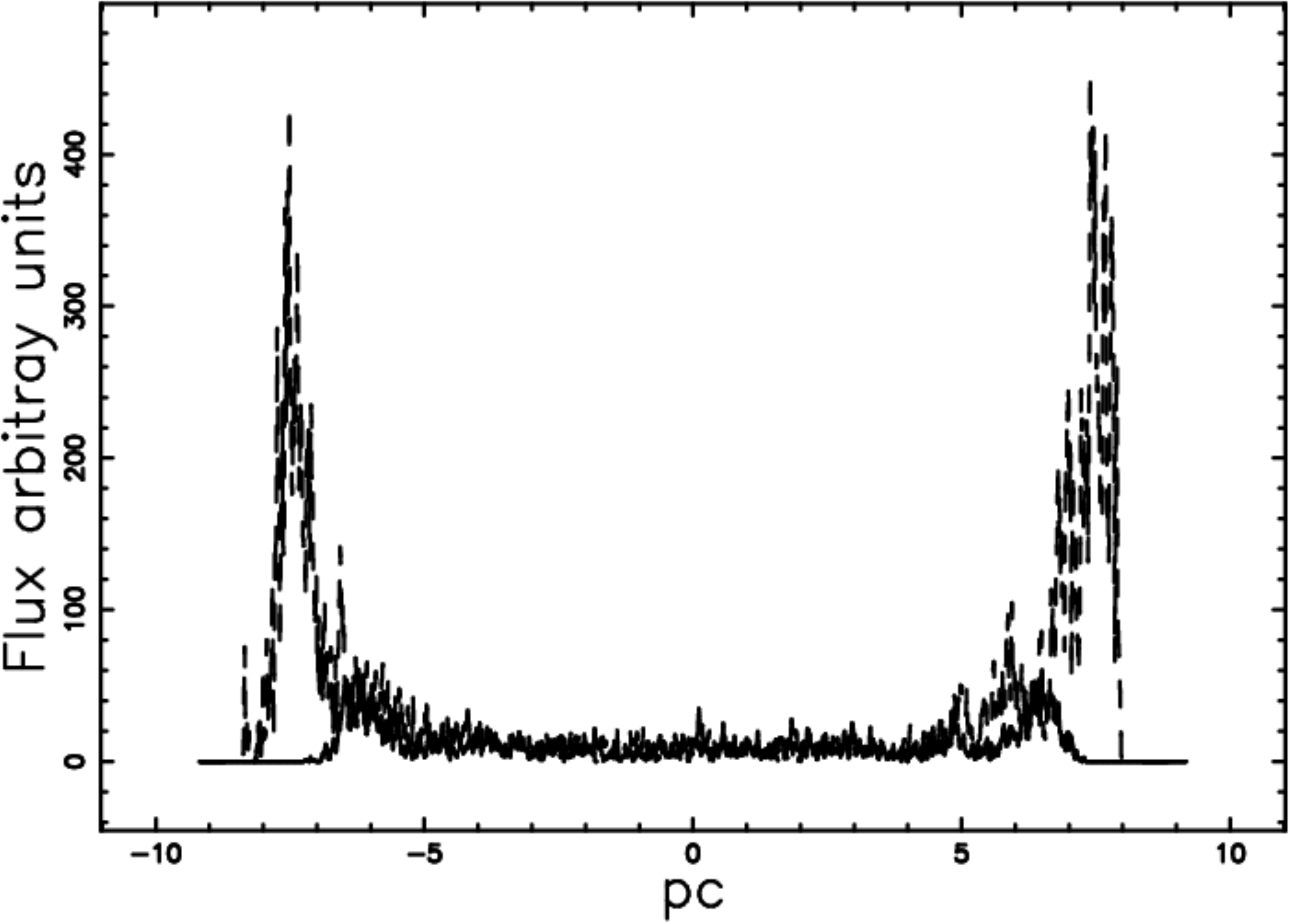}
\end {center}
\caption
{
 Two cut  along  perpendicular lines  of
 {\it I }  for SNR  \sn1006 
 reported  in Figure~\ref{sn1006_colore}~.
Optically thin layer.
}
\label{sn1006_2lines}
    \end{figure*}

Assuming that the flux reversed in the 
non--thermal emission follows a similar law
through the parameter $\chi_X$ ( the efficiency in the X--region)
the effect of velocity is simulated through the following
algorithm.
Once  the IP are  spatially  generated, the number of times 
NTIMES 
over which to repeat the cycle is given
by
\begin{equation}
NTIMES = 1 +NTIMES_{MAX} *
( \frac {v - v_{min}}{v_{max}-v_{min}})^3  
\quad, 
\end {equation}
where  $NTIMES_{MAX}$
 is the maximum  of the allowed values
of  NTIMES minus 1,   and v is  the velocity associated
to each IP.
The asymmetric contour map obtained when
the spatial step is $\approx$ 2 * gyro--radius  
is  reported  in Figure~\ref{sn1006_colore}
and  the cut along two perpendicular lines
of the projection grid in 
Figure~\ref{sn1006_2lines}.
%

\begin{figure*}
\begin{center}
\includegraphics[width=10cm,angle=-90]{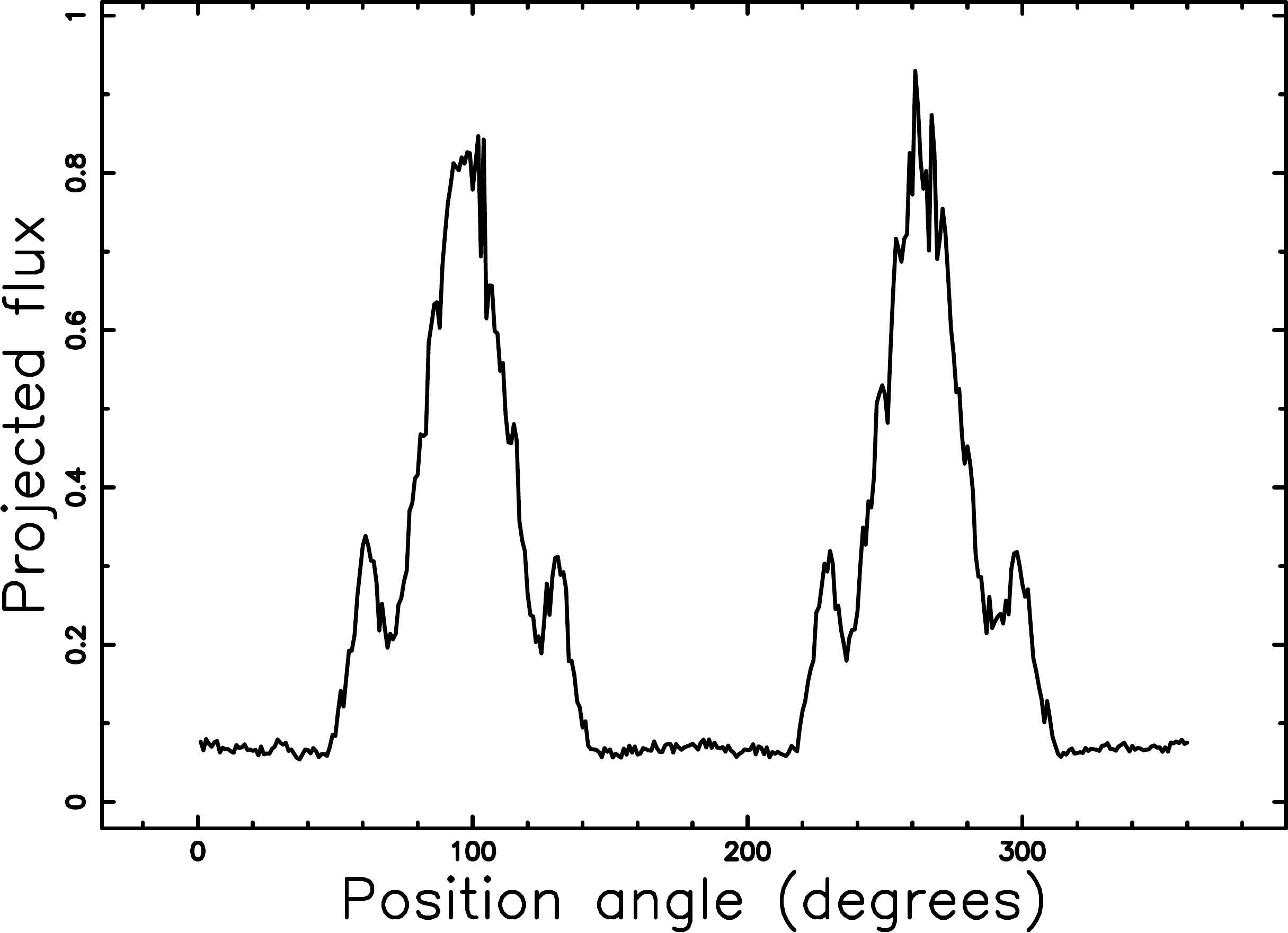}
\end {center}
\caption
{
Azimuthal maximum of the profiles of 
intensity as function of the position angle
in degrees for SNR  \sn1006.
Same  parameters as in Figure~\ref{sn1006_colore}.
}
\label{sn1006_360}
    \end{figure*}

In Figure~\ref{sn1006_2lines}   
the asymmetry both in the peak to peak distance 
and the difference in the two maximum
is evident.
The ratio between the X-ray emission 
in the bright limbs (NE or SW) and toward the 
northwest or southeast  
( at 2 keV) is around 10 , see  Figure 5 top right
in 
\cite{Rothenflug2004}. 
Conversely  our theoretical  ratio , see 
Figure~\ref{sn1006_2lines}, is 9.84.
It is  also  possible  to plot the maximum of  the theoretical 
intensity as  function  of the position angle ,
see  Figure~\ref{sn1006_360}.
The reader can make a comparison
with the observational counterpart
represented by Figure~5 top right, dashed line in 
\cite{Rothenflug2004}.
\label{velocities}

\section{Conclusions}

{\bf Law of motion}
The law  of motion in the case of a symmetric motion 
can be modeled by 
a  power  law  solution ,
the Sedov Solution or the radial momentum
conservation.
These three  models allow to determine the approximate 
age of A39   which is 8710 $yr$ for the Sedov solution and 50000 $yr$
for the radial momentum conservation.
In presence of gradients as given  , for example ,
by an exponential behavior, the solution is deduced 
through the radial momentum conservation.
The comparison with the astronomical data is now more complicated
and the single and multiple efficiency  in the radius determination
have  been introduced.
When , for example ,  MyCn 18   is considered ,
the multiple efficiency over 18 directions is $90.66\%$
when the age of 2000 $yr$ is adopted.
In the  case  of  \etacar  
the multiple efficiency over 18 directions is $85\%$
for a  fixed age of  158 $yr$ in  an exponentially  varying medium.
In the  case  of the weakly asymmetric 
 \sn1006   the  efficiency is  $94.9\%$ in the polar direction 
and  $92.5\%$  in the equatorial direction 
for a  fixed age of  974 $yr$ in   an exponentially  varying medium.

{\bf Diffusion} 

The number density in a thick layer 
surrounding the ellipsoid of expansion can be considered 
constant or variable from a maximum value to a minimum
value with the growing or diminishing radius in respect 
to the expansion position.
In the case of a variable number density the framework
of the mathematical diffusion has been adopted,
see formulas~(\ref{cab}) and ~(\ref{cbc}).
The  case of diffusion with drift has been analytically
solved , see formulas~(\ref{cab_drift}) and ~(\ref{cbc_drift}),
and the theoretical formulas have been compared
with values generated by Monte Carlo simulations.

{\bf Formation of the image}

The intensity of the image of a symmetrical PN  or SNR 
in the case of optically thin medium can be computed
through
an analytical evaluation of lines of sight 
when the number density is constant between two spheres
see formula~(\ref{irim}).
The case  of  a  symmetrical 
diffusive  process  which  is  built in 
presence  of three  spheres 
we  should  distinguish  between  
\begin{itemize}
\item
intensity of emission  proportional
to the square of the number density corresponding to the case of PN ,
see 
formulas (~\ref{I_1}), (~\ref{I_2}) and  (~\ref{I_3}).
\item
intensity of emission  proportional
to  the number density corresponding to the case of SNR,
see 
formulas (~\ref{I_1l}), (~\ref{I_2l}) and  (~\ref{I_3l}).
\end{itemize}

In the case of complex  morphologies  
assuming an optically
thin medium,
it is
possible to make a model image
once two hypotheses are made:
\begin{enumerate}
\item
The thickness of the emitting layer, $\Delta R$,
is the same everywhere
$\Delta R = 0.03 R_{max} $, where $R_{max}$
 is the maximum radius of expansion.
\item
The density of the emitting layer is constant
everywhere
\end{enumerate}
A 2D image  of the
PNs  Ring nebula  and MyCn 18
, 
the  hybrid  Homunculus/\etacar  nebula 
and  the  weakly asymmetric SNR \sn1006  
are 
 shown  respectively  in  Figures 
\ref{ring_heat},
\ref{mycn18_heat},
\ref{eta_heat}  and
\ref{sn1006_colore}.


\end{document}